\documentclass[a4paper,fleqn,usenatbib]{mnras}

\usepackage{amssymb}
\usepackage{amsmath}
\usepackage{graphicx}
\usepackage{natbib}
\usepackage{hyperref}
\usepackage{pdflscape}
\usepackage{longtable}
\usepackage{multicol}
\usepackage{multirow}
\usepackage{url}
\usepackage{graphicx}
\usepackage{color}
\usepackage{txfonts}

\DeclareRobustCommand{\ION}[2]{%
\relax\ifmmode
\ifx\testbx\f@series
{\mathbf{#1\,\mathsc{#2}}}\else
{\mathrm{#1\,\mathsc{#2}}}\fi
\else\textup{#1\,{\mdseries\textsc{#2}}}%
\fi}
\newcommand{\lam}{$\lambda$}

\newcommand{\nii}{[\ION{N}{ii}]}

\newcommand{\oii}{[\ION{O}{ii}]}
\newcommand{\oiii}{[\ION{O}{iii}]}
\newcommand{\sii}{[\ION{S}{ii}]}

\newcommand{\Ha}{$\rm{H}\alpha$}
\newcommand{\Hb}{$\rm{H}\beta$}

\newcommand{\temp}{T$_{\rm e}$}

\title[MZ relation by SAMI]{The SAMI Galaxy Survey: Exploring the gas-phase Mass-Metallicity Relation} 

\author[S.F.S\'anchez et al.]{S.F. S\'anchez$^{1}$, 
J.K. Barrera-Ballesteros$^{2,1}$, 
C. L\'opez-Cob\'a$^{1}$,
S. Brough$^{3,4}$,
J. J. Bryant$^{4,5,7}$,
\newauthor 
J. Bland-Hawthorn$^{4,5}$,
S. M. Croom$^{4,5}$,
J. van de Sande$^{4,5}$,
L. Cortese$^{4,6}$, 
M. Goodwin$^{8}$,
\newauthor
J.S. Lawrence$^{8}$,
A. R. L\'opez-S\'anchez$^{8,9}$,
S. M. Sweet$^{4,10}$,
M. S. Owers$^9$, 
S. N. Richards$^{11}$,
\newauthor 
C. J. Walcher$^{12}$,
and the SAMI Team
\\
$^1$Instituto de Astronom\'ia, Universidad Nacional Aut\'onoma de  M\'exico, A.~P. 70-264, C.P. 04510, M\'exico, D.F., Mexico \\
$^2$Department of Physics \& Astronomy, Johns Hopkins University, Bloomberg Center, 3400 N. Charles St., Baltimore, MD 21218, USA\\
$^3$ School of Physics, University of New South Wales, NSW 2052, Australia\\
$^4$ ARC Centre of Excellence for All Sky Astrophysics in 3 Dimensions (ASTRO 3D)\\
$^5$ Sydney Institute for Astronomy (SIfA), School of Physics, The University of Sydney, NSW, 2006, Australia\\
$^6$ International Centre for Radio Astronomy Research, University of Western Australia, 35 Stirling Highway, Crawley WA 6009, Australia\\
$^7$ Australian Astronomical Optics, AAO-USydney, School of Physics, University of Sydney, NSW 2006, Australia\\
$^8$ Australian Astronomical Optics, AAO-Macquarie, Faculty of Science and Engineering, Macquarie University, 105 Delhi Rd, North Ryde, NSW 2113, Australia\\
$^9$ Department of Physics and Astronomy Macquarie University NSW 2109 Australia\\
$^{10}$ Centre for Astrophysics and Supercomputing, Swinburne University of Technology,PO Box 218, Hawthorn, VIC 3122, Australia \\
$^{11}$  SOFIA Operations Center, USRA, NASA Armstrong Flight Research Center, 2825 East Avenue P, Palmdale, CA 93550, USA\\
$^{12}$ Leibniz-Institut f\"ur Astrophysik Potsdam (AIP), An der Sternwarte 16, D-14482 Potsdam, Germany\\ 
}

\date{Accepted XXX. Received YYY; in original form ZZZ}

\pubyear{2016}

\begin{document}
\label{firstpage}
\pagerange{\pageref{firstpage}--\pageref{lastpage}}
\maketitle

\begin{abstract}
We present a detailed exploration of the stellar mass vs. gas-phase metallicity 
relation (MZR) using integral field spectroscopy data obtained from $\sim$1000 
galaxies observed by the SAMI Galaxy survey. These spatially resolved spectroscopic 
data allow us to determine the metallicity within the same physical scale 
($\mathrm{R_{eff}}$) for different calibrators. The shape of the MZ relations 
is very similar between the different calibrators, while there are large offsets 
in the absolute values of the abundances. We confirm our previous results derived using
the spatially resolved data provided by the CALIFA and MaNGA surveys: (1) 
we do not find any significant secondary relation of the MZR with either the 
star formation rate (SFR) nor the  specific SFR (SFR/M$_*$) for any of the 
calibrators used in this study, based on the analysis of the { individual} 
residuals; (2) if there is a dependence with the SFR, it is weaker than the 
reported one ($r_c\sim -$0.3), it is confined to the low mass regime 
(M$_*<$10$^9$M$_\odot$) or high SFR regimes, and it does not produce any 
significant improvement in the { description of the average population of 
galaxies. The aparent disagreement with published results based on single fiber 
spectroscopic data could be due to (i) the interpretation of the secondary 
relation itself; (ii) the lower number of objects sampled at the low mass 
regime by the current study; or (iii) the presence of extreme star-forming 
galaxies that drive the secondary relation in previous results}.
\end{abstract}
\section{Introduction}

Metals are the product of thermonuclear reactions that occur as stars 
are born and die. They enrich the gas in the interstellar medium, where the 
enrichment is modulated by gas inflows and outflows. Observationally, gas-phase 
oxygen abundance has the largest importance, as the most frequent metal element. 
oxygen is mostly created by core-collapse supernovae associated with star-formation 
events. It produces strong emission lines in the optical wavelength range when
ionized and it is a particularly good tracer of the abundance in the inter-stellar 
medium. Therefore, it is a key element to understand the matter cycle of 
stellar evolution, death and metal production.

For these reasons oxygen has been used as a probe of the evolution of galaxies. 
For example, the presence of a { negative} oxygen abundance gradient in spiral 
galaxies \citep{sear71,comte74} and the Milky Way \citep{peim78}, recurrently 
confirmed with updated observations using larger surveys of galaxies 
\citep[e.g.][]{sanchez13,laura16,belf17,laura18} and \ion{H}{ii} regions in our 
galaxy \citep[e.g.,][]{esteban18}, is one of the key pieces of evidence for the 
inside-out scenario of galaxy growth \citep[e.g.][]{matt89,bois99}. Several 
different scaling relations and patterns have been proposed between the oxygen 
abundance and other properties of galaxies { , like (i) the luminosity 
\citep[e.g.][]{leque79,Skillman:1989p1592}, (ii) the surface brightness 
\citep[e.g.][]{VilaCostas:1992p322,zaritsky94}, (iii) the stellar mass 
\citep[e.g.][]{tremonti04,kewley08} and (iv) the gravitatinal potential 
\citep{eduge18}. Some properties derived from the abundance, like the effective 
yield, correlate with global properties of galaxies too 
\citep[e.g.][]{Garnett:2002p339}. Beyond the integrated or average properties 
of galaxies, the spatially resolved oxygen abundances also present scaling 
relations with the local surface brightness \citep[e.g.][]{P14} and the 
stellar mass density \citep[e.g.][]{rosales12,jkbb16}.} All of them provide 
strong constraints on how galaxies evolve, connecting different products 
of stellar evolution, like stellar mass and luminosity, or tracers of the 
dynamical stage, like velocity and gravitational potential, with oxygen abundance.

A particularly important relation is the mass-metallicity relation
(MZ-relation), since it connects the two main products of stellar
evolution. This relation has been known for decades
\citep[e.g.][]{1992MNRAS.259..121V}, however, it was not explored in
detail using a statistically significant and large sample until more
recently: \citet[][T04 hereafter]{tremonti04} show that these two
parameters exhibit a tight correlation with a dispersion of $\sim$0.1
dex over $\sim$4 orders of magnitudes in stellar mass. This
correlation presents a similar shape at very different redshifts
\citep[e.g.][]{erb06,erb08,henry13,saviane14,salim15}, but shows a clear
evolution that reflects the change of the two involved parameters
with cosmic time \citep[e.g.][]{mari11,mous11}. The relation also 
presents a very similar shape irrespective of the oxygen abundance
calibrator, with an almost linear trend for M$_*<$10$^{10}$M$_\odot$,
and then a bend and flattening towards an asymptotic value for larger
stellar masses \citep[e.g.][]{kewley08}. On the other hand, the normalization 
and scale of the abundances depend strongly on the calibrator
\citep[e.g.,][]{2008ApJ...681.1183K,sanchez17,bb17}. Finally, a similar 
shape is derived independently of whether single aperture or spatially 
resolved spectroscopic data are being used 
\citep[e.g.,][]{2012ApJ...756L..31R,sanchez14,bb17}.

The MZ-relation presents two different regimes that are interpreted in
a very different way. For the low mass regime, the
linear relation between the stellar mass and the oxygen abundance can
be interpreted as the consequence of the star-formation history in
galaxies {, without involving major dry mergers}. Since both
stellar mass and oxygen abundance are the consequence of
star-formation, both of them should grow in a consistent way,
co-evolving. { Under the assumption of no inpflow/outflow of gas, the
so-called {\it closed-box}, the chemical enrichment would be fully
consistent with the integral of the star-formation rate over cosmic
times (star-formation histories), and therefore a linear relation
between oxygen abundance and stellar mass would be expected. It has been 
known for a long time that this simple enrichment model cannot
reproduce the metallicity distribution in the disk of
galaxies \citep[e.g.][]{erb08,belf16a,jkbb18}. Therefore, infall and 
leaking of gas is required \citep[e.g. G-dwarf problem, ][]{2000AmJPh..68...95B}, 
as well as an equilibrium between in-/outflows and the reservoir of gas 
in galaxies \citep[e.g.][]{matt86,lilly13}.}

%

At high masses, in the asymptotic regime, the gas metallicity seems to reach a
saturation value independent of stellar mass. That saturation value 
depends on the adopted calibrator. T04
interpreted that saturation as a consequence of galactic outflows that
regulate the metal content. A priori it was assumed that outflows are
stronger for galaxies with higher star-formation rates, that are at the same 
time also the more massive galaxies (among those forming stars). 
  { This assumption seems to be 
  confirmed observationally \citep{Heckman2001},
  despite the fact that the loading factor decreses with stellar mass
  too \citep[e.g.][]{Peeples2011}. } In this hypothesis an
equilibrium is reached between the oxygen production and the metals
expelled by outflows \citep[e.g.][]{dave11,lilly13,belf16a,wein17}. 
As outflows from galaxies are global processes, this interpretation 
requires that outflows affect the global metallicity 
in galaxies. A caveat would then be that outflows are more frequently 
found in the central regions of galaxies \citep[e.g.][L\'opez-Cob\'a et
  al. submitted]{carlos16}. 

Another interpretation not involving
outflows is that the asymptotic value is a natural consequence of the
maximum amount of oxygen that can be produced by stars, i.e., the
yield. Irrespective of the inflows or outflows of gas, oxygen
abundance cannot be larger than the theoretical limit of production of
this element { at a considered gas fraction (f$_{gas}$). In the
  case of the {\it closed-box} model metallicity is proportional to
  the yield times the natural logarithm of the inverse of
  f$_{gas}$. Therefore, if all gas is consumed in a galaxy
  (f$_{gas}=$0), the oxygen abundance diverges, becoming infinite (in
  essence, not due to the production of more metals, but due to the
  lack of hydrogen). However, all galaxies with measured gas phase
  oxygen abundance have a certain amount of gas, being f$_{gas}$
  clearly non zero. Indeed, \citet{pily07} show that the current
  asymptotic value is compatible with a {\it closed-box} model for a
  f$_{gas}\sim$5-10\%. In the presence of gas flows, the asymptotic
  value would correspond to a different f$_{gas}$, as the theoretical
  yield is modified by the corresponding effective one.} In this
scenario metal enrichment is dominated by local processes, with a
limited effect of the outflows and only requiring gas accretion to
explain not only the global mass-metallicity relation by its local
version, the so-called $\Sigma_*$-Z relation
\citep{rosales12,sanchez13,2016MNRAS.463.2513B}, and even the
abundance gradients observed in spiral galaxies
\citep[e.g.][]{sanchez14,laura16,laura18,poet18}. The detailed shape
of the MZ-relation is therefore an important constraint for the two
proposed scenarios.

In the last decade it has been proposed that the MZ-relation exhibits a 
secondary relation with the star-formation rate (SFR), first reported by 
\citet{elli08}. This secondary relation was proposed (i) as a modification 
of the dependence of the stellar mass with a parameter that includes both 
this mass and the SFR \citep[the so-called Fundamental Mass Metallicity 
relation, or FMR;][]{mann10}; (ii) as a correlation between the three 
involved parameters \citep[the so-called Fundamental Plane of 
Mass-SFR-Metallicity, or FP;][]{lara10a}; or (iii) as a correlation between 
the residuals (after subtraction of the the MZ-relation) with either the SFR 
or the specific star-formation rate, sSFR \citep{2014ApJ...797..126S}. This 
relation was proposed based on the analysis of { single-fiber} spectroscopic 
data provided by the Sloan Digital Sky Survey \citep[SDSS;][]{york2000}.

The existence of this secondary relation is still debated based on the analysis 
of new integral field spectroscopic (IFS) data. The analysis of the data 
provided by the CALIFA survey \citep{sanchez12a} for a sample of 150 
nearby galaxies did not find a secondary relation with the SFR 
\citep{sanchez13}. \cite{2013A&A...550A.115H} found similar results with 
integrated values provided by a drift-scan observational setup. Also, T04 
explored the residuals once subtracted the MZ relation, and they did not find 
a depedency with the EW(\Ha), a tracer of the sSFR \citep[e.g.][]{sanchez13,belf17}. 
This result has been confirmed using larger IFS datasets by \citet{bb17} 
and \citet{sanchez17}. They used nearly $\sim$2000 galaxies extracted from 
the MaNGA survey \citep{manga} and the updated dataset of $\sim$700 galaxies 
provided by the CALIFA survey \citep{sanchez12a}, respectively. More recent 
results, using single aperture spectroscopic data, found that the presence 
of an additional dependence with the SFR strongly depends on the adopted 
calibrator, disappearing for some calibrators \citep[e.g.][]{kash16}, or 
being weaker than previously found \citep[e.g.][]{telf16}. { Finally, 
re-analysis of the IFS dataset by \citet{2014ApJ...797..126S} emphasized the complexity 
of providing the correct parametrization of the Mass-SFR-Metallicity plane, 
finding that the dependence on SFR is strongest for the galaxies with 
highest sSFR.}

In this article we explore the MZ relation and its possible dependence on 
SFR using the IFS data extracted from the on going SAMI galaxy survey 
\citep{sami}. The article is distributed in the following sections: the 
sample of galaxies, a summary of the reduction and the overall dataset 
used in this article are presented in Section \ref{sample}; the analysis 
performed over this dataset is described in Section \ref{sec:ana}, including 
a description of how the different involved parameters (stellar mass, 
star-formation rate and the oxygen abundance from different calibrators) 
are derived; a description of the Mass-Metallicity (MZ) relation for the 
analyzed galaxies is presented in Section \ref{sec:MZ}, exploring the 
proposed dependence with the SFR in Section \ref{sec:residuals}; a detailed 
analysis of the Fundamental Mass-Metallicity Relation \citep{mann10} and 
Fundamental Plane \citep{lara10a} are described in Section \ref{sec:FMR} 
and Section \ref{sec:FP}, respectively. The proposed dependence of the 
residuals after subtraction of the MZ-relation and the residuals of the 
SFR have removal of the dependence with stellar mass are included in Section 
\ref{sec:salim}. Section \ref{sec:dis} includes the discussion of the results 
of all these analyses, with a summary of the conclusions included in Section 
\ref{sec:con}. Throughout this article we assume the standard $\Lambda$ Cold 
Dark Matter cosmology with the parameters: H$_0$=71 km/s/Mpc, $\Omega_M$=0.27, 
$\Omega_\Lambda$=0.73.

\section{Sample and data}
\label{sample}

The selection of the SAMI Galaxy Survey sample is described in detail
in \citet{bryant2015}, with further details in \citet{owers2017}. A 
comparison with the sample selection of other integral field spectroscopic 
surveys was presented in \citet{sanchez17}.

In short, the SAMI Galaxy Survey sample consists of two separate sub-samples:
(i) a sub-sample drawn from the Galaxy And Mass Assembly (GAMA) survey
\citep{driver2011} and (2) an additional cluster sample. The SAMI-GAMA
sample consists of a series of volume-limited sub-samples, in which
the covered stellar mass increases with redshift. It includes galaxies
in a wide range of environments, from isolated up to massive groups,
but it does not contain cluster galaxies. For this reason a second
sub-sample was selected, by selecting galaxies from eight different
galaxy clusters in the same redshift foot-print of the primary sample
(i.e., $z \leq 0.1$) as described in \citet{owers2017}. A stellar mass
selection criteria was applied to the cluster sample, with different
lower stellar mass limits for clusters at different
redshifts. Finally, a small subset of filler targets were included to
maximize the use of the multiplexing capabilities of the SAMI
instrument \citep{sharp06,joss11,bryant2011,sami}.

The sample analyzed here consists of a random sub-set of the foreseen
final sample of SAMI targets (just over 3,000 objects once the survey
is completed). It comprises the { 2307 galaxies observed by August 2017,
included in the internal SAMI v0.10 distribution}. These galaxies cover
the redshift range between 0.005$<z<$0.1, with stellar masses between
$\sim$10$^7$-10$^{11.5}$ M$_\odot$, and with a extensive coverage of
the colour-magnitude diagram, as already shown in \citet{sanchez17}.
{ This sub-sample can be volume corrected following the prescriptions
described in \citet{sanchez18b}, finding that it is complete in the range
between $\sim$10$^{8.5}$-10$^{11.2}$ M$\odot$. }

\subsection{Data reduction}
\label{sec:redu}

The data reduction is described in detailed in \citet{allen2015} and
\citet{sharp2015}, and it is similar to the one adopted for the SAMI
DR1 \citet{green18} and DR2 \citet{scot18}. We present here a brief
summary.

The first steps of the reduction comprise the overscan subtraction,
spectral extraction, CCD flat-fielding, fibre throughput and
wavelength calibration and finally sky subtraction. The result of that
steps is the standard RSS-frame \citep{sanchez06a}. These steps are
performed using the {\sc 2dfDR} data reduction
package\footnote{https://www.aao.gov.au/science/software/2dfdr}.

Then each RSS-frame is spectrophotometricaly calibrated and corrected
for telluric absorption. Finally, the resulting RSS-frames are
combined into a 3D datacube by resampling them onto a regular grid,
with a spaxel size of 0.5$\arcsec$$\times$0.5$\arcsec$. This procedure
involves a spatial registration of the different dithered positions, a
correction of the differential atmospheric refraction and final
zero-point absolute flux recalibration. All these final procedures are
performed using the {\sc SAMI Python} package described in
\citet{allen2014}. The final result of the data reduction is a single
datacube for each observed target and each wavelength regime.

The SAMI observational setup uses the two arms of the spectrograph,
one in the blue, covering the wavelength range between
$\sim$3700\AA\ and $\sim$5700\AA\, with a resolution of $\sim$173 km
s$^{-1}$ (FWHM), and one in the red, covering the wavelength range
between $\sim$6250\AA\ and $\sim$7350\AA, with a resolution of
$\sim$67 km s$^{-1}$ (FWHM). Therefore, the standard data-reduction
produces two different datacubes, with two different wavelength
ranges, spectral resolutions and spectral sampling. In this analysis
we combined those two datacubes into a single cube, following a
similar procedure as that adopted for the CALIFA datacubes
\citep[e.g.][]{DR3}, by (i) degrading the resolution of the red-arm
(R$\sim$4300) datacubes to that of the blue-arm (R$\sim$1800), (ii)
re-sampling the full spectra to a common sampling of 1\AA, and (iii)
correcting them for Galactic extinction, using the attenuation curves
provided by the SAMI data-reduction. { The degradation and
  resampling of the spectra is required for our adopted analysis scheme
  (described below).} The final datacubes cover a wavelength range
between $\sim$3700\AA\ and $\sim$7350\AA, with a {\it gap} between
$\sim$5700\AA\ and $\sim$6250\AA, in a homogeneous way in terms of
spectral resolution. These cubes cover a wavelength range similar to
the one covered by the V500 datacubes of the CALIFA survey 
\citep[e.g.][]{sanchez12a}, apart from the {\it gap}, with a
resolution similar to that of the MaNGA datacubes
\citep[e.g.][]{law15}. The final format of this COMBO datacubes is
similar to the one adopted for the cubes in the CALIFA galaxy survey
\citep[e.g.][]{rgb15}. We make use of those COMBO datacubes in the
current analysis, since we consider that they provide the best
compromise between spectral resolution and largest wavelength coverage
that can be provided by the SAMI dataset. This is particularly
important to optimize the removal of the stellar continuum in the best 
possible way before analysis of the stellar continuum. These combined data 
cubes would not be suitable for kinematic analysis. 

{  The final spectrophotometric precision and accuracy of the analyzed 
data were explored by \citet{green18}. They state clearly that the accuracy 
of the photometry is of the order of 4\%, when compared to that of the SDSS 
imaging survey, with a typical precision of 10\%. This precision
is slightly worse that that derived for other IFS surveys \citep[3-4\%][]{DR3},
however, is partially due to the lower S/N of the SAMI data in the continuum. 
A similar calculation for the SDSS itself indicate that for point-like sources 
the typical precision is of $\sim$4\%, without knowing the exact value for 
extended targets \footnote{\url{https://www.sdss.org/dr14/algorithms/spectrophotometry/}},
although it is expected to be larger. This precision should not affect significantly
any of the parameters analyzed in the current study, as reported based on the comparison
between different IFU surveys in Sec. \ref{sec:results}. If any, it may increase 
the dispersion in the reported correlations, but not in a large way, as shown in 
the present article. From the original observed sample sample we exclude
53 galaxies for which there was an obvious error in the spectrophotometric
calibration between the blue and red arms, and/or they present a S/N clearly 
lower than the average. Due to the limited number of excluded galaxies, that 
exclusion should not affect the results.}

\section{Analysis}
\label{sec:ana}

The analysis of the datacubes was done making use of {\sc Pipe3D}\citep{Pipe3D_II}, 
a pipeline designed to (i) fit the stellar continuum with synthetic population 
models and (ii) extract the information from the ionized gas emission lines of 
Integral Field Spectroscopy (IFS) datacubes. {\sc Pipe3D} uses {\sc FIT3D } 
algorithms as the basic fitting routines 
\citep{Pipe3D_I}\footnote{\url{http://www.astroscu.unam.mx/~sfsanchez/FIT3D/}}. 
The adopted implementation of {\sc Pipe3D} analyzes the stellar populations by 
decomposing them using the GSD156 single stellar population (SSP) library. This 
library, first described in \citet{cid-fernandes13}, comprises 156 SSP templates, 
that sample 39 ages (1 Myr to 14 Gyr, in an almost logarithm scale), and 4 
different metallicities (Z/Z$\odot$=0.2, 0.4, 1, and 1.5), adopting the Salpeter 
Initial Mass Function \citep[IMF][]{Salpeter:1955p3438}. These templates have 
been extensively used in previous studies 
\citep[e.g.][]{perez13,rosa14,ibarra16,sanchez18,sanchez18b}. { The degradation 
and resampling of the spectra described in the previous section was therefore 
mandatory since (i) the resolution of the adopted SSP library is worse than that 
of the red arm, and (ii) we require to fit both wavelength ranges 
simultaneously to find a good overall model.}

We include here a brief summary of the procedure. For further details, 
including the adopted attenuation law, and uncertainties of the overall process 
and the derived parameters we refer the reader to \citet{Pipe3D_I,Pipe3D_II}. 
First, spatial binning is performed in each datacube to reach a
goal S/N of 20 across the FoV. This is slightly lower than the S/N requirement 
adopted by {\sc Pipe3D} in the analysis of the CALIFA and MaNGA datasets, and 
it is tuned due to the lower S/N of the SAMI data in the stellar continuum. 
Then, the stellar continuum of the co-added spectra corresponding to each 
spatial bin/voxel was fitted with a multi-SSP model. Once the best model for the stellar 
population in each bin was derived, the model is adapted for each spaxel. This is 
done by re-scaling the model in each bin to the continuum flux intensity at the 
considered spaxel, as described in \citet{cid-fernandes13} and \citet{Pipe3D_I}. 
Finally, the stellar mass density is derived for each spaxel using this model, 
following the procedures outlined in \citet{mariana16}. The integrated stellar 
mass is derived for each galaxy by co-adding the stellar mass density within 
the FoV of the corresponding datacube. The typical error on the integrated 
stellar mass, as estimated by \citet{Pipe3D_II}, is of the order of $\sim$0.15 dex. 
No aperture correction was applied to the provided stellar masses, that are 
therefore limited to the 16$\arcsec$/diameter aperture of the SAMI datacubes. 
However, we should note that a direct comparison with the stellar masses provided 
by the SAMI survey, based on non-aperture limited broad-band photometry 
(Scott et al., submitted), shows almost a one-to-one relation with a standard 
deviation of $\sim$0.18 dex, once corrected for the differences in cosmology 
and IMFs. We found 57 outliers, for which stellar masses deviate more than 
2$\sigma$ from the average distribution. Those galaxies have been removed 
from the analysis. As a sanity check, we repeated the full analysis shown 
along this article using the photometric-based stellar masses without any 
significant change of the results. Therefore, aperture effects are not 
affecting our analysis in a significant way. A comparison between both 
estimates of the stellar mass is included in Appendix \ref{sec:Mass}.

Once performed the analysis of the stellar population we derive the required 
parameters (flux intensity, equivalent width and kinematic properties) for 
the ionized gas emission lines. To that end we create a cube that comprises 
just the information from these emission lines by subtracting the best fit 
stellar population model, spaxel-by-spaxel, from the original cube. This 
pure gas emission cube inherits the noise and the residuals from the best 
fit multi-SSP model, too. These residuals are propagated to the errors of 
the pure gas cube. Finally, the considered parameters for each individual 
emission line within each spectrum at each spaxel of this cube are extracted 
using two different procedures: (i) a parametric fitting, assuming a single 
Gaussian function, and (ii) a weighted momentum analysis, following the 
procedures described in \citet{Pipe3D_II}. { This 2nd procedure is as 
standard as the previous one, being broadly adopted in the analysis of 
emission lines in particular in radio astronomy and even in the extraction 
of multi-fiber spectroscopic data. It is based on a pure momentum or 
summation of the flux intensities, weighted by an assumed guess shape, and 
an iteration to converge. Both procedures produce similar results, in 
particular for star-forming regions. However, the second one is more general, 
since it requires less assumptions (i.e., the shape of the emission line 
profile), and produces better results in highly assymmetric emission lines.} 
In this particular analysis we adopted the results from the second procedure. 
More than 50 emission lines are included in the analysis, however, for 
the current study we make use of the flux intensities and equivalent 
widths of the following ones: \Ha, \Hb, \oii\ \lam3727,
\oiii\ \lam4959, \oiii\ \lam5007, \nii\ \lam6548,
\nii\ \lam6583, \sii \lam6717 and \sii \lam6731. None of the considered emission 
lines are located in the spectral {\it gap} of the COMBO datacubes for any of 
the galaxies analyzed here, due to the redshift range of the sample. The final 
product of this analysis is a set of maps comprising the spatial distributions 
of the emission line flux intensities and equivalent widths. These maps are 
then corrected for dust attenuation. We derive the spatial distribution of 
the \Ha/\Hb\, ratio. Then we adopt 2.86 as the canonical value 
\citep{osterbrock89} and use a Milky-Way-like extinction law \citep[][ with 
R$_{\rm V}$=3.1]{cardelli89}. For spaxels with values of the \Ha/\Hb\ ratio 
below 2.86 we assume no dust attenuation. In any case these spaxels represent 
just 2-3\%\ of the total number. The SAMI survey also provides a different 
estimate of the emission line fluxes, produced by the LZIFU pipeline 
\citep[][]{ho2016b}. The differences between the two procedures and 
the effects on the considered emission lines are discussed in 
Appendix \ref{sec:LZIFU}.

\begin{figure*}
 \minipage{0.99\textwidth}
 \includegraphics[width=\linewidth]{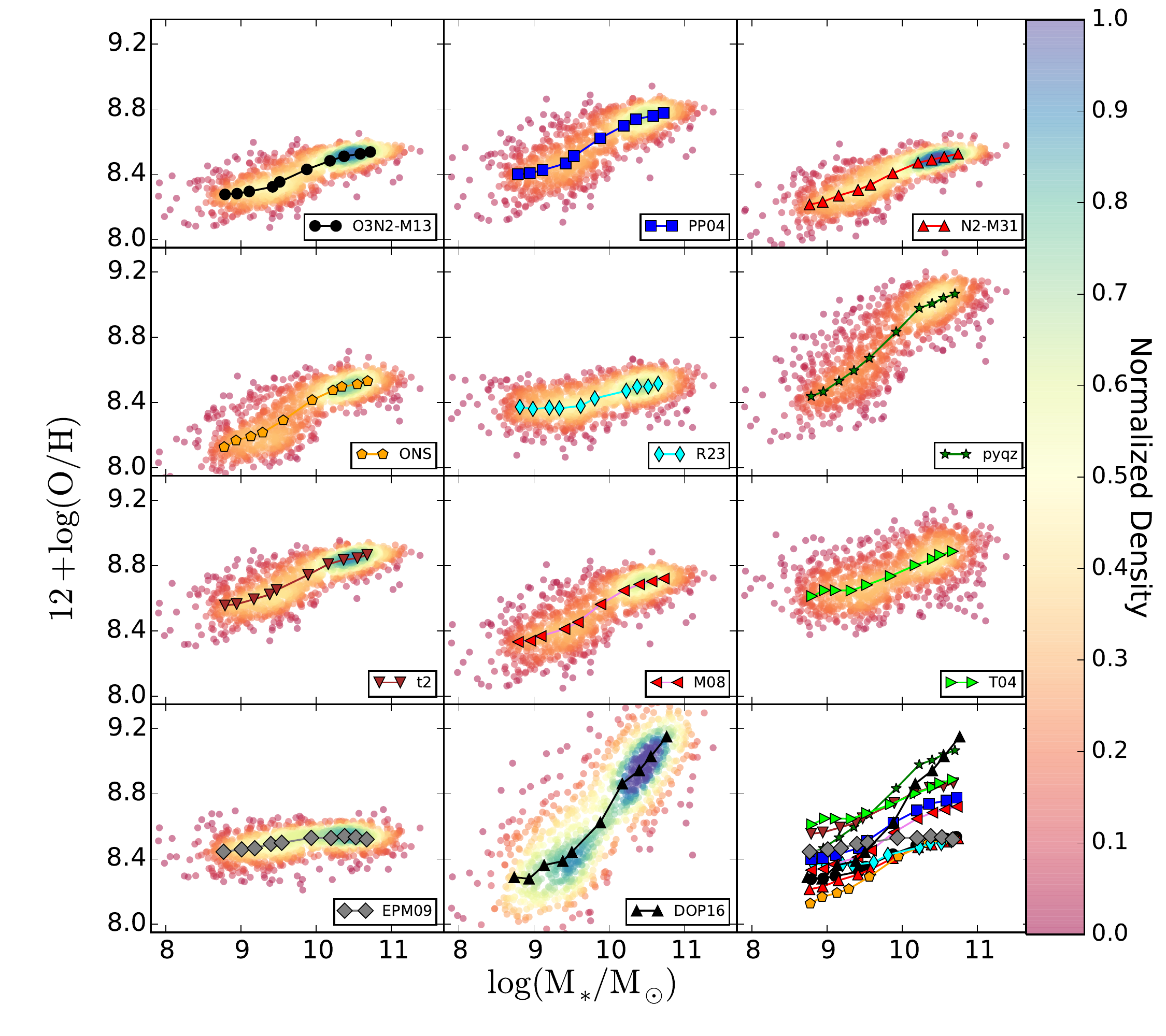} 
 \endminipage
\caption{Mass-metallicity relation for the set of eleven oxygen abundance calibrators used in the present study for the sample of 1044 galaxies extracted from the SAMI galaxy survey. Each calibrator is shown in an individual panel, with the adopted name for the calibrator written in the corresponding inset. The colored solid circles in each panel indicate the individual values of the stellar mass and oxygen abundance at the effective radius, color-coded by the density of points (with blue meaning more dense areas, and red less dense ones). The single-color symbols connected with a solid-black line represent the median value of the considered parameters at a set of bins described in the text. For comparison purposes, the median values for all the considered calibrators are included in the bottom-right panel. The different calibrators are described in the text.} 
 \label{fig:MZ}
\end{figure*}

The usual oxygen abundance calibrators can only be applied for those spaxels in 
which ionization is dominated by star-formation (i.e, young stars). To select 
the spaxels compatible with this ionization we follow \citet{mariana16} 
and \citet{laura18}, excluding those regions above the \citet{kewley01} demarcation 
line in the \oiii/\Hb\, vs \nii/\Ha\ diagnostic diagram \citep{baldwin81} 
and all regions with equivalent width of \Ha\ lower than 6\,\AA \citep{sanchez14}. 
We consider that these combined criteria exclude most of the possible 
contaminators to the ionization, selecting regions in which ionization is 
compatible with young stars \citep{lacerda18}. The spatial distribution of 
the luminosity of \Ha\ is obtained by considering the comological distance, 
and applying the corresponding transformation to the dust corrected \Ha\ 
intensity maps. Finally, the SFR, spaxel-by-spaxel, was estimated based on 
the \citet{kennicutt98} calibration \citep[using the ][IMF]{Salpeter:1955p3438}. 
Like in the case of the stellar mass, the integrated SFR was estimated by 
co-adding it across the entire FoV of the datacube for each galaxy. Therefore, 
the SFR is restricted to the aperture of the SAMI COMBO cubes too. No 
aperture correction was applied.

The SFR was derived without using the strict selection criterion outlined before, 
in particular the EW(\Ha) cut. Therefore, our derivation of the SFR includes the 
contribution of different sources of ionization, even the diffuse gas. It is 
known that for retired galaxies or regions in galaxies 
\citep[e.g.][]{stas08,sign13,mariana16} the diffuse gas is most probably ionized 
by old stars, either post-AGBs \citep[e.g.][]{sarzi10,sign13,2016A&A...586A..22G} 
or HOLMES \citep[e.g.][]{mori16}. However, there could be a non negligible
contribution by photon leaking from star-forming areas, although its relevance is 
still not clearly determined \citep[e.g.][]{rela12,sanchez14,mori16}. Both 
contributions may affect the SFR estimation. In the case of the post-AGB/HOLMES 
ionization, its contribution to the total \Ha\, luminosity (and to the SFR) is 
of a few percent, based on the location of retired galaxies in the sequence 
of star-forming galaxies \citep[e.g.][]{2015A&A...584A..87C,mariana16}. This was 
recently confirmed by Bitsakis et al. (submitted). On the other hand, the 
contribution of leaked photons from \ION{H}{ii} should be in principle added 
to the total budget, to reproduce correctly the SFR. Therefore, applying the 
severe cut described before, that guarantees a correct estimation of the oxygen 
abundance, could produce a bias in the derived SFR. Finally, we should indicate 
that a signal-to-noise cut of 3$\sigma$ was applied in the \Ha\, fluxes derived 
spaxel-by-spaxel. This cut was relaxed to 1$\sigma$ for the remaining emission 
lines. The former cut ensures a reliable detection of ionized gas in a spaxel, 
and the later one imposes a not very restrictive limit in the error of the d
erived parameters. Errors are properly propagated in the derivation of the 
oxygen abundances and SFRs, based on a Monte Carlo iteration. Although the 
cut in S/N may seem not sufficiently restrictive on a spaxel-by-spaxel basis, 
we should remind the reader here that the final parameters derived for each 
galaxy involve hundreds of spaxels, and therefore, the final errors are very 
small in general. Final errors are dominated by the systematics in the adopted 
calibrators rather than by errors in the individual measurements.

\begin{table*}
\caption[Fitted parameters for the MZR and its scatter.]{Best fitted parameters for the two functional forms adopted to characterize the MZR and its scatter for the different estimations of the oxygen abundance analyzed here. For each different calibrator the following is listed (i) the standard deviation of the original set of oxygen abundance values ($\sigma_{log(O/H)}$); (ii) the fitted parameters $a$ and $b$ in Eq.\ref{eq:fit} to the MZR;  (iii) the standard deviation ($\sigma$ MZR) of the residuals after subtracting the best fitted curve from the oxygen abundances; and (iv) the coefficients of the polynomial function adopted in Eq. \ref{eq:poly}, defined as the pMZR relation, together with the $\sigma$ pMZR, the standard deviation of the residuals once subtracted the best polynomial function. The third decimal in the parameters is included to highlight any difference, despite the fact that we do not consider that any value beyond the 2nd decimal could be significant.}
\label{tab:val}
\begin{tabular} {c c c c c c c c c c }
\hline
Metallicity   & $\sigma_{log(O/H)}$& \multicolumn{2}{c}{MZR Best Fit} & $\sigma_{\rm MZR}$ & \multicolumn{4}{c}{pMZR Polynomial fit} &  $\sigma_{pMZR}$\\
 Indicator & (dex) & $a$  & $b$ & (dex) & $p_0$  & $p_1$ & $p_2$ & $p_3$  & (dex) \\
\cline{3-4}
\cline{6-9}
O3N2-M13 & 0.120  & 8.51 $\pm$ 0.02 & 0.007 $\pm$ 0.002 & 0.102 &  8.478 $\pm$ 0.048 & -0.529 $\pm$ 0.091 & 0.409 $\pm$ 0.053 & -0.076 $\pm$ 0.010 & 0.077 \\
PP04     & 0.174  & 8.73 $\pm$ 0.03 & 0.010 $\pm$ 0.002 & 0.147 &  8.707 $\pm$ 0.067 & -0.797 $\pm$ 0.128 & 0.610 $\pm$ 0.074 & -0.113 $\pm$ 0.013 & 0.112 \\
N2-M13   & 0.133  & 8.50 $\pm$ 0.02 & 0.008 $\pm$ 0.001 & 0.105 &  8.251 $\pm$ 0.047 & -0.207 $\pm$ 0.088 & 0.243 $\pm$ 0.051 & -0.048 $\pm$ 0.009 & 0.078 \\
ONS      & 0.168  & 8.51 $\pm$ 0.02 & 0.011 $\pm$ 0.001 & 0.138 &  8.250 $\pm$ 0.083 & -0.428 $\pm$ 0.159 & 0.427 $\pm$ 0.093 & -0.086 $\pm$ 0.017 & 0.101 \\
R23      & 0.102  & 8.48 $\pm$ 0.02 & 0.004 $\pm$ 0.001 & 0.101 &  8.642 $\pm$ 0.076 & -0.589 $\pm$ 0.150 & 0.370 $\pm$ 0.092 & -0.063 $\pm$ 0.018 & 0.087 \\
pyqz     & 0.253  & 9.02 $\pm$ 0.04 & 0.017 $\pm$ 0.002 & 0.211 &  8.647 $\pm$ 0.088 & -0.718 $\pm$ 0.171 & 0.682 $\pm$ 0.101 & -0.133 $\pm$ 0.019 & 0.143 \\
t2       & 0.139  & 8.84 $\pm$ 0.02 & 0.008 $\pm$ 0.001 & 0.115 &  8.720 $\pm$ 0.065 & -0.487 $\pm$ 0.124 & 0.415 $\pm$ 0.072 & -0.080 $\pm$ 0.013 & 0.087 \\
M08      & 0.206  & 8.88 $\pm$ 0.03 & 0.010 $\pm$ 0.001 & 0.169 &  8.524 $\pm$ 0.070 & -0.148 $\pm$ 0.134 & 0.218 $\pm$ 0.080 & -0.040 $\pm$ 0.015 & 0.146 \\
T04      & 0.150  & 8.84 $\pm$ 0.02 & 0.007 $\pm$ 0.001 & 0.146 &  8.691 $\pm$ 0.102 & -0.200 $\pm$ 0.204 & 0.164 $\pm$ 0.126 & -0.023 $\pm$ 0.024 & 0.123 \\
EPM09    & 0.077  & 8.54 $\pm$ 0.01 & 0.002 $\pm$ 0.001 & 0.074 &  8.456 $\pm$ 0.044 & -0.097 $\pm$ 0.085 & 0.130 $\pm$ 0.051 & -0.032 $\pm$ 0.010 & 0.071 \\
DOP09    & 0.348  & 8.94 $\pm$ 0.08 & 0.020 $\pm$ 0.004 & 0.288 &  8.666 $\pm$ 0.184 & -0.991 $\pm$ 0.362 & 0.738 $\pm$ 0.217 & -0.114 $\pm$ 0.041 & 0.207 \\
\hline
\end{tabular}
\end{table*}

We consider a set of 11 calibrators in our derivation of the oxygen abundance 
spaxel-by-spaxel. This approach is similar to the one described in \citet{sanchez17} 
and \citet{bb17}, and it is aimed to reduce and explore the systematics in the 
results due to the differences in the adopted calibrators \citep[e.g.][]{poet18}. 
For doing this calculations we used the dust-corrected intensity maps of the 
ionized gas emission lines indicated above. To characterize the oxygen abundance 
of each galaxy we estimate the value at the effective radius $\rm{R_{eff}}$. 
\citet{sanchez13} demonstrated that this value agrees with the average abundance 
across the entire optical extent of the considered galaxies (at least up to 2.5 
$\rm{R_{eff}}$), within  $\sim$0.1 dex. Even more, when dealing with aperture 
limited IFS data, that do not cover the full optical extent of galaxies, like 
in the current dataset, this value is more representative of the average 
abundance than the mean value across the FoV, since this later one is biased 
towards the values in the inner regions. This oxygen abundance is derived by 
means of a linear regression to the oxygen abundance gradient, derived by 
azimuthal averages following concentric elliptical rings. 
The inner region ($\rm{R<0.5 R_{eff}}$) was excluded based on the deviations 
from the mean radial gradient described in several previous studies 
\citep{sanchez14,laura16,belf17,laura18}. The effective radius was extracted 
from the SAMI catalog \citep{bryant2015}, measured from the SDSS $r$-band 
images when available ($\sim$80\% of the targets). For the remaining galaxies we
estimated them from synthetic $r-$band images created from the datacubes, as 
the elliptical aperture (considering the position angle and ellipticity of 
the galaxies) encircling half the flux. A comparison between the values for 
the 80\% of the galaxies for which the two estimates are available indicate 
that they agree within a standard deviation of 20\%. { Only  27\% of the 
SAMI galaxies are covered up to 2.0$\rm{R_{eff}}$, and in general the fitted 
regime covers the range between 0.5$\rm{R_{eff}}$ and the limit of the FoV. 
This limit correponds to 1.7$\rm{R_{eff}}$ on average, with 71\% of the galaxies 
covered up to 1.5$\rm{R_{eff}}$ (and all of them up to 1$\rm{R_{eff}}$). All 
SAMI galaxies have an effective radius larger than the PSF size, and in only 
14\% the PSF is larger than 0.5$\rm{R_{eff}}$. Therefore, the PSF size is 
not an issue in the derivation of the abundance gradients in the current dataset. }

Finally we could estimate the oxygen abundance at the $\rm{R_{eff}}$, fulfilling 
the different criteria indicated before, for a total of 1044 objects, that 
comprise our final sample. As expected most of these galaxies are late-type, 
star-forming galaxies, since we have selected objects that have starforming 
regions within their optical extension. 

There is a long standing debate about the absolute scale of the oxygen abundance, 
described in detail in the literature \citep[e.g.][]{kewley08,angel12,sanchez17}, 
that is beyond the scope of the current study. To minimize the biases of 
selecting a particular abundance calibrator and to describe the shape of the 
MZ distribution in the most general way we did not adopt a single oxygen 
abundance calibrator, as indicated before, but a set of eleven different ones. 
They were selected to be representative of four different families of calibrators: 
(i) those based or anchored to estimations done using the {\it Direct Method} 
\citep[DM hereafter, e.g.][]{epm17}; (ii) those that consider the inhomogeneities 
in the electron temperature in ionized nebulae, known as the $t_2$ correction 
\citep{peim67}; (iii) those calibrators that use the values anchored to the 
DM only for the low-abundance regime, using values derived form photoionization 
models for the high-abundance regime (i.e. mixed calibrators); and finally 
(iv), those anchored to the predictions by photoionization models. The maximum 
values of the oxygen abundances and the dynamical range increases from the 
first ones (based on the DM) to the last ones (based on photoionization 
models). To represent the first family we included the calibrators based on 
the O3N2 and N2 indicators, described by \citet[][O3N2-M13 and N2-M13 
hereafter]{marino13}, the calibrator based on R23 \citep{2004ApJ...617..240K}, 
following the parametrization by \citet{rosales11}, modified to produce 
abundances that match those ones derived by the DM (R23 hereafter). The 
ONS calibrator \citep{pilyugin10}. And finally, the version of the O3N2 
calibrator that corrects for relative abundance from nitrogen to oxygen 
\citep[][EPM09 hereafter]{epm09}. To consider the second family, we applied 
the $t_2$ correction proposed by \citet{2012ApJ...756L..14P}, for the average 
oxygen abundance estimated from the four calibrators included in the first 
family. The third family of calibrators are represented by the calibrator 
based on the O3N2 indicator by \citet{pettini04} (PP04 hereafter) and the 
one based on a combination or R23 and the previous one proposed by 
\citet{maio08} (M08 hereafter). Finally, those based on photo-ionization 
models only are represented by three calibrators: (a) The oxygen abundance 
provided by the {\tt pyqz} code \citep{vogt15}, with the following set 
of emission line ratios: O2, N2, S2, O3O2, O3N2, N2S2 and O3S2, as described 
by \citet{dopita13} (pyqz hereafter); (b) The calibrator described by 
\citet{dopita16cal}, based on the N2/S2 and N2 line ratios (DOP16 hereafter); 
and finally, (c) the calibrator based on R23 described in \citet{tremonti04} 
(T04 hereafter). Table \ref{tab:val} presents the full list of adopted 
calibrators described here. For more details on the adopted calibrators 
we refer the reader to \citet{sanchez17}.

We did not try to be complete in the current exploration of the possible 
oxygen abundance calibrators. However, we consider that we have a good 
representative family of the different types of calibrators, thus minimizing 
the effects of selecting a particular one or a particular type or family 
of calibrator.

\begin{figure*}
\minipage{0.99\textwidth}
\includegraphics[width=\columnwidth]{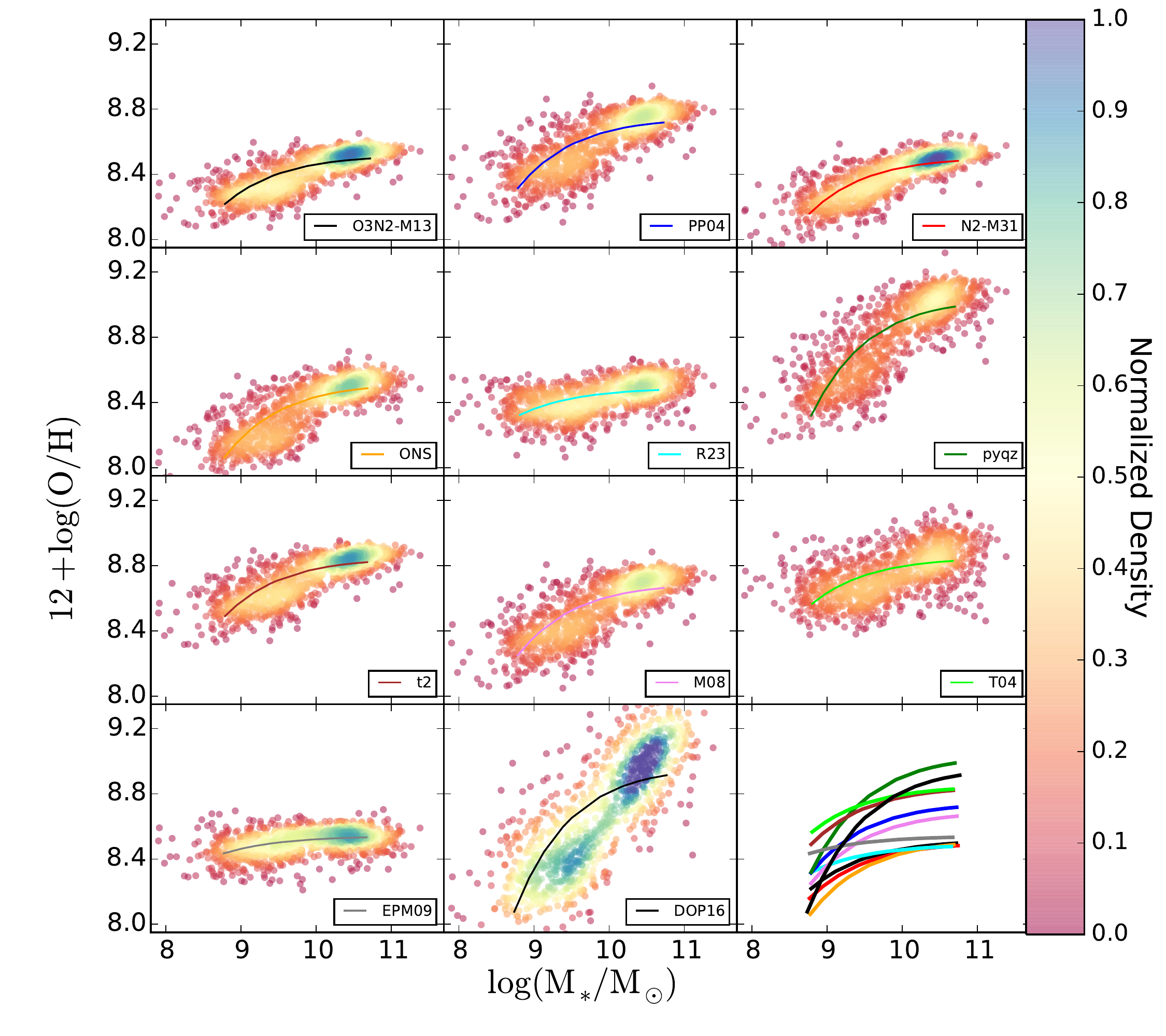}
\endminipage
\caption{Best fit MZR model for the different analyzed calibrators. The colored solid circles, in each panel, correspond to the values derived using each abundance estimator. Colors indicate the density of points. The solid lines with different colors represent the best fit models, with each color corresponding to one of the considered abundance calibrators. The best fit models for all calibrators are shown in the bottom-right panel for comparison purposes.}
\label{fig:FIT}
\end{figure*}

\begin{figure*}
\minipage{0.99\textwidth}
\includegraphics[width=\columnwidth]{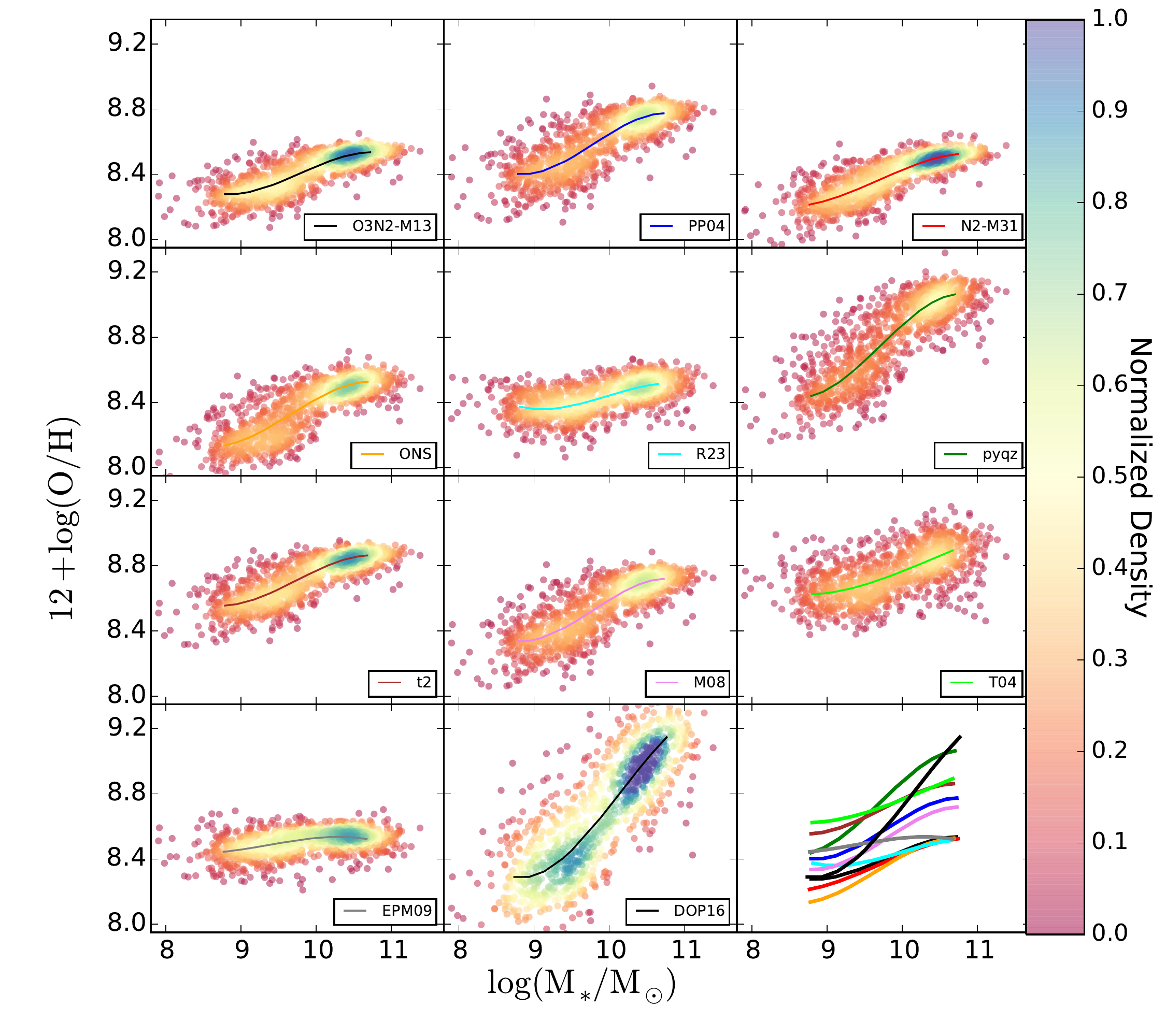}
\endminipage
\caption{Best fit $p$MZR model for the different analyzed calibrators. The colored solid circles, in each panel, correspond to the values derived using each abundance estimator. Colors indicate the density of points. The solid lines with different colors represent the best fit models, with each color corresponding to one of the considered abundance calibrators. The best fit models for all calibrators are shown in the bottom-right panel for comparison purposes.} 
\label{fig:pFIT}
\end{figure*}

\section{Results}
\label{sec:results}

In the previous section we described how we extracted the relevant
parameters involved in the current exploration (stellar masses,
star-formation rates and the characteristic oxygen abundance) for the
1044 galaxies extracted from the SAMI survey explored in here. In the
current section we describe the shape of the MZ relation for the
present dataset and its possible dependence on the star-formation
rate.

\subsection{The MZ relation}
\label{sec:MZ}

In Fig.\ref{fig:MZ} we show the distribution of individual oxygen abundances over 
stellar mass, together with the median abundances in different stellar mass bins 
for our set of calibrators. 
There is a clear increase of the oxygen abundance with the stellar mass for 
$\rm M>10^{9.5}\,M_{\odot}$, reaching an asymptotic value for more massive 
galaxies \citep[the {\it equilibrium} value in the nomenclature of][]{belf15}. 
For masses below $\rm M<10^{9.25}\,M_{\odot}$, contrary to the results found 
in previous studies \citep[e.g.][]{tremonti04}, we do not find a steady decline 
in the oxygen abundance. Instead there seems to be a plateau that was not 
observed in previous spatially resolved analysis either \citep[e.g.][]{sanchez17,bb17}. 
It is important to note that in those studies there was a lack of statistics at 
low stellar masses, with the samples only being complete at $M>10^{9-9.5}\,M_{\odot}$, 
in both cases. In the case of SAMI we cover lower stellar masses, as shown by 
\citet{sanchez17}. 

It is clear that the absolute scale of the MZR depends on the adopted
calibrator, as already described in many previous studies
\citep[e.g.][]{kewley08,angel12,sanchez17,bb17}. In general,
calibrators based on the DM cover a smaller range of abundances
(thus, smaller scatter) and lower values of the abundance on average than
those based on photoionization models. Finally, mixed calibrators lie
in between both of them. Indeed, as already noticed by
\citet{sanchez17} and \citet{bb17}, the dispersions around the average
value in each mass bin are significantly larger for
photoionization-model based calibrators, making them less suitable to
explore possible secondary relations with other parameters if this
dispersion is intrinsic to the calibrator itself.  Finally, the 
{\it t2} correction \citep{peim67} modifies the abundances based on the DM,
increasing their values such that they are similar to those from photoionization
models. This does not imply that the {\it t2} makes the
two abundances compatible. { Some authors have suggested that
a non-thermalized distribution of electrons in the nebulae,
a physical process that produces similar results as a {\it t2} distribution,
could make both derivations of the oxygen abundance compatible 
\citep[e.g.][]{nich12,nich13,dopita13}. This was refuted recently by 
\citet{ferland16}. We consider that the nature of the {\it t2} correction} is
different than the reason why the abundances based on the direct
method and those based on photoionization models disagree {
\citep[following][]{mori16}. Therefore, from our point of view } 
it is a pure numerical coincidence { that both abundances agree.}

We derive the median abundances within stellar mass bins of a minimum of 
0.25 dex, ranging between $10^{8.5}\,M_{\odot}$ and $10^{11}\,M_{\odot}$, 
as shown in Fig. \ref{fig:MZ}. Due to the non-homogeneous sampling of the 
stellar masses by the current dataset, the size of the stellar mass bins 
is increased from the minimum value until there are at least 50 galaxies 
per bin. Then, we fit them adopting the same functional form used by 
\citet{sanchez13}, \citet{sanchez17} and \citet{bb17}, to characterize the MZR:
\begin{equation}
y=a+b(x-c)\exp(-(x-c))
\label{eq:fit}
\end{equation}
where $y = 12+\log(\mathrm{O/H})$ and $x=\log(M_*/ M_{\odot})-8.0$. 
This relation is physically motivated, describing a scenario in which 
the oxygen abundance rises almost linearly with stellar mass until a 
certain asymptotic value. The fitting coefficients, $a$, $b$ and $c$ 
represent the asymptotic value of the metallicity at high masses, the 
strength of the curvature and, finally the stellar mass at which the 
metallicity reaches the asymptotic value. Following \citet{sanchez17} 
and \citet{bb17}, $c$ is fixed (to 3.5 in this case). This implies that 
all calibrators reach the asymptotic value at the same stellar mass 
of $M_*=10^{11.5}$ $M_{\odot}$ (a reasonable assumption based on the 
exploration of the distributions). Changing this value within a range 
between 3-5 does not significantly modify the results. It would affect 
the actual value of the $b$ parameter. However the shape of the relation 
and the dispersion around it is not affected. Table \ref{tab:val} lists 
the estimated parameters for the different calibrators based on the fitting 
procedure. As indicated before, \temp-based calibrators present the lowest 
values for the asymptotic metallicity. The curvature depends slightly on 
the adopted calibrator, being larger for the {\sc pyqz} and DOP16 calibrators, 
both based on photoionization models, which are the ones that present larger 
dispersions around the mean distribution. In general the values of this 
parameter are similar to the ones reported in the literature for different 
calibrators \citep[e.g.][]{sanchez13,sanchez17,bb17}. In particular, 
the asymptotic metallicities agree within 2$\sigma$ with the reported 
values for both CALIFA \citep[][Tab. 1]{sanchez17} and MaNGA 
\citep[][ Tab. 1]{bb17}, for those calibrators that are in common. We 
illustrate the similarities between the MZ-relations reported for the 
three different IFS galaxy surveys with Figure \ref{fig:comp_IFS}, discussed 
in Appendix \ref{sec:comp_IFS}. This result stresses the importance of using 
the same fitting procedures, measure the stellar masses and abundances in a 
consistent way, and adopting the same calibrators, when performing this kind 
of comparison.

Fig. \ref{fig:FIT} shows the best fit MZR model together with the distributions of 
data points for the different calibrators, as shown in Fig. \ref{fig:MZ}. It 
is clear that this first adopted functional form, although it is physically 
motivated, cannot reproduce in detail the shape of the observed distribution 
for some of the calibrators. In particular, it cannot reproduce the possible 
flattening observed at low stellar masses in the oxygen abundance distribution. 
{ In general, this flattening was not reported neither in previous analysis 
using IFS galaxy surveys \citep[e.g.][]{sanchez13,bb17,sanchez17}, nor in studies 
based on single-fiber spectroscopic datasets \citep[e.g.][]{tremonti04,mann10,lara13}, 
not even in those that explore the MZR in the low mass regime 
\citep[e.g.][]{lee06,wien13}. It is not obvious in all calibrators neither. 
Therefore, we consider that it could be a combined effect of sample selection, 
S/N cut in line intensities, and their effect in the calibrators. Indeed, for 
some calibrators this shape was reported by previous studies using SDSS 
spectroscopic data \citep[e.g.][]{kewley08}. In any case, irrespectively of 
its origin, not taking it into account} may introduce an artificial dispersion 
around the mean value. We repeat the analysis using a 4th-order polynomial 
function \citep[as in][]{kewley08,mann10}, to constrain how much the results 
depends on the actual functional form that it is adopted. Therefore, 
we consider the following additional functional form:
\begin{equation}
y= \Sigma^4_{i=0} p_i\ x^i
\label{eq:poly}
\end{equation}
where $y$ and $x$ represent the same parameters as in Eq.\ref{eq:fit}. 
Hereafter we will refer to this functional form as the pMZR and to the one 
described in Eq. \ref{eq:fit} as the MZR, to distinguish them from each other. 
The results of this polynomial fitting are listed in Tab.\ref{tab:val}, 
including the four coefficients of the function and the standard deviation 
around the best fitting curve. Contrary to what was reported for the CALIFA 
and MaNGA datasets, the polynomial pMZR fitting produces a lower scatter in the 
residuals for all of the adopted calibrators as compared to the MZR functional 
form described before. This can be seen in Figure \ref{fig:pFIT},  where the 
best fitting polynomial function is shown together with the data points for the 
different calibrators. This difference is maybe due to (i) the contribution 
of the galaxies in the low-mass range, not covered by other surveys in general, 
(ii) the differences in the sample selection or (iii) the aperture differences 
between the different surveys. While the oxygen abundance should not be affected 
by this aperture bias since we have selected this parameter at a characteristic 
radius of the galaxy (R$_{eff}$), it is still possible that it affects the 
stellar masses. In the present work these are restricted to the masses within the 
FoV of the SAMI instrument. In order to test this later possibility we have 
repeated the analysis using the photometric stellar masses provided by the SAMI 
collaboration \citep{bryant14}. We found no significant differences neither here 
nor in the further analysis. Thus, we consider that the first two reasons 
discussed before should be the source of the discrepancy. { Indeed, cutting 
the SAMI sample to the range were the CALIFA sample is complete 
\citep[$>$10$^{9.5}$][]{walcher14}, makes the pMZR functional form irrelevant.}

\begin{figure*}
\minipage{0.495\textwidth}
\includegraphics[width=\columnwidth]{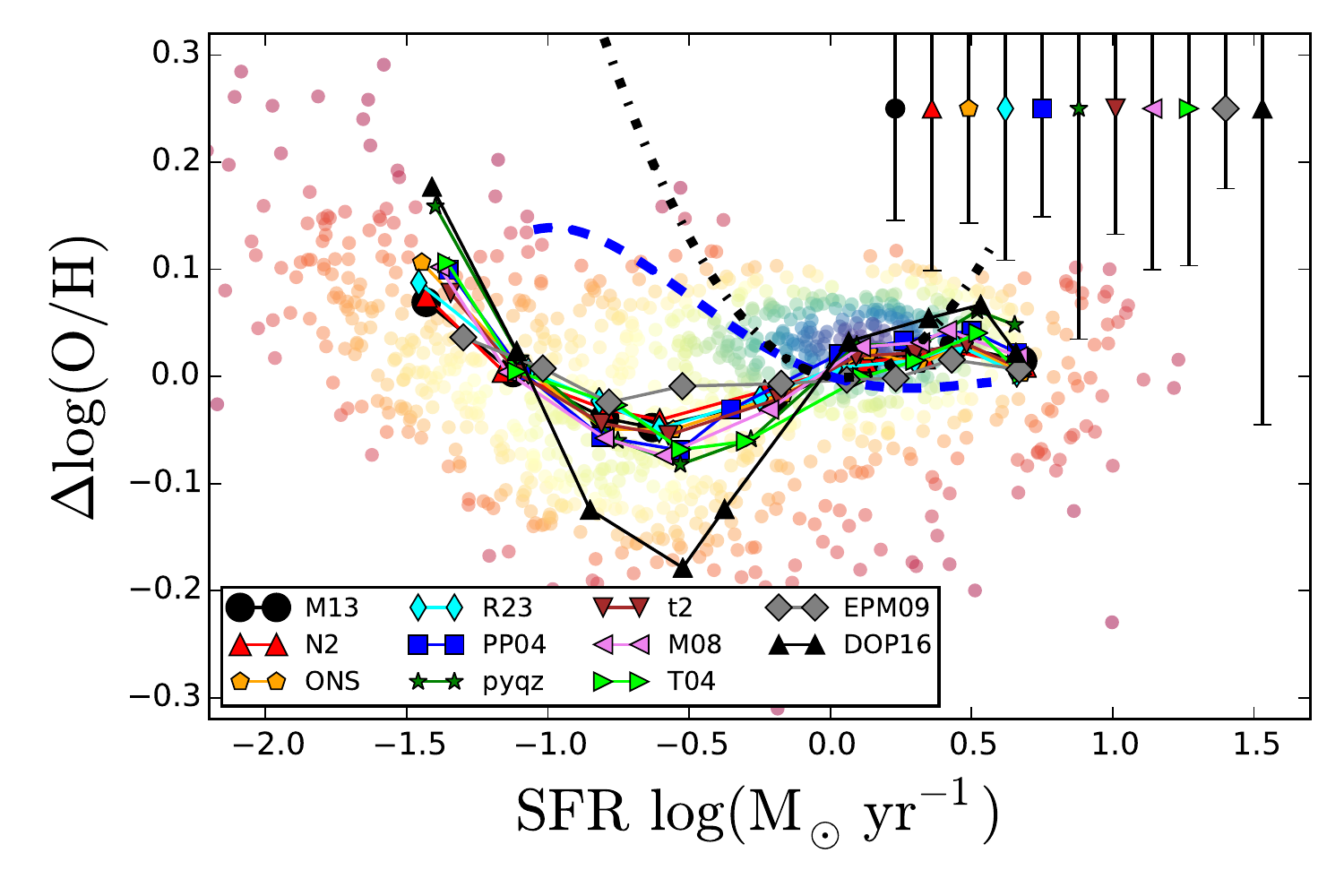}
\endminipage
\minipage{0.495\textwidth}
\includegraphics[width=\columnwidth]{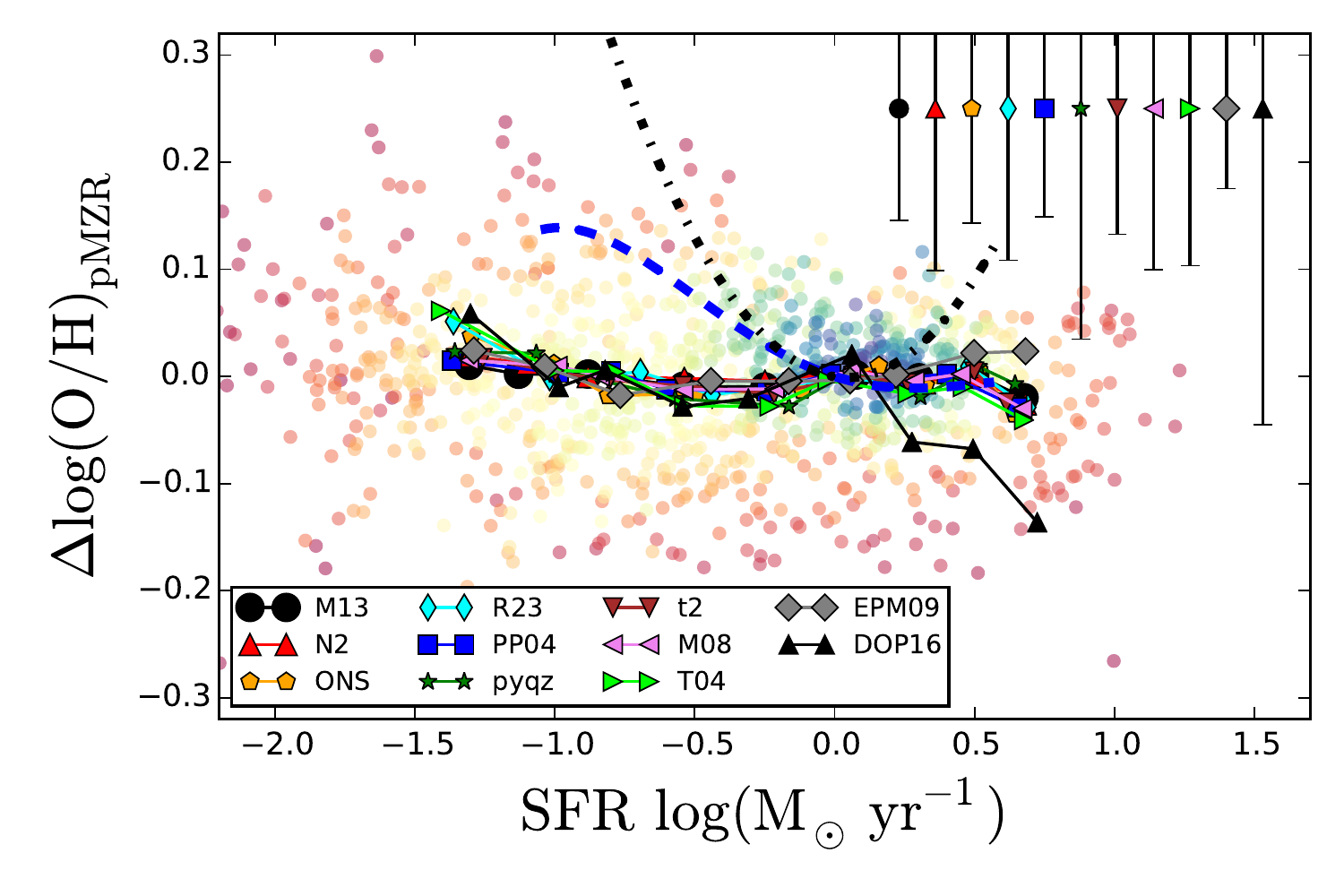}
\endminipage
\caption{Residuals of the MZR (left-panel) and the pMZR (right-panel) from the different analyzed calibrators against the SFR. For the cloud of data points we show the values estimated by the PP04 abundance calibrator, for simplicity. Each solid circle corresponds to an individual galaxy, color coded by density of points. The solid-lines, connecting different symbols, represent the median values in each SFR bin with each color and symbol corresponding to a different calibrator, following the nomenclature shown in Fig. \ref{fig:MZ}. The average standard deviation of the residuals for each calibrator is represented by an error bar at the top-right of each panel. In both panels, we show the expected residuals as a function of the SFR based on the relations proposed by \citet{mann10} (blue-dashed line), and by \citet{lara10a} (black-solid line).} 
\label{fig:dMZ}
\end{figure*}

\begin{table*}
\caption[]{Results of the linear fit of the residuals vs. SFR for the two characterizations of the MZ-relation described in Sec. \ref{sec:MZ}, MZR and pMZR. For the different metallicity calibrators we include the Pearson correlation coefficients between the two parameters (r$_c$), together with the zero-points ($\alpha$) and slopes ($\beta$) of the proposed linear relation, and the standard deviation around the best fit regression ($\sigma$).}
\label{tab:D_OH_SFR}
\begin{tabular} {r r r r r r r r r }
\hline
Metallicity   & \multicolumn{4}{c}{$\Delta$MZR vs. SFR} & \multicolumn{4}{c}{$\Delta$pMZR vs. SFR} \\
Indicator &  r$_c$ & $\alpha$  & $\beta$ & $\sigma$ &  r$_c$ & $\alpha$  & $\beta$ & $\sigma$\\ 
\cline{3-5}
\cline{7-8}
O3N2-M13 &  -0.22  &  0.008 $\pm$ 0.010 & 0.012 $\pm$ 0.013 & 0.105  &  -0.10 &  -0.007  $\pm$ 0.005 &  -0.011  $\pm$ 0.007  & 0.077 \\
PP04     &  -0.22  &  0.012 $\pm$ 0.015 & 0.019 $\pm$ 0.019 & 0.151  &  -0.09 &  -0.010  $\pm$ 0.007 &  -0.016  $\pm$ 0.010  & 0.112 \\
N2-M13   &  -0.22  &  0.009 $\pm$ 0.008 & 0.012 $\pm$ 0.011 & 0.107  &  -0.17 &  -0.006  $\pm$ 0.003 &  -0.015  $\pm$ 0.005  & 0.077 \\
ONS      &  -0.22  &  0.007 $\pm$ 0.010 & 0.018 $\pm$ 0.016 & 0.142  &  -0.13 &  -0.007  $\pm$ 0.006 &  -0.013  $\pm$ 0.010  & 0.100 \\
R23      &  -0.20  &  0.006 $\pm$ 0.013 & 0.003 $\pm$ 0.017 & 0.101  &  -0.08 &  -0.008  $\pm$ 0.006 &  -0.012  $\pm$ 0.008  & 0.087 \\
pyqz     &  -0.25  &  0.024 $\pm$ 0.020 & 0.021 $\pm$ 0.029 & 0.215  &  -0.17 &  -0.008  $\pm$ 0.006 &  -0.010  $\pm$ 0.008  & 0.142 \\
t2       &  -0.24  &  0.007 $\pm$ 0.012 & 0.012 $\pm$ 0.016 & 0.117  &  -0.12 &  -0.011  $\pm$ 0.006 &  -0.015  $\pm$ 0.009  & 0.086 \\
M08      &  -0.14  &  0.013 $\pm$ 0.015 & 0.023 $\pm$ 0.021 & 0.172  &  -0.16 &  -0.014  $\pm$ 0.007 &  -0.019  $\pm$ 0.009  & 0.145 \\
T04      &  -0.21  &  0.001 $\pm$ 0.021 & 0.002 $\pm$ 0.025 & 0.147  &  -0.21 &  -0.019  $\pm$ 0.007 &  -0.026  $\pm$ 0.009  & 0.120 \\
EPM09    &  -0.15  &  0.005 $\pm$ 0.005 & 0.005 $\pm$ 0.007 & 0.075  &  -0.04 &   0.011  $\pm$ 0.006 &   0.010  $\pm$ 0.008  & 0.072 \\
DP09     &  -0.17  &  0.010 $\pm$ 0.033 & 0.040 $\pm$ 0.043 & 0.295  &  -0.23 &  -0.066  $\pm$ 0.013 &  -0.087  $\pm$ 0.018  & 0.202 \\
\hline
\end{tabular}
\end{table*}

\subsection{Dependence of the residuals of the MZR on SFR}
\label{sec:residuals}

We explore any possible secondary dependence of the MZR on SFR
studying the correlations of the residuals of the two different
characterizations of the MZ-distribution with this parameter. Then, we
explore if introducing this dependence reduces the scatter in a
significant way. Our reasoning is that any possible secondary relation
that does not decrease significantly the scatter is not needed to
describe the observed distributions.

In Figure \ref{fig:dMZ} we show the residuals of the oxygen for each
calibrator as a function of the SFR, once subtracted the best fitted
MZR relation ($\Delta\log(\mathrm{O/H}_{MZR})$, left panel) and pMZR
relation ($\Delta\log(\mathrm{O/H}_{pMZR})$, right panel). In both
panels we present the median values in SFR bins of 0.3
$\log(M_{\odot} \, \mathrm{yr^{-1}})$ width covering a range between
-1.7 and 0.9 $\log(M_{\odot} \, \mathrm{yr^{-1}})$. These bins were
selected to include more than 20 objects in each bin, to guarantee
robust statistics.

There is considerable agreement in the median of the residuals for
most of the calibrators over the considered range of SFR. The one
that deviates most is the DOP16. Despite these differences, the
distribution of residuals is compatible with zero for all
calibrators, taking into account the standard deviation of each
individual bin ($\sigma$).  In the case of the pMZR-residuals, the
mean values are also compatible with zero, considering the error of
the mean (i.e., $\sigma$/$\sqrt{n}$). The largest differences are
found for the regime of lower SFRs
($\log{\mathrm{SFR}}<-0.5$). However, the different does not seem to
be significant in an statistical way. For those galaxies that lie on
the star-formation main sequence (SFMS), this SFR corresponds to
stellar masses lower or of the order of 10$^{9.5}$M$_\odot$
\citep[e.g.][]{mariana16}. In this regime we found a possible plateau
in the MZ distribution (Fig. \ref{fig:MZ}). Curiously, the trend
described for the MZR-residual is different than the one found for the
pMZR. In the previous one galaxies with SFR between 0.1-0.4 M$_\odot
yr^{-1}$ present a slightly lower oxygen abundance, contrary to the
reported trends in the literature \citep[e.g.][]{mann10}, rising again
for SFRs lower than 0.05 M$_\odot yr^{-1}$ . For the pMZR residuals
the possible dependence with the SFR is even weaker, with a slight
trend to higher abundances in the low SFR regime
($\log{\mathrm{SFR}}<-1.5$), but compatible with no dependence at any
SFR. 

\citet{mann10} and \citet{lara10a} presented two different functional
forms to describe the secondary relation that they found with the
SFR. Both relations are quite different. The so called FMR relation
\citep{mann10} proposes a correction to the stellar-mass as
independent parameter in the MZ-relation. On the other hand, the
relation described by \citet{lara10a}, considers that the three
parameters are located in a plane in the Mass-Metallicity-SFR space
(the so-called MZ-SFR Fundamental-Plane). We compare with their
predictions adopting the following procedure: (i) following
\cite{mann10}, we derive the average residual curve subtracting the
MZR relation published by them, using their functional form and
parameters, without a SFR dependence (i.e., $\mu_0$ of their Eq.4) to
the same relation, including this dependency (i.e., $\mu_{0.32}$ of
their Eq.4). When doing so, we assume that the SFR follows the stellar
mass in a strict SFMS relation \citep[by adopting the one published in
][]{mariana16}. For this reason, this is just an average trend, and
not the exact prediction of the FMR, that we will explore in following
sections. { The validity of the proposed average trend to describe
the distribution of galaxies analyzed by \citet{mann10} is discussed
in detail in Appendix \ref{sec:SDSS}.} This average trend is shown as a dashed-blue line in
Fig. \ref{fig:dMZ}. { If the secondary relation found by \cite{mann10} was
present for the average propulation of star-forming galaxies, this population
should follow this dashed-blue line. However, if the driver for the FMR was the
presence of a particular process in a particular group of galaxies (e.g., outflows
present in extreme star-forming galaxies), the average population is not required
to follow it. Therefore, we should clarify that the fact that they do not match
with the data is not a definitive proof or disproof of the FMR, but a clue for its 
driver}; (ii) For the MZ-SFR Fundamental Plane proposed by
\citet{lara10a} we just subtract the MZR estimated by them from their
proposed FP (Eq. 1 of that article). In this case there is no
assumption on the relation between the SFR and the stellar mass, and
therefore, the prediction is exact, not a first order approximation,
like the previous one. This trend is shown as a black-dotted line in
Fig. \ref{fig:MZ}. By construction the ($\Delta\log(\mathrm{O/H})$
predicted by both relations is zero when the $\log(SFR/M_{\odot} \,
\mathrm{yr^{-1}})=0$. In both cases the effect of introducing a
secondary relation is more evident at low SFRs, that corresponds to
lower regime of stellar masses \citep[Fig. 1 of][]{mann10}. A visual
inspection of both figures shows that we cannot reproduce the
predicted trends for any of the proposed relations, neither adopting
the residuals of the MZR nor the pMZR ones. We should keep in mind
that the disagreement is larger with the FP-relation suggested by
\citet{lara10a}, for which we have not done any approximation.
If the effects of the FMR are due to the residuals of the SFR with 
respect to the SFMS, there is still a possibility that our data are 
compatible with the proposed relation. { We thus insist at this point, 
that this section is not a disproof of the FMR, but evidence that it 
is not driven by the main population of star-forming galaxies.}

Based on a visual inspection we find no clear trend between the
residuals of the two analyzed functional forms of the MZ relation and
the SFR. Nevertheless, we attempt to quantify whether a potential
secondary relation may reduce the scatter around the mean
distributions by performing a linear regression between the two
parameters. The outcome of this analysis is included in Table
\ref{tab:D_OH_SFR}, including the Pearson correlation coefficient, the
best-fit parameters (slope and zero-point), and the standard
deviation once the fitted linear relation was removed ($\Delta$MZ-res),
for both residuals, $\Delta\log(\mathrm{O/H}_{MZR})$ and
$\Delta\log(\mathrm{O/H}_{pMZR})$, shown in Fig. \ref{fig:dMZ}. The
correlation coefficients show that there is a very weak negative trend
between the first residual and the SFR ($r_c\sim -$0.2), whose
strength depends slightly on the calibrator, ranging from $\sim -$0.15
to $\sim -$0.25 for the EPM09 and {\sc pyqz} calibrators
respectively. For the second residual, there is an
even weaker trend on average, with correlation coefficients ranging
between $\sim -$0.05 and $\sim -0$.23 for the EPM09 and the DOP16
calibrators respectively. In agreement with the very low correlation
coefficients, the derived zero-points of the relation are very near to
zero for all considered calibrators, and statistically compatible with
zero in many of them. Finally, the scatter of the residuals when
considering a secondary dependence with the SFR, quantified by the
standard deviation around the best fit linear regression, is not
improved for any of the analyzed calibrators and adopted functional
forms. In other words, the inclusion of this secondary relation does
not provide a better representation of the data. Similar results were
found by all similar analyses performed using up-to-date IFS data
\cite[e.g][]{sanchez13,sanchez17,bb17}.

\begin{figure*}
\minipage{0.495\textwidth}
\includegraphics[width=\columnwidth]{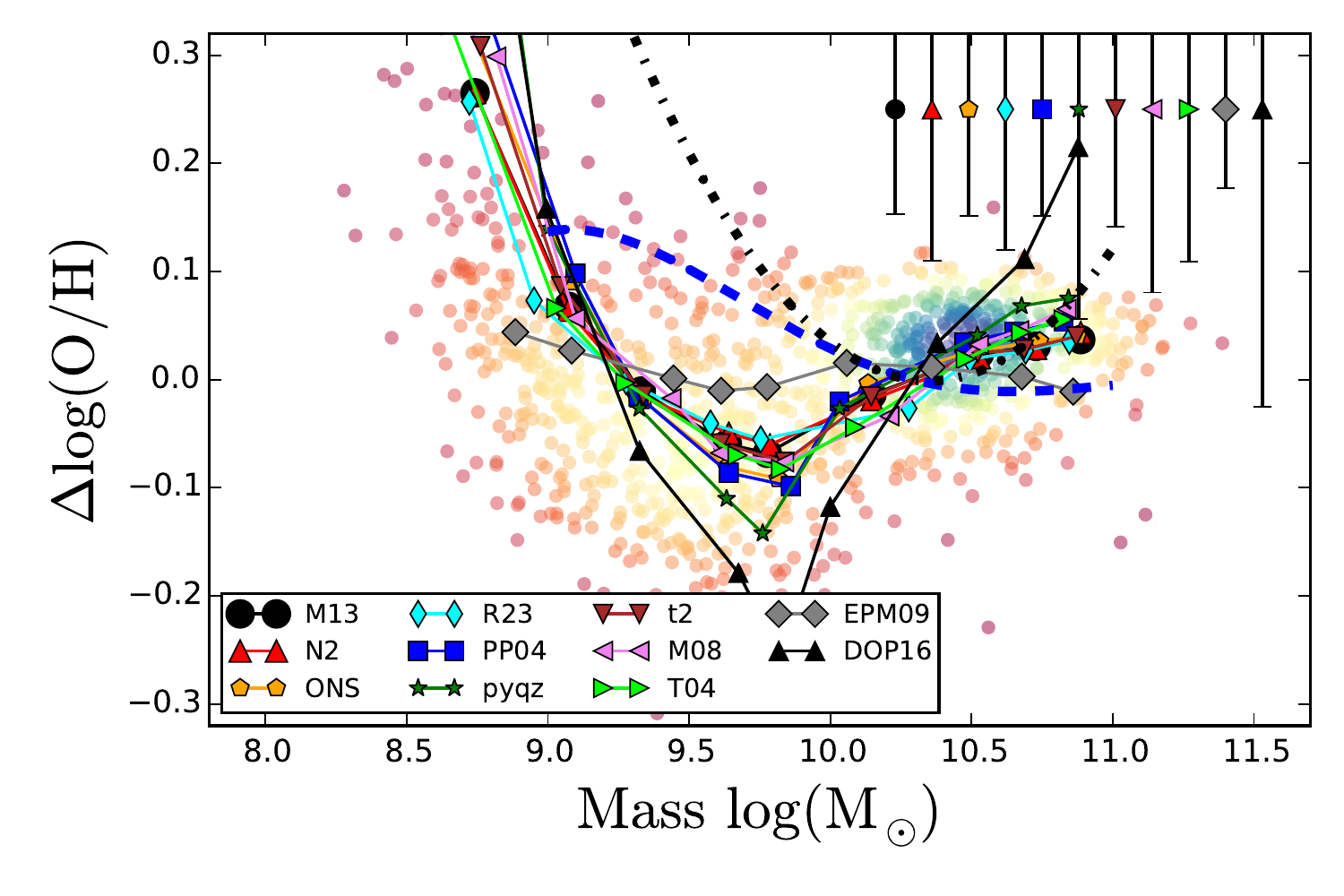}
\endminipage
\minipage{0.495\textwidth}
\includegraphics[width=\columnwidth]{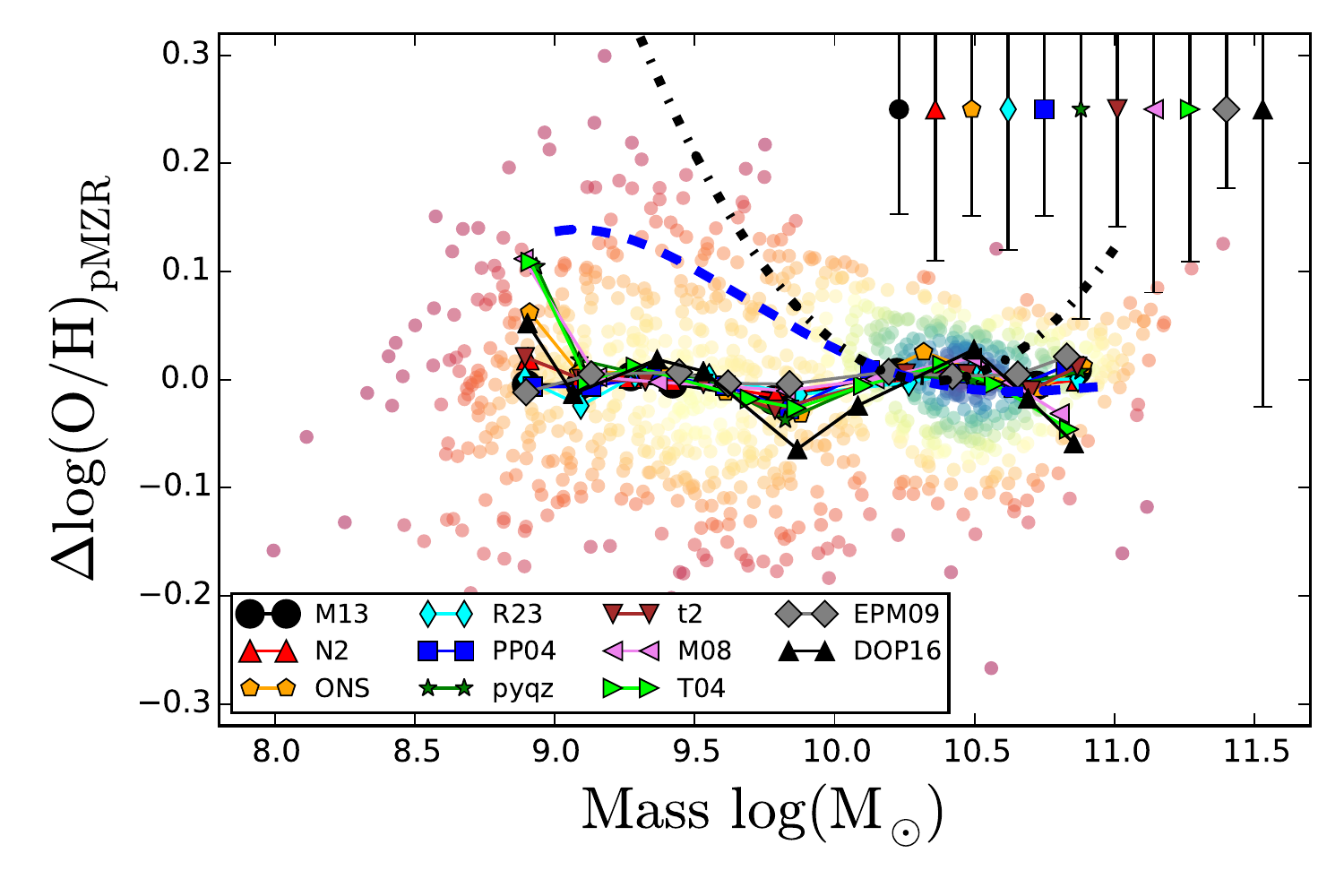}
\endminipage
\caption{ Residuals of the MZR (left-panel) and the pMZR (right-panel) for the different analyzed calibrators against the stellar mass. For the data, we show the values estimated by the PP04 abundance calibrator, for simplicity. Each solid circle corresponds an individual galaxy, color coded by density of points. The solid lines, connecting different symbols, represent the median values in each SFR bin with each color and symbol corresponding to a different calibrator, following the nomenclature shown in Fig. \ref{fig:dMZ}. The average standard-deviation of the residuals for each calibrator is represented by an error bar at the top-right of each panel. In both panels, we show the expected residuals as a function of the SFR based on the relations proposed by \citet{mann10} (blue-dashed line), and by \citet{lara10a} (black-solid line).} 
\label{fig:dMZ_M}
\end{figure*}

\begin{table*}
\caption[]{ Results of the linear fit of the residuals of the two characterizations of the MZ-relation described in Sec. \ref{sec:MZ}, MZR and pMZR, vs. the stellar mass. For the different metallicity calibrators we include the Pearson correlation coefficients between the two parameters (r$_c$), together with the zero-points ($\alpha$) and slopes ($\beta$) of the proposed linear relation, and the standard deviation around the best fit regression ($\sigma$).}
\label{tab:D_OH_Mass}
\begin{tabular} {r r r r r r r r r }
\hline
Metallicity   & \multicolumn{4}{c}{$\Delta$MZR vs. Mass} & \multicolumn{4}{c}{$\Delta$pMZR vs. Mass} \\
Indicator &  r$_c$ & $\alpha$  & $\beta$ & $\sigma$ &  r$_c$ & $\alpha$  & $\beta$ & $\sigma$\\ 
\cline{3-5}
\cline{7-8}
O3N2-M13 &  -0.06  &   0.253  $\pm$ 0.359 & -0.023   $\pm$ 0.036  & 0.097  &   0.07 & -0.050  $\pm$ 0.037 &   0.005  $\pm$ 0.004  & 0.077 \\
PP04     &  -0.06  &   0.345  $\pm$ 0.513 & -0.031   $\pm$ 0.051  & 0.140  &   0.07 & -0.074  $\pm$ 0.054 &   0.007  $\pm$ 0.005  & 0.111 \\
N2-M13   &  -0.02  &   0.184  $\pm$ 0.350 & -0.016   $\pm$ 0.035  & 0.099  &   0.02 &  0.019  $\pm$ 0.044 &  -0.002  $\pm$ 0.004  & 0.078 \\
ONS      &  -0.10  &   0.498  $\pm$ 0.481 & -0.046   $\pm$ 0.048  & 0.130  &  -0.01 &  0.065  $\pm$ 0.104 &  -0.006  $\pm$ 0.010  & 0.101 \\
R23      &  -0.05  &   0.452  $\pm$ 0.386 & -0.044   $\pm$ 0.039  & 0.099  &   0.06 & -0.057  $\pm$ 0.050 &   0.006  $\pm$ 0.005  & 0.087 \\
pyqz     &  -0.11  &   0.622  $\pm$ 0.764 & -0.057   $\pm$ 0.077  & 0.194  &  -0.06 &  0.168  $\pm$ 0.127 &  -0.016  $\pm$ 0.013  & 0.142 \\
t2       &  -0.08  &   0.287  $\pm$ 0.405 & -0.026   $\pm$ 0.040  & 0.109  &   0.05 & -0.029  $\pm$ 0.066 &   0.003  $\pm$ 0.007  & 0.086 \\
M08      &   0.02  &  -0.058  $\pm$ 0.443 &  0.008   $\pm$ 0.044  & 0.170  &  -0.02 &  0.173  $\pm$ 0.152 &  -0.017  $\pm$ 0.015  & 0.145 \\
T04      &  -0.03  &   0.481  $\pm$ 0.504 & -0.048   $\pm$ 0.052  & 0.141  &  -0.09 &  0.334  $\pm$ 0.180 &  -0.034  $\pm$ 0.019  & 0.121 \\
EPM09    &  -0.08  &   0.126  $\pm$ 0.078 & -0.012   $\pm$ 0.008  & 0.073  &   0.07 & -0.082  $\pm$ 0.040 &   0.009  $\pm$ 0.004  & 0.072 \\
DP09     &   0.06  &   0.430  $\pm$ 1.163 & -0.038   $\pm$ 0.119  & 0.275  &  -0.08 &  0.228  $\pm$ 0.169 &  -0.024  $\pm$ 0.017  & 0.201 \\
\hline
\end{tabular}
\end{table*}

\subsection{Dependence of the residuals of the MZR on stellar mass}
\label{sec:res_Mass}

{ \citet{mann10} suggested that the secondary relation could be
driven by galaxies of low-mass and large sSFR, and not only or
mostly by the main population of star-forming galaxies. Similar
results were presented by \citet{2010ApJ...715L.128A} and
\citet{telf16}. For this reason, it is important to explore the
residuals of the MZR relation versus stellar
mass. Fig. \ref{fig:dMZ_M} shows the residuals of the oxygen
abundance data after subtraction of the MZR and pMZR relations (i.e., the same
values as shown in Fig. \ref{fig:dMZ}), but this time as a function of stellar
mass.  In both panels we present the median values in
stellar-mass bins of 0.3 $\log(M_{\odot})$ width covering a range
between 10$^{8}$ and 10$^{11}$ M$_{\odot}$. Like in the case of the analysis
of the residuals as a function of the SFRs, we quantified the possible
additional relation with the stellar mass by exploring the Pearson
correlation coefficient, performing a linear regression between the two
parameters, and deriving the standand deviations after removal of the best
fit linear regression. The results of this analysis are included in
Table \ref{tab:D_OH_Mass}.

  If the adopted functional form for the MZ relation was
  representative of the distribution, there should not be any significant
  additional correlation with the stellar mass of the residuals of that
  relation. Indeed the correlation coefficients between the residuals
  and the stellar mass are very low, of the order of $\pm$0.1, and
  the slopes of the derived linear regressions are in general consistent
  with zero. The introduction of an additional dependence with the mass
  does not decrease the scatter of the residuals neither. Like in the case
  of the distribution of residuals against the SFR, the largest differences
  are found for the MZR relation, in the low-Mass regime. Indeed, even
  for the pMZR relation, the scatter seems to increase slightly at lower
  stellar masses, where we found the larger number of outliers. Thus,
  it is still possible that the trends described by \citet{mann10} can
  only be found for low-mass galaxies.
 }

\subsection{Studying the FMR in detail}
\label{sec:FMR}

We have shown that the residuals of the MZR and pMZR relations do not show a 
clear dependence on the SFR for any of the analyzed calibrators, and only a 
possible weak trend for low stellar masses. As indicated before, we used the 
assumption of a linear trend with SFR for this exploration. Therefore, strictly 
speaking, our results do not agree with a hypothetical {\it linear} secondary 
relation with the SFR, i.e., the relation proposed by \citet{lara10a}.

\begin{figure*}
 \minipage{0.98\textwidth}
 \includegraphics[width=\linewidth]{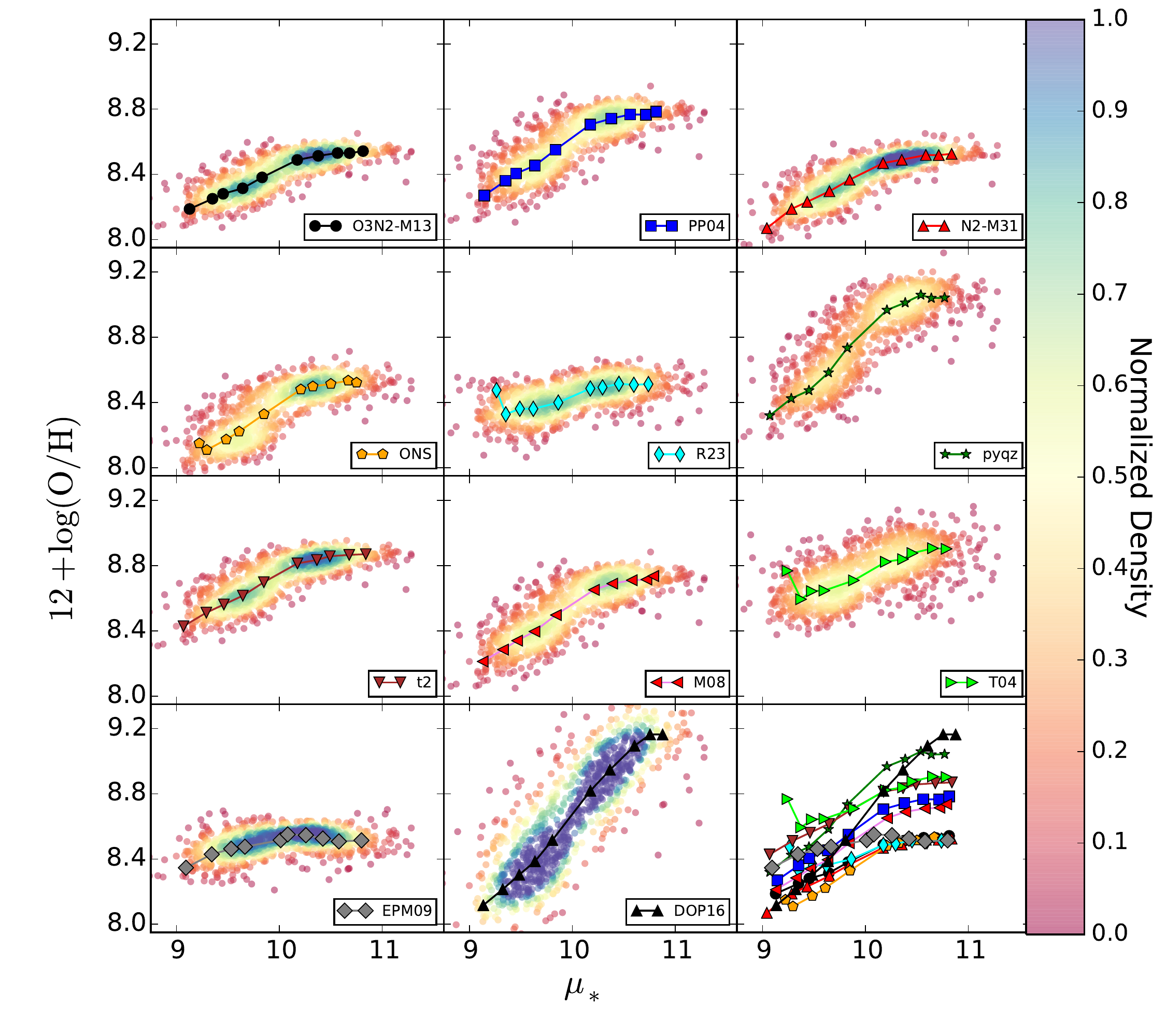}
 \endminipage
\caption{Application of the FMR relation proposed by \citet{mann10}to the sample 
of galaxies and the different metallicity calibrators introduced in the present work. 
Symbols have the same meaning as the ones used in  Figure \ref{fig:MZ}.} 
 \label{fig:FMR}
\end{figure*}

\begin{table*}
\caption[Fitting parameters for the FMR and its scatter.]{Best fit parameters for the two functional forms adopted to characterize the FMR and their scatter for the different abundance estimations considered in our analysis. We include, for each different calibrator, (i) the standard deviation of the values of the oxygen abundance ($\sigma_{log(O/H)}$); (ii) the best-fit $a$ and $b$ parameters as described in Eq.\ref{eq:fit}, applied to the FMR (i.e., modifying the mass by the $\mu_{*,0.32}$ parameter); (iii)  the standard deviation of the abundances once subtracted the best fit curve, characterized by Eq. \ref{eq:fit} and the parameters indicated before ($\sigma_{\rm FMR}$); (iv) the coefficients of the polynomial function adopted in Eq. \ref{eq:poly}, defined as the pFMR relation (i.e., modifying the mass by the $\mu_{*,32}$ parameter) and finally, (v) the standard deviation once subtracted the best fit polynomial function, $\sigma_{\rm pFMR}$. The values of the different parameters are shown up to the third decimal to highlight the differences. However, we consider that only the first two decimals are significant.}
\label{tab:FMR_val}
\begin{tabular} {r r r r r r r r r }
\hline
Metallicity   & \multicolumn{2}{c}{FMR Best Fit} & $\sigma_{\rm FMR}$ & \multicolumn{4}{c}{pFMR Polynomial fit} &  $\sigma_{pFMR}$\\
 Indicator &  $a$  & $b$ & (dex) & $p_0$  & $p_1$ & $p_2$ & $p_3$  & (dex) \\
\cline{2-3}
\cline{5-8}
O3N2-M13   & 8.55 $\pm$ 0.01 & 0.016 $\pm$ 0.002 & 0.086 &  8.219 $\pm$ 0.236 & -0.392 $\pm$ 0.372 & 0.421 $\pm$ 0.188 & -0.086 $\pm$ 0.031 & 0.071 \\
PP04       & 8.80 $\pm$ 0.02 & 0.024 $\pm$ 0.002 & 0.125 &  8.268 $\pm$ 0.356 & -0.503 $\pm$ 0.556 & 0.582 $\pm$ 0.279 & -0.121 $\pm$ 0.045 & 0.103 \\
N2-M13     & 8.54 $\pm$ 0.01 & 0.017 $\pm$ 0.001 & 0.085 &  7.681 $\pm$ 0.155 &  0.306 $\pm$ 0.252 & 0.112 $\pm$ 0.130 & -0.041 $\pm$ 0.022 & 0.075 \\
ONS        & 8.56 $\pm$ 0.02 & 0.022 $\pm$ 0.002 & 0.117 &  9.179 $\pm$ 0.482 & -2.244 $\pm$ 0.754 & 1.444 $\pm$ 0.377 & -0.260 $\pm$ 0.061 & 0.102 \\
R23        & 8.51 $\pm$ 0.03 & 0.007 $\pm$ 0.003 & 0.090 & 10.996 $\pm$ 0.961 & -4.270 $\pm$ 1.498 & 2.218 $\pm$ 0.756 & -0.362 $\pm$ 0.124 & 0.114 \\
pyqz       & 9.08 $\pm$ 0.04 & 0.032 $\pm$ 0.003 & 0.167 &  9.241 $\pm$ 0.344 & -2.262 $\pm$ 0.567 & 1.653 $\pm$ 0.298 & -0.312 $\pm$ 0.051 & 0.140 \\
t2         & 8.89 $\pm$ 0.02 & 0.018 $\pm$ 0.001 & 0.094 &  8.358 $\pm$ 0.180 & -0.267 $\pm$ 0.286 & 0.408 $\pm$ 0.145 & -0.088 $\pm$ 0.024 & 0.080 \\
M08        & 8.97 $\pm$ 0.01 & 0.026 $\pm$ 0.001 & 0.153 &  7.961 $\pm$ 0.293 &  0.115 $\pm$ 0.458 & 0.301 $\pm$ 0.229 & -0.077 $\pm$ 0.037 & 0.139 \\
T04        & 8.88 $\pm$ 0.04 & 0.012 $\pm$ 0.003 & 0.130 &  11.572 $\pm$ 1.122 & -4.755 $\pm$ 1.748 & 2.449 $\pm$ 0.880 & -0.390 $\pm$ 0.144 & 0.143 \\
EPM09      & 8.55 $\pm$ 0.01 & 0.007 $\pm$ 0.001 & 0.073 &  7.750 $\pm$ 0.267 &  0.767 $\pm$ 0.450 &-0.217 $\pm$ 0.240 &  0.014 $\pm$ 0.041 & 0.073 \\
D0P09      & 9.12 $\pm$ 0.06 & 0.047 $\pm$ 0.005 & 0.254 &  9.050 $\pm$ 0.239 & -2.226 $\pm$ 0.377 & 1.549 $\pm$ 0.190 & -0.264 $\pm$ 0.031 & 0.188 \\
\hline
\end{tabular}
\end{table*}

To explore if our data are consistent with the FMR we will follow \citet{mann10} 
and introduce the $\mu_*$ parameter, that depends on the SFR, defined as 
\begin{equation}
\mu_* = \log(M/M_\odot) + \alpha\log(SFR)
\label{eq:FMR}
\end{equation}
with $\alpha=-0.32$. Substituting the stellar mass by $\mu_*$ in the MZR 
(Eq. \ref{eq:fit}) and pMZR (Eq. \ref{eq:poly}) relations, we determine if there 
is a decrease in the scatter around the mean distribution by introducing this 
parameter.

The distribution of the oxygen abundances over the $\mu_*$ parameter for the 
different calibrators is shown in Fig. \ref{fig:FMR} (following the same 
nomenclature as in Fig. \ref{fig:MZ}). The shape of the different distributions 
for each calibrator is very similar, with the same pattern already described in 
Sec.~\ref{sec:MZ}. Like the Mass-Metallicity relation, we characterize the 
$\mu_*$-Metallicity relation using the same parametrization as described in 
Eq.~\ref{eq:fit} and Eq.~\ref{eq:poly}, substituting the stellar mass by 
$\mu_*$. We will refer them as the FMR and pFMR, respectively. Table \ref{tab:FMR} 
lists the best fit parameters for both functional forms adopted to characterize 
the shape of the relation between $\mu_*$ and the oxygen abundance, together 
with the corresponding  standard deviations around the best fit curves. 

In the case of the FMR parametrization the values for the asymptotic oxygen 
abundance ($a$) are very similar, irrespectively of whether we use M$_*$ or $\mu_*$. 
The parameter that defines the strength of the bend of the distribution, $b$ is 
of the same order for both distributions. For the pFMR parametrization the 
coefficients are more difficult to interpret, although in general they are 
of the same order as those found for the pMZR. Finally, we do not find any 
significant decrease in the standard deviation of the distribution of the 
oxygen abundance residuals when introducing the proposed secondary 
dependence on the SFR, as can be verified by comparing the values in 
Table \ref{tab:FMR_val} with the corresponding ones in Table \ref{tab:val}.

\begin{table*}
\caption{Parameters derived for the generalized FMR$_d$ relation derived 
by the adopted fitting process, when the $d$ parameter in Eq.\ref{eq:NEW_FMR} 
is not fixed to the value reported in the literature. We report the three 
parameters included in Eq.\ref{eq:FMR} ($a$, $b$ and $d$), together with 
the standard deviation of the oxygen abundances after subtraction of the 
best fit FMR$_d$ relation ($\sigma$ MZ-res). 
The best fit $d$ parameter derived when adopting a polynomial functional 
form for the FMR (pFRM$_d$ formalism), is reported too. In this particular 
case we adopted  Eq. \ref{eq:FMR}, substituting the stellar mass by the 
$\mu_{*,d}$ parameter. Finally, we list the standard deviations after subtraction
of the best fit pFRM$_d$ model. All these parameters are included for the 
different estimations of oxygen abundance included in this study. Like in 
the previous tables some parameters are presented including the third 
decimal, to highlight the differences, although we consider that the values 
are significant only to the second decimal.}
\label{tab:FMR}
\begin{tabular} {c c c c c c c}
\hline
Metallicity   & \multicolumn{3}{c}{Generalized FMR Best Fit} & $\sigma_d$ FMR-res & pFMR best fit &  $\sigma_d$ pFMR-res \\
 Indicator &  $a$ (dex)  & $b$ (dex $/ \log(M_{\odot}))$ & $d$ (dex$/\log(M_{\odot}/yr)$ & (dex)  & $d$ (dex$/\log(M_{\odot}/yr)$ & (dex)\\
\cline{2-4}
\cline{6-6}
O3N2-M13 & 8.55 $\pm$ 0.02 & 0.020 $\pm$ 0.003 & -0.498 $\pm$ 0.054 & 0.077 & -0.395 $\pm$ 0.063 & 0.069\\
PP04     & 8.80 $\pm$ 0.02 & 0.029 $\pm$ 0.004 & -0.498 $\pm$ 0.045 & 0.111 & -0.395 $\pm$ 0.052 & 0.100\\
N2       & 8.54 $\pm$ 0.02 & 0.021 $\pm$ 0.003 & -0.470 $\pm$ 0.050 & 0.079 & -0.388 $\pm$ 0.059 & 0.071\\
ONS      & 8.54 $\pm$ 0.02 & 0.022 $\pm$ 0.004 & -0.421 $\pm$ 0.058 & 0.110 & -0.285 $\pm$ 0.076 & 0.097\\
R23      & 8.54 $\pm$ 0.02 & 0.017 $\pm$ 0.003 & -0.631 $\pm$ 0.065 & 0.083 & -0.541 $\pm$ 0.071 & 0.081\\
pyqz     & 9.06 $\pm$ 0.03 & 0.034 $\pm$ 0.004 & -0.419 $\pm$ 0.045 & 0.160 & -0.277 $\pm$ 0.057 & 0.136\\
t2       & 8.88 $\pm$ 0.02 & 0.022 $\pm$ 0.003 & -0.479 $\pm$ 0.050 & 0.086 & -0.389 $\pm$ 0.059 & 0.078\\
M08      & 8.98 $\pm$ 0.02 & 0.034 $\pm$ 0.004 & -0.511 $\pm$ 0.044 & 0.141 & -0.448 $\pm$ 0.048 & 0.135\\
T04      & 8.89 $\pm$ 0.02 & 0.022 $\pm$ 0.004 & -0.577 $\pm$ 0.062 & 0.124 & -0.429 $\pm$ 0.078 & 0.117\\
EPM09    & 8.56 $\pm$ 0.02 & 0.008 $\pm$ 0.003 & -0.548 $\pm$ 0.122 & 0.069 & -0.602 $\pm$ 0.085 & 0.067\\
DP09     & 9.08 $\pm$ 0.03 & 0.057 $\pm$ 0.005 & -0.502 $\pm$ 0.032 & 0.226 & -0.375 $\pm$ 0.039 & 0.185\\
\hline
\end{tabular}
\end{table*}

Furthermore, following \citet{sanchez17}, we repeated the analysis by letting 
the $\alpha$ parameter in Eq. \ref{eq:FMR} (the one that controls the secondary 
dependence on the SFR) free in the fit, instead of fixing the value to the one 
suggested by \citet{mann10}. We study that option by fitting the data with the 
following equation:
\begin{equation}
\mathrm{y}=a+b(\mathrm{x}+d\mathrm{s}-c)\exp(-(\mathrm{x}+d\mathrm{s}-c))
\label{eq:NEW_FMR}
\end{equation}
where $y$, and $x$ are the same parameters as in Eq. \ref{eq:fit}, and $s$ is 
the logarithm of the SFR. The $c$ parameter is set to 3.5. This is the same 
procedure used by \citet{mann10} to derive the FMR. In this case, $d$ 
corresponds to the $\alpha$ parameter in \citet{mann10}. We will label this 
parametrization as the FMR$_d$ relation, hereafter. Tab. \ref{tab:FMR} includes 
the results of this analysis, too. In summary, we found a stronger relation 
with the SFR than that reported by \citet{mann10}. These values agree with 
those reported by \citet{sanchez17}. However, even considering this stronger 
dependence we do not find a significant improvement in the global standard 
deviation of the residuals once the best fit function is subtracted from the 
original distribution of oxygen abundances. Only for some calibrators is 
there marginal improvement, of the order of $\sim$0.02 dex. However, this 
improvement is always of the order or even below that produced by modifying 
the functional form from the one shown in Eq. \ref{eq:fit} by a polynomial 
function, without requiring consideration of a secondary dependence with the 
SFR (see values in Tab. \ref{tab:val}. Indeed, introducing this generalized 
secondary dependence and using a polynomical function, i.e., modifying $x$ in Eq. 
\ref{eq:poly} by a parameter $x'=x-d*s$, with $s$ being the logarithm of the 
SFR, we do not find any significant improvement in the reported scatter for 
any of the calibrators. We include the values derived for $d$ together with 
the standard deviation of the corresponding residuals in Tab. \ref{tab:FMR}. 
We do not reproduce the full list of derived polynomial coefficients for 
clarity, since thez do not add any new information.

%
\begin{figure*}
 \minipage{0.995\textwidth}
 \includegraphics[width=\linewidth]{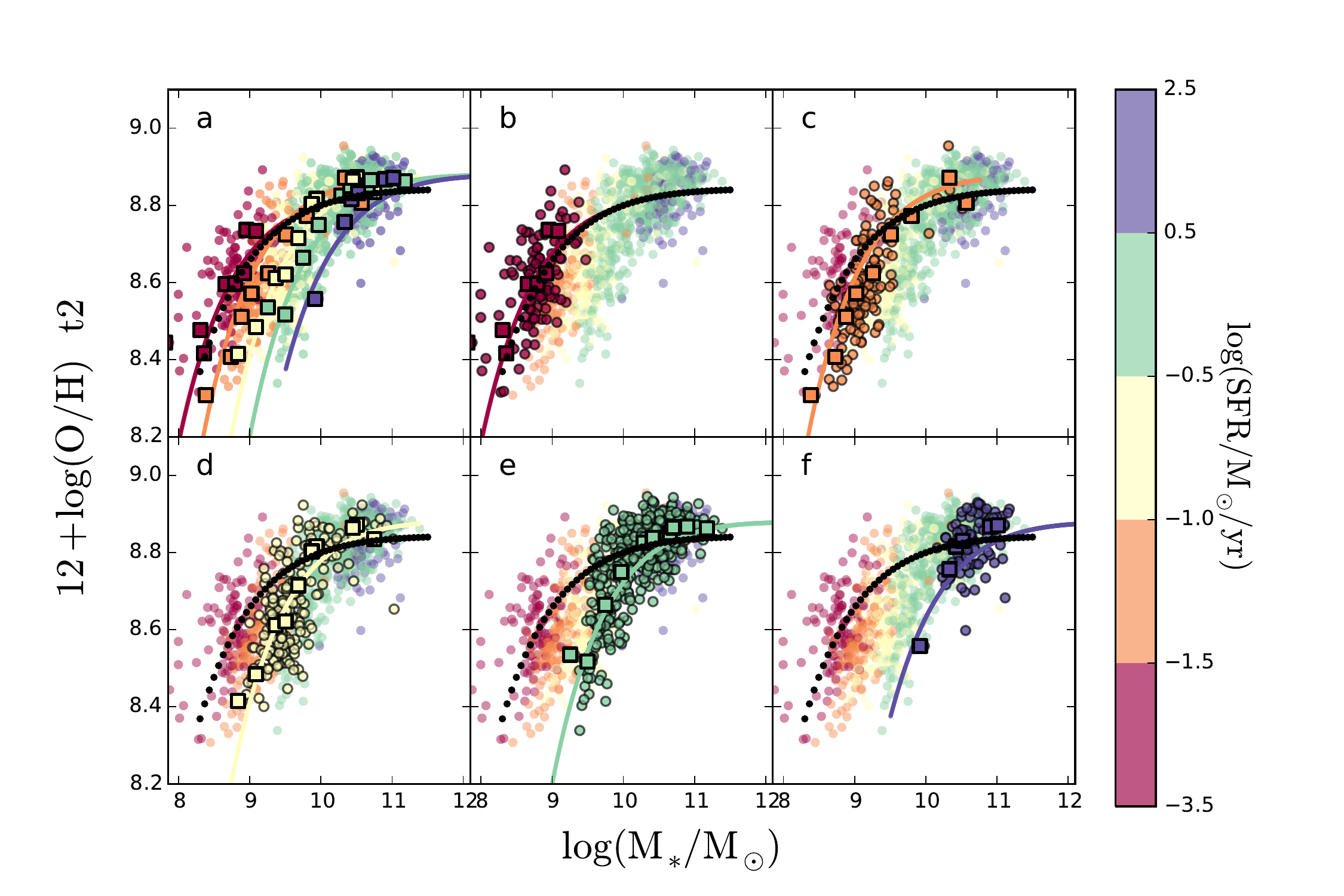}
 \endminipage
\caption{ Mass-metallicity relations using the oxygen abundances derived using the $t2$ calibrator for the sample of galaxies analyzed in this study. Solid circles represent the values for each single galaxy, color coded by the SFR as indicated by the colourbar. The solid squares represent the average values in the same stellar mass bins as shown in Fig. \ref{fig:MZ}. Panel (a) includes the full dataset, together with the average $\mu_{*,d}$Z relations derived for each SFR bin. These models correspond to the parametrization described in Eq. \ref{eq:NEW_FMR}, using the values listed in Tab. \ref{tab:FMR}. The black dotted line corresponds to the original MZ relation, not considering the possible relation with the SFR, i.e., to the parametrization described in Eq. \ref{eq:fit}, using the values reported in Tab. \ref{tab:val}. Each individual panel from (b) to (f) corresponds to a different SFR bin, ranging from the lowest to the highest values of the star-formation rate. }
\label{fig:NEW_FMR}
\end{figure*}

So far, we do not find a decrease in the global standard deviation in the same way 
as the one reported by \citet{mann10}, neither using and fitting for the proposed 
parametrization for the dependence on SFR nor directly using the reported values 
for the proposed correction. Therefore, we cannot confirm their results with the 
current analysis and dataset in this respect.
However, it is still possible that there is no global improvement in
the scatter over the total range of masses and star-formation rates but the
introduction of a secondary relation could improve the representation
of the data in certain mass or star-formation ranges. In this
regard previous results \citep[e.g.][]{sanchez17} were not totally
conclusive. For some calibrators it seems that the dispersion decreases when 
introducing the secondary relation at low masses (M$_*<$10$^{9.5}$M$_\odot$). 
In order to perform this test we split the dataset in
different ranges of stellar masses and star-formation rates and
compared the mean value and the standard deviations of both the best
fit FMR$_d$ and MZR for each considered dataset.

The result of this analysis is illustrated by Figure
\ref{fig:NEW_FMR}, where we show the distribution for one particular
calibrator ($t2$). The same qualitative results are derived using any
of the considered calibrators, but we describe in detail the results for
only one just for clarity. The figure shows the individual $t2$ oxygen
abundances for each galaxy as a function of stellar mass, separated in 
five SFR bins. { Following \citet{mann10} and
\citet{andr13}, } in addition we represent the average FMR$_d$
relation derived for each bin, and the derived MZR
(Tab. \ref{tab:val}). Thus, the relation assuming $d$=0. A visual
inspection of the Figure gives the impression that when including an
additional dependence with the SFR, with the functional form presented
in \citet{mann10}, a better description of the data is obtained.
{ This is even more clear for the mean values in bins of stellar
  masses, derived using the same scheme adopted in Fig. \ref{fig:MZ}
  and Sec. \ref{sec:MZ}, shown as solid squares in
  Fig. \ref{fig:NEW_FMR}. }

However, when comparing the differences between the global MZR and the
FMR$_d$ in different SFR ranges, at the range of stellar masses sampled by 
the corresponding points, it is evident that the differences { in the 
representation of the full population of galaxies} are not large. 
Table \ref{tab:FMR_t2} shows a detailed comparison of the different residuals 
after subtraction of the derived FMR$_d$, MZR and pMZR  relations, described 
in the previous sections. In
general, the residuals around the FMR$_d$ relation are more consistent
with zero, with slightly lower scatter, than those of the
MZR. However, contrary to what is predicted by the proposed FMR, the
residuals of the MZR do not present a systematic trend with larger
abundances for lower SFRs and smaller abundances for higher SFRs. For
example, for very low SFRs ($<$10$^{-1.5} M_\odot/yr$) there is a
positive deviation in the residual of $\sim$0.04$\pm$0.14 dex,
however, for SFRs between 10$^{-1.5}-$10$^{-1} M_\odot/yr$, the
deviation is negative ($\sim$-0.05$\pm$0.09 dex: i.e., contrary to that
proposed by the FMR picture). On the other hand, for the explored
ranges in the stellar mass there is no general improvement in the
residuals, being appreciable only for stellar masses below 10$^{9.5}
M_\odot$. Even more, both residuals show a rather similar and
consistent trend to mis-represent the data at low masses. This is
consistent with a polynomial function
representing the data better than the assumed functional form. Indeed, in
these mass regime the residuals of the pMZR are smaller than those
of the FMR$_d$. For the pMZR the introduction of a secondary
dependence with the SFR only slightly reduces the scatter in the residuals 
for the more massive galaxies. Thus, the trend with the SFR
is different if we represent the MZ distribution using two different
functional forms.

Finally, the standard deviation of the distribution of the residuals
indicates that there is no significant improvement for
all the explored ranges of SFRs and stellar masses. For the MZR the reduction 
ranges between 0.008 and 0.022 dex for the SFR bins and between
0.000 and 0.027 dex for the stellar mass ones, although in general the
reduction is of the order of 0.01 dex. On the other hand, for the pMZR
the reduction of the scatter, if any, is at the third decimal of
the standard deviation. In summary, if there is an
improvement by introducing a secondary relation with the SFR (a) it
seems to be limited to some particular ranges of SFRs or stellar masses, and 
(b) its quantitative effect depends on the adopted parametrization of the 
MZ-relation and the adopted calibrator. In general, the data points are well 
described with the pMZR model, without assuming any secondary dependence on SFR.

\begin{table*}
\caption{Comparison of the residuals from the FMR$_d$ and MZR relations, based on the adopted fitting procedure (i.e., using the values listed in Tab. \ref{tab:FMR} and Tab. \ref{tab:val}), in five bins of SFRs and stellar masses, respectively. We list (i) the SFRs and stellar masses covered, (ii) the mean SFR ($<$SFR$>$), with its standard deviation, and finally (iii) the mean values and standard deviations of the residuals for the two relations explored here. We include the third decimal in the parameters to highlight the differences. However, we remind the reader that we do not consider significant those values beyond the second decimal.}
\label{tab:FMR_t2}
\begin{tabular} {c r r r r}
\hline
$\log(SFR/M_\odot/yr)$   &$<\log(SFR)>$& FMR$_d$-res & MZR-res & pMZR-res \\
range & $log(M_\odot/yr)$ & (dex) & (dex) & (dex) \\
\hline
$[$-3.5,-1.5$]$ & -1.80  $\pm$ 0.27 & -0.001  $\pm$ 0.113 &  0.038  $\pm$ 0.135 &  0.023  $\pm$ 0.124\\
$[$-1.5,-1$]$   & -1.23  $\pm$ 0.15 & -0.043  $\pm$ 0.082 & -0.075  $\pm$ 0.090 &  0.004  $\pm$ 0.098\\
$[$-1,-0.5$]$   & -0.74  $\pm$ 0.15 & -0.042  $\pm$ 0.085 & -0.086  $\pm$ 0.098 & -0.013  $\pm$ 0.084\\
$[$-0.5,0.5$]$  & -0.01  $\pm$ 0.28 &  0.018  $\pm$ 0.072 & -0.022  $\pm$ 0.091 & -0.012  $\pm$ 0.073\\
$[$0.5,2.5$]$   &  0.75  $\pm$ 0.17 &  0.042  $\pm$ 0.062 & -0.001  $\pm$ 0.068 & -0.022  $\pm$ 0.067\\
\hline
$\log(M_*/M_\odot)$   & $<\log(SFR)>$& FMR$_d$-res & MZR-res & pMZR-res \\
range & $log(M_\odot/yr)$ & (dex) & (dex) \\
\hline
$[$8,9.5$]$   & -1.26  $\pm$ 0.45  & -0.040  $\pm$ 0.088 & -0.068  $\pm$ 0.115 & -0.001  $\pm$ 0.102\\
$[$9.5,10$]$  & -0.42  $\pm$ 0.33  & -0.023  $\pm$ 0.097 & -0.084  $\pm$ 0.104 & -0.015  $\pm$ 0.101\\
$[$10,10.5$]$ &  0.01  $\pm$ 0.42  &  0.036  $\pm$ 0.056 &  0.004  $\pm$ 0.056 & -0.002  $\pm$ 0.055\\
$[$10.5,11$]$ &  0.28  $\pm$ 0.44  &  0.030  $\pm$ 0.050 &  0.018  $\pm$ 0.050 & -0.011  $\pm$ 0.051\\
$[$11,12.5$]$ &  0.30  $\pm$ 0.52  & -0.002  $\pm$ 0.068 &  0.010  $\pm$ 0.066 &  0.024  $\pm$ 0.074\\
\hline
\end{tabular}
\end{table*}

\subsection{Exploring in detail the Mass-SFR-Z plane}
\label{sec:FP}

\citet{lara10a} proposed a linear relation between the three parameters involved
in the current analysis (oxygen abundance, stellar mass and SFR) in the form
of a plane in this three dimensional space that they called 
the Fundamental Plane (FP). The three parameters
would then be related following the functional form:
\begin{equation}
\mathrm{\log(M_*/M_\odot)}=\alpha\mathrm{\log(SFR/M_\odot/yr)}+\beta\mathrm{(12+log(O/H))}+\gamma
\label{eq:ll_FP}
\end{equation}
They found that the distribution of data points is well represented by the following
parameters: $\alpha=1.122$, $\beta=0.474$ and $\gamma=-0.097$, and
that by taking into account the SFR the new proposed relation would
present a clearly lower scatter ($\sim$0.16 dex) than the MZ relation
(for which they reported a scatter of 0.26 dex). A more recent update
of these results \citep{lara13}, using a combination of SDSS and GAMA data,
confirms the presence of this relation, although the reduction of the scatter is less 
strong (0.2 dex for the FP relation, compared to 0.15 
dex of the MZR and 0.35 for the SFMS). In both cases the reported dispersions are
larger than the ones reported for the MZR using similar datasets in
the literature (e.g., T04), and much larger than the one of the
proposed FMR. Actually, the proposed relation seems to be more
directly connected to the well known SFMS of galaxies
\citep[e.g.][]{brinchmann04}, since the parameter connecting the SFR
and the stellar mass corresponds very nearly to the slope of this relation 
\citep[i.e., $\sim$0.8,][]{mariana16}. In this regard, the introduction of the 
oxygen abundance does not seem to produce a significant reduction of the scatter 
in the SFMS. The scatter in general is around $\sim$0.2-0.3 dex over a wide redshift 
range \citep[e.g.~][]{Speagle14}.

Contrary to previous proposed relations between the stellar mass and the oxygen abundance, 
these authors propose that the relation between
both parameters is linear, departing from the usual shape shown
before. In Sec. \ref{sec:residuals} we have already shown that we cannot
reproduce the observed distribution of the residuals of the MZR and
pMZR with the relation proposed by \citet{lara10a} (Fig. \ref{fig:dMZ}), and
that imposing a linear relation of those residuals with the SFR
does not improve its scatter. However, since we have adopted a different
calibrator for the oxygen abundance and we use a significantly
different sample, maybe the actual parameters for the proposed functional
form could be different. Therefore, we have fitted equation \ref{eq:ll_FP} to
our dataset to provide the best representation of the data considering the
functional form suggested by \citet{lara10a}.

\begin{table}
\caption{Parameters derived for the linear M-SFR-Z relation, when considering the functional form suggested by \citet{lara10a} (Eq.\ref{eq:ll_FP}), based on the described fitting procedure. We include the best fit $\alpha$, $\beta$ and $\gamma$ parameters of the equation for each different calibrator, together with the standard deviation after subtraction of the derived model ($\sigma$ FP-res). As in previous tables some values are represented to the third decimal to show any possible difference, although we consider that the values are significant only to the second decimal.}
\label{tab:FP}
\begin{tabular} {c c c c c}
\hline
Metallicity   & \multicolumn{3}{c}{FP Best Fit} & $\sigma$ FP-res\\
 Indicator & $\alpha$  & $\beta$  & $\gamma$  & (dex) \\
\cline{2-4}
O3N2 & 0.19 $\pm$ 0.02 & -0.054 $\pm$ 0.022 & 8.044 $\pm$ 0.054 & 0.074\\
PP04 & 0.27 $\pm$ 0.03 & -0.078 $\pm$ 0.027 & 8.062 $\pm$ 0.065 & 0.106\\
N2   & 0.21 $\pm$ 0.02 & -0.059 $\pm$ 0.022 & 7.972 $\pm$ 0.055 & 0.076\\
ONS  & 0.23 $\pm$ 0.03 & -0.041 $\pm$ 0.027 & 7.916 $\pm$ 0.066 & 0.102\\
R23  & 0.13 $\pm$ 0.03 & -0.063 $\pm$ 0.024 & 8.155 $\pm$ 0.058 & 0.084\\
pyqz & 0.35 $\pm$ 0.03 & -0.056 $\pm$ 0.031 & 8.117 $\pm$ 0.077 & 0.144\\
t2   & 0.21 $\pm$ 0.03 & -0.057 $\pm$ 0.024 & 8.302 $\pm$ 0.058 & 0.083\\
M08  & 0.31 $\pm$ 0.03 & -0.105 $\pm$ 0.031 & 8.123 $\pm$ 0.076 & 0.143\\
T04  & 0.19 $\pm$ 0.03 & -0.078 $\pm$ 0.029 & 8.357 $\pm$ 0.070 & 0.121\\
EPM09& 0.06 $\pm$ 0.02 & -0.017 $\pm$ 0.022 & 8.391 $\pm$ 0.054 & 0.071\\
DOP09& 0.59 $\pm$ 0.04 & -0.200 $\pm$ 0.037 & 7.488 $\pm$ 0.089 & 0.194\\
\hline
\end{tabular}
\end{table}

Table \ref{tab:FP} shows the result of this analysis, including
the best fit parameters for each calibrator together with
the standard deviation of the residuals. The comparison of these
standard deviations with the ones shown in Tab. \ref{tab:val}, 
shows that in general the proposed functional form decreases the
dispersion marginally for all calibrators when compared
with the MZR. However, it does not provide any improvement
with respect to the pMZR relation, that, as indicated before, does 
not include the SFR as a third parameter. Further, 
assuming just a linear dependence on stellar mass, the 
reported scatter of the residuals is very similar to the one
reported here. In summary, we cannot reproduce the results
found by \citet{lara10a}, assuming their proposed parametrization. 

\begin{figure*}
\includegraphics[width=\columnwidth]{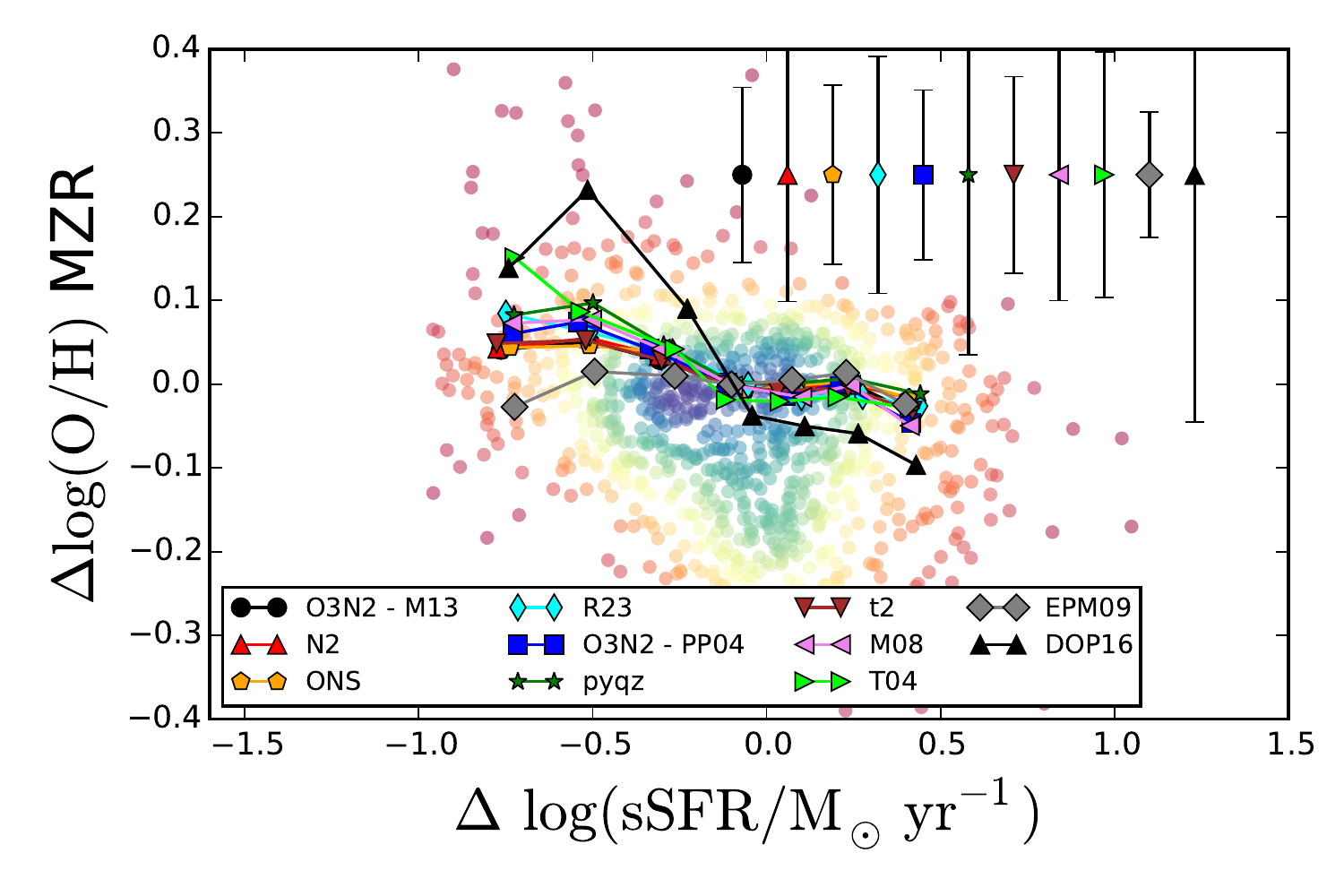}\includegraphics[width=\columnwidth]{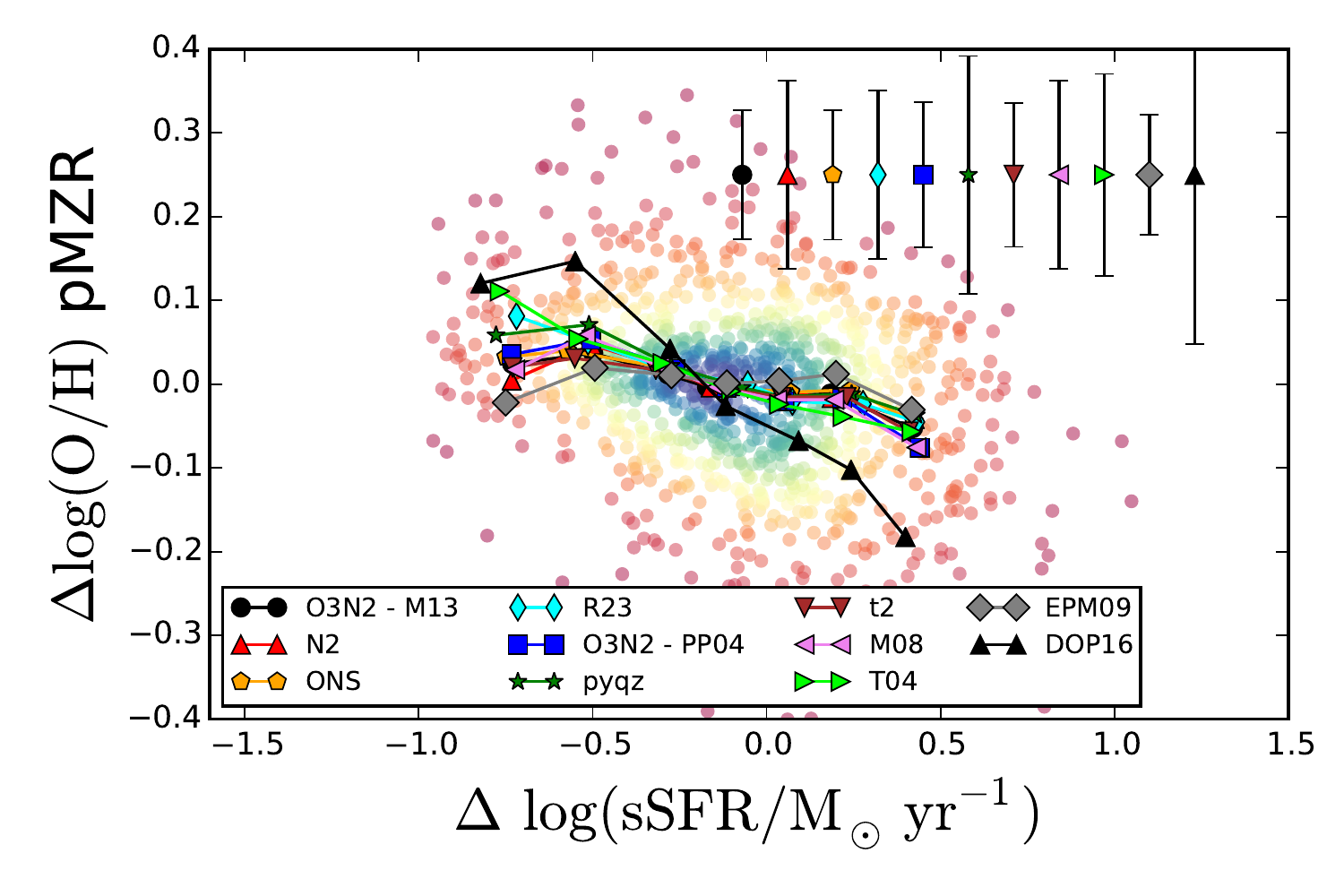}
\caption{Distribution of the residuals of the oxygen abundance after removal of the dependence with the stellar mass (the MZR), as a function of the residuals of the sSFR after removal of the dependence on stellar mass (the SFMS). Solid circles represent the individual values when adopting the PP04 calibrator (color codded by the density of points), and the line-connected symbols represent the median value of the oxygen abundance residuals as a function of the median value of the sSFR in a set of bins (in this later parameter). The symbols are the same as the ones adopted in Fig.\ref{fig:dMZ}. } 
\label{fig:dMZ_M}
\end{figure*}

\begin{table*}
\caption{Results of the study of the possible relation between the residuals of the oxygen abundance and the sSFR after removal of the corresponding dependences on stellar mass (i.e., the MZ and SFMS relations), for the different estimations of the oxygen abundances shown in Fig. \ref{fig:dMZ_M}. We include the Pearson correlation coefficient between the two parameters (r$_c$), the zero-point ($\alpha$) and the slope ($\beta$) for the best fit linear relation, and the standard deviation when this relation is subtracted ($\sigma$-res). We report the values for the residuals of both functional forms adopted for the mass-metallicity relation, the MZR and the $p$MZR parametriz}
\label{tab:salim2}
\begin{tabular} {c c c c c c c c c }
\hline
Metallicity   & r$_c$ & \multicolumn{2}{c}{$\Delta\log{O/H}$-MZR vs. $\Delta\log{(sSFR)}$} & $\sigma$-res & r$_c$ & \multicolumn{2}{c}{$\Delta\log{O/H}$-pMZR vs. $\Delta\log{(sSFR)}$} & $\sigma$-res  \\
 Indicator & & $\alpha$ (dex) & $\beta$ (dex $/ \log(yr^{-1}))$ & (dex) & & $\alpha$ (dex) & $\beta$ (dex $/ \log(yr^{-1}))$ & (dex) \\
\cline{3-5}
\cline{7-9}
O3N2-M13 & -0.287 &  0.01 $\pm$ 0.01 & -0.06 $\pm$ 0.01 & 0.096 & -0.290 & -0.01 $\pm$ 0.01 & -0.06 $\pm$ 0.01 & 0.074 \\
PP04     & -0.286 &  0.01 $\pm$ 0.01 & -0.08 $\pm$ 0.01 & 0.139 & -0.288 & -0.02 $\pm$ 0.01 & -0.09 $\pm$ 0.02 & 0.108 \\
N2-M13   & -0.297 &  0.01 $\pm$ 0.01 & -0.05 $\pm$ 0.01 & 0.098 & -0.323 & -0.01 $\pm$ 0.01 & -0.05 $\pm$ 0.02 & 0.074 \\
ONS      & -0.220 &  0.01 $\pm$ 0.01 & -0.05 $\pm$ 0.01 & 0.136 & -0.235 & -0.01 $\pm$ 0.01 & -0.06 $\pm$ 0.01 & 0.099 \\
R23      & -0.310 & -0.01 $\pm$ 0.01 & -0.10 $\pm$ 0.01 & 0.095 & -0.278 & -0.01 $\pm$ 0.01 & -0.11 $\pm$ 0.01 & 0.084 \\
pyqz     & -0.224 &  0.02 $\pm$ 0.01 & -0.09 $\pm$ 0.01 & 0.206 & -0.237 & -0.01 $\pm$ 0.01 & -0.08 $\pm$ 0.01 & 0.138 \\
t2       & -0.289 & -0.01 $\pm$ 0.01 & -0.07 $\pm$ 0.01 & 0.108 & -0.299 & -0.01 $\pm$ 0.01 & -0.06 $\pm$ 0.01 & 0.082 \\
M08      & -0.287 &  0.01 $\pm$ 0.01 & -0.10 $\pm$ 0.01 & 0.141 & -0.300 & -0.02 $\pm$ 0.01 & -0.09 $\pm$ 0.02 & 0.108 \\
T04      & -0.263 &  0.01 $\pm$ 0.01 & -0.16 $\pm$ 0.03 & 0.141 & -0.248 & -0.02 $\pm$ 0.01 & -0.14 $\pm$ 0.01 & 0.119 \\
EPM09    & -0.136 & -0.01 $\pm$ 0.01 &  0.01 $\pm$ 0.02 & 0.073 & -0.142 & -0.01 $\pm$ 0.01 & -0.01 $\pm$ 0.02 & 0.070 \\
DOP09    & -0.316 & -0.01 $\pm$ 0.02 & -0.24 $\pm$ 0.05 & 0.273 & -0.348 & -0.06 $\pm$ 0.01 & -0.26 $\pm$ 0.03 & 0.194 \\
\hline
\end{tabular}
\end{table*}

\subsection{The [O/H]-SFR relation after subtraction of the stellar mass dependences}
\label{sec:salim}

The putative secondary relation of the oxygen abundance with the SFR can be explored
after removal of the primary dependence of the two involved parameters on stellar mass. 
Following \citet{2014ApJ...797..126S}, we select
those galaxies that are located within the so-called SFMS to perform this analysis,
removing the retired or partially retired galaxies from the sample. In doing so
we adopted the parametrization presented by \citet{mariana16} for the
relation between the SFR and the stellar mass. Thus, we select only
those galaxies that fulfill the following criteria:

\begin{equation}
\log{(SFR/M_\odot/yr^{-1})}>-9.58+0.835\log{(M/M_\odot)}
\label{eq:sSFR_cut}
\end{equation}

This selection has a limited effect on the sample, that now comprises 1020 galaxies. Indeed,
not making that selection does not change significantly the results, but we prefer to keep it
as it was the procedure adopted in previous studies. 

Once we have performed that selection we remove the dependence of the sSFR on stellar mass 
by adopting the SFMS relation derived by \citet{mariana16}. We have checked that it 
is a good representation of the star-forming galaxies
in the current sample. This defines the parameter:
\begin{equation}
\Delta\log{(sSFR)}=\log{(sSFR/yr^{-1})}+8.34+0.19\log{(M/M_\odot)}
\label{eq:Delta_sSFR}
\end{equation}
as the residual of the sSFR with respect to the SFMS. By construction this parameter does 
not retain any dependence on stellar mass. Therefore, to do this analysis adopting either 
the sSFR or the SFR does not produce any significant change in the results, as has been 
shown for other samples \citep[e.g.][]{sanchez17}.

Then, we remove the dependence of the oxygen abundance on the
stellar mass by adopting a similar procedure. However, in this case we have two possible
parametrizations for this dependence, as discussed in Sec. \ref{sec:MZ}, that we 
defined as MZR (Eq.\ref{eq:fit}) and pMZR (Eq.\ref{eq:poly}), using the residuals discussed
in Sec. \ref{sec:residuals}, defined as $\Delta$MZR and $\Delta$pMZR. Adopting this procedure, 
we totally remove the dependence of the two (oxygen abundances and SFRs) on stellar mass. 
We should note that selecting particular ranges of masses and looking for possible secondary 
dependences with the SFR does not guarantee that the primary dependence with the mass is 
totally removed. This procedure was adopted in previous sections and implemented in 
several previous studies \citep[e.g.][]{2014ApJ...797..126S}. However, the selection of a
range of parameters, even a narrow one, may retain part of the dependence with the stellar mass.

Figure \ref{fig:dMZ_M} shows the distribution of the two considered residual abundances,
$\Delta$MZR and $\Delta$pMZR, over the residuals of the specific star-formation rate 
$\Delta\log{sSFR}$, for the different calibrators discussed in this article. The cloud 
of points corresponds to the distribution for the O3N2-PP04 calibrator, with the
distribution of the remaining calibrators represented by the average
value in a set of bins in $\Delta\log{sSFR}$ of 0.3 dex width. We selected this calibrator 
just because it is at the average location of the remaining ones, due to its mixed 
nature \citep[e.g.][]{sanchez17}, but beyond that it is irrelevant which calibrator 
is shown in the figure. For the $\Delta$MZR most of the calibrators show no clear 
trend between both residuals, and there is
only a very mild trend in the case of the $\Delta$pMZR residuals. The calibrator
for which the trend seems to be stronger is the DOP16 one, this particular trend being 
consistent with the predictions of the FMR.

We quantify the possible correlations between both parameters by (1) deriving the
Pearson correlation coefficient (r$_c$) between both parameters for each calibrator,
and (2) by performing a linear regression and determining if the scatter is reduced 
with respect to the one observed in the original MZR distribution. The results of this
analysis are listed in  Table \ref{tab:salim2}, including the r$_c$ coefficient together 
with the parameters of the linear fitting and the standard deviations of the residuals after 
subtraction of the best fit model. The reported correlation coefficients indicate that 
across all the calibrators there is either a weak correlation ($r_c\sim -$0.35) or no 
correlation ($r_c\sim -$0.14) between the residuals. This result has a very high 
significance level, with a probability larger than 99.99\% in all
the cases. Consistently, the slopes reported by the linear regression are very small,
and there is little or no reduction of the scatter. 

In summary, our results do not show a clear secondary relation between the oxygen
abundance and the SFR after removal of the dependence of both on stellar mass.

\begin{figure}
\includegraphics[width=\columnwidth]{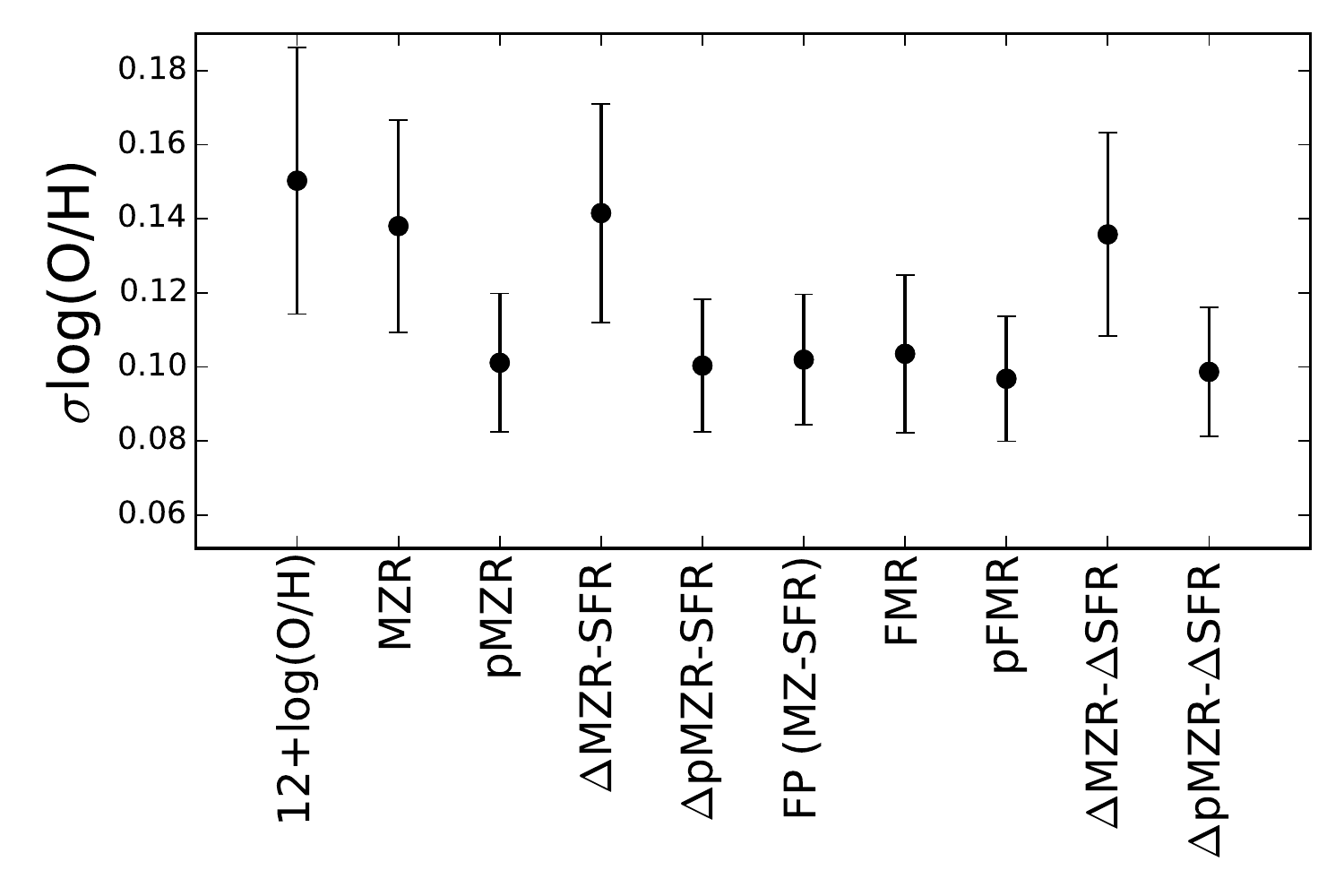}
\caption{Average standard deviations of the distributions of oxygen abundances for each of the considered calibrators for (1) the original distribution of abundances, (2) the residuals from the best fit MZR and pMZR parametrization, (3) the residuals from the linear regression of the residuals after subtraction of the best fit MZR and pMZR models, (4) the residuals of the best fit Fundamental Plane, (5) the residuals of the best fit FMR and pFMR models, and (5) the residuals of the best linear regression of the analysis of the distribution of the oxygen abundance against the SFR after subtraction of the dependence on stellar mass. The error bars indicate the standard deviation around the mean values derived for each calibrator in each of the considered analysis. This figure summarizes the content of all the results in the present study.} 
\label{fig:std}
\end{figure}

\section{Discussion}
\label{sec:dis}

We explore the mass-metallicity relation based on the most recent dataset provided by the 
SAMI survey comprising a total of 2273 of galaxies. For 1044 of those we derive the stellar 
mass, the oxygen abundance and the star-formation rate. In the current
analysis we followed the procedures applied on previous IFU surveys, for a sample with 
a size between the one explored by \citet{sanchez17} for the CALIFA survey and the one 
explored by \citet{bb17} for the MaNGA survey. Adopting a set of different abundance 
calibrators we determine the shape of the MZ relation in the most general way,
exploring the possible dependence with the star-formation rate. The previously reported 
trend of this relation described in many different publications \citep[e.g.][]{tremonti04} 
is confirmed. We found that this shape is described well, for most of the adopted calibrators, 
by a linear+exponential functional form like the one adopted by \citet{sanchez14b} (MZR 
parametrization). However, it is much better described by a fourth order polynomial function, 
similar to the one adopted by \citet{2008ApJ...681.1183K} (pMZR parametrization). The main 
difference between the different abundances calibrators is found in the asymptotic value 
found at the high mass end, and in the scatter around the best reported relations. The 
reported relation is as tight as the one usually reported for single-aperture spectroscopic 
surveys \citep[$\sim$0.1 dex][]{tremonti04} for most of the adopted calibrators,
but not as tight as the one found in previous IFU surveys 
\citep[$\sim$0.05 dex][]{sanchez13,sanchez17,bb17}. Those calibrators based on 
photoionization models present the larger scatter around the mean distribution over stellar 
mass, as already noticed by \citet{sanchez17}, although the differences are much larger 
than in the case of previous IFU studies. The smaller field-of-view and somewhat more 
coarse spatial sampling of the SAMI data may introduce a larger uncertainty in the oxygen 
abundance derivation based on the radial profile when compared to the CALIFA ones. However, 
the spatial sampling in SAMI is better than in MaNGA, in general, and the FoV is similar 
for at least 1/3 of the sampled objects. Therefore, this should not be a major driver 
for potential differences. Indeed, the main trends remain very similar among the three 
datasets, as shown in Fig. \ref{fig:comp_IFS}.

We study the different scenarios proposed in the literature to describe a possible secondary 
relation of the oxygen abundance and the mass-metallicity relation with the SFR in the broadest 
possible way. We assume that the abundance depends primarily on stellar mass, and explored 
(i) if the residuals of the best fit MZ relations adopting two different functional forms 
do indeed correlate with the SFR, (ii) how well the data are represented by so-called 
Fundamental Mass-Metallicity Relation proposed by \citep[][]{mann10}, or a relation with 
a similar functional form and different fitted parameters, (iii) how well the data are 
represented by a functional form similar to the so-called Fundamental Plane of 
Mass-SFR-Z, proposed by \citet{lara10a}, and (iv) if there is a relation between the 
residuals of both the oxygen abundances and the SFRs once considered their respective 
dependence on stellar mass. In none of these cases did we find either a strong or 
moderate secondary correlation with the SFR or a clear improvement in the scatter when 
introducing this third parameter. This result is summarized in Figure \ref{fig:std}, 
where the standard deviation of the oxygen abundance around the mean value is shown together 
with the standard deviations of the residuals for each adopted functional form. We represent 
the mean value (central point) and the standard deviation (error bars) among the values 
reported for the eleven calibrators considered along this study. The figure illustrates 
clearly that there is no reduction in the scatter when by introducing any of the analyzed 
secondary relations with the SFR neither for the MZR nor for the pMZR parametrizations. 
If there is any improvement, it is marginal, in particular if the MZR-parametrization is 
adopted, and for certain calibrators, and/or in some particular ranges of the analyzed 
parameters. This is the case of the calibrators with the larger initial scatter, which 
may indicate that they present secondary trends with other parameters, like the ionization 
strength, somehow correlated with the SFR. 

\citet{sanchez17} reported a possible dependence with the SFR confined to the regime 
of low masses (M$<$10$^{9.5}$M$_\odot$). We cannot confirm that result. We only find 
a mild improvement in this regime ($\sim 0.02$ dex in the mean value and $\sim$0.02 dex 
in the dispersion), when comparing a generalized FMR relation (allowing to vary the 
parameter that introduces the dependence with the SFR) with the MZR-parametrization. 
However, when introducing a polynomial function (the pMZR-parametrization), the improvement 
is mimimized (no improvement in the mean value and $\sim$0.02 dex in the dispersion). 
Since this latter functional form was not discussed in detail in this previous article 
we cannot determine if the nature of the difference is due to the use of this new 
parametrization of the relation between the oxygen abundance and the stellar mass. 
However, we can claim that if there is a secondary relation this would be confined to 
the low stellar mass regime.

In general our results are consistent with those reported by previous IFU surveys 
\citep{sanchez13,bb17,sanchez17}. These results thus support the claim by \citet{sanchez13} 
that there is no need to introduce the SFR as a secondary parameter to describe the 
distribution of oxygen abundances. Other authors are not able to reproduce this secondary 
relation either \citep[e.g.][]{kash16}, or found a much weaker ones \citep[e.g.][]{telf16} 
in different datasets. The current analysis stresses the need for a good parametrization 
of the relation between oxygen abundance and its primarily tracer, the stellar mass.
Being constrained to a particular functional form, like the MZR-parametrization, may 
lead to the wrong impression that there is a possible secondary dependence with the SFR 
for some particular calibrators (although the effect is not significant). The introduction 
of different functional forms, like the polynomial pMZR-parametrization blurs the effect 
of the SFR. This is particularly important since the dependence of the oxygen abundance 
on stellar mass is not linear, but the one between the SFR and the mass is. Indeed, in 
some cases the stronger secondary correlation between the oxygen abundance and the SFR 
is described at the location of the bend in the MZ distribution 
\citep[e.g., $\sim$10$^{9.5}$-10$^{10}$M$_\odot$][]{2014ApJ...797..126S}, 
or in the low mass range \citep[e.g.][]{2010ApJ...715L.128A,sanchez17}

There is no clear or single reason why IFS analyses consistently report a lack of a global 
secondary relation between the oxygen abundance and the SFR while analyses based on single 
fiber spectroscopic surveys clearly found it. Among the discussed reasons in the literature 
are: (i) the observational setups, (ii) the analyzed samples, (iii) how the different 
quantities are derived, (iv) which are the functional forms that describe better the 
main relations, and finally (iv) what we understand to be a secondary relation. Most of 
these reasons have been discussed extensively in previous articles. We revise them in 
light of our new results.

\subsection{Aperture effects}

The effects of aperture in the SDSS derivation of extensive and intensive parameters is 
a key issue in the different discussions. Even what is understood to be an aperture effect 
is different. Recent studies \citep[e.g.][]{eduge18}, attempt to minimize the effect of 
not sampling the complete optical extension of galaxies using single aperture spectra by 
selecting only those galaxies that fit within the aperture for a certain scale (like the 
effective radius). However, recent explorations of the variation of the different involved 
properties for different apertures, in particular the star-formation rate and the oxygen 
abundance, have shown that this selection is not enough due to the intrinsic variations 
of the profiles and the dependence with other properties, in particular morphology 
\citep[e.g.][]{iglesias16,duarte17}. Therefore, the effect of using aperture-fixed 
measurements in the analysis of secondary relations between parameters is far from 
being understood. However, it is clear that it may have an effect in this kind of 
exploration. Indeed, \citet{sanchez13} already showed that simulating SDSS-like single 
aperture spectroscopic data using spatially resolved IFS data may lead to a secondary 
relation between the oxygen abundance and the SFR similar to the one reported by 
\citet{mann10}. This effect was also reported by more recent results \citep{telf16}, 
although it was not discussed in detail.

\subsection{Sample Selection}

Analyzing the possible relations using different samples may lead to
different results, if the selection effects are not clearly
understood. It is this basic reason why we have embarked in a similar
exploration using similar techniques on different datasets. In
\citet{bb17} and \citet{sanchez17} we perform a similar analysis as
the one described here for the MaNGA and CALIFA IFS datasets,
respectively. Despite the differences in the samples
\citep[e.g.][]{sanchez17}, different selection criteria
\citep{wake17,walcher14}, and observational setup and strategy
\citep{sanchez12a,law15}, both results are very similar. Even
more, the reported results are very consistent with the one found
here. Therefore, we consider that this approach to the problem has
minimized or at least limited the effects of sample
selection. However, it is still the case that samples observed using
spatial resolved spectroscopy \citep[e.g.][]{manga} are an order of
magnitude less numerous than those observed by single aperture
spectroscopy \citep[e.g.][]{york00,liske2015}. Thus, it may be
the case that the proposed secondary relations are significant only
when a much larger number of galaxies is sampled. We will require an
IFS survey with a similar number of galaxies as the SDSS or GAMA
surveys to explore that possibility.

\subsection{Derivation of the involved parameters}

The derivation of oxygen abundances has systematic uncertainties that have been widely 
explored in the literature \citep[e.g.][just cite a few]{kewley08,angel12,sanchez17}. 
These systematic differences may induce secondary relations if the adopted abundance 
indicators correlate with other properties, like the ionization strength, that indeed 
may correlate with the SFR \citep[e.g.][]{poet18}. It is clearly beyond the scope of 
this article to disentangle the long standing discussion on oxygen abundance derivation. 
However, since this could be one of the reasons for the reported differences, in the 
current article we explore a broad range of oxygen abundance calibrators. In general, 
the results derived for all calibrators are very similar. Of the eleven adopted calibrators 
there is only one (DOP16) for which we could report a very weak secondary trend with 
the SFR. Therefore, we consider that differences in the calibrators are not the main 
driver for the differences in the results.

In addition to the intrinsic uncertainties when deriving the oxygen abundance, this 
derivation is affected by the possible mix of ionization conditions in the considered apertures. 
This effect has been addressed by different authors \citep[e.g.][]{davies16}, showing 
how the degradation of the resolution and the mix with ionization conditions for which 
the calibrators are not valid increase the dispersion in the derived abundances 
\citep[e.g.][]{zhang16}, and change the shape of the abundance distribution in 
galaxies \citep[e.g.][]{mast14}. This is particularly relevant since oxygen abundances 
present gradients \citep[e.g.][]{sanchez13}, whose shape may depend on the stellar 
masses \citep[e.g.][]{2016A&A...587A..70S,belf16a}. Aperture corrections 
\citep[e.g.][]{2016ApJ...826...71I} or aperture matching between different 
galaxy types \citep[e.g.][]{eduge18} cannot account for these effects.

The adopted analysis is based on the exploration of the secondary relation using as 
characteristic abundance of the whole galaxy that measured at the effective radius. 
This is the same approach adopted by \citet{sanchez13}, \citet{bb17} and \citet{sanchez17}. 
This is unfeasible using single aperture spectroscopic data, despite the efforts of doing 
so \citep[e.g.][]{eduge18}. This may also introduce a potential difference, since those 
analyses are basically sampling the central oxygen abundance in galaxies (or at least 
weighting it much more, since single aperture are light-weight averaged). We know that 
there is a drop or flatenning in the central oxygen abundances in certain galaxies, and 
this drop is more evident for more massive galaxies \citep{2016A&A...587A..70S,belf17,laura18}. 
It is possible that the secondary relation with the SFR found in single-aperture 
spectroscopic surveys is an effect of this central drop, which it is not present across 
the whole disk, as traced by the current analysis. Therefore, where the oxygen abundance 
is measured within a galaxy may have an implication on the results too.

Beyond the differences in the derivation of the oxygen abundances, there are also 
significant differences in the derivation of both the SFR and the stellar masses. 
Most studies adopt the H$\alpha$ luminosity, dust corrected based on the H$\alpha$/H$\beta$, 
assuming a certain ionization condition and particular extinction law, to trace the 
star-formation rate assuming certain calibrators. \citep[e.g.][]{kennicutt89,2015A&A...584A..87C}. 
In this derivation there are strong differences between single-aperture and 
spatially resolved analyses. In the first case all ionizations within the measured 
apertures are mixed \citep[e.g][]{binn94,binn09,sarzi10,sign13,mast14,belf16a,davies16}. It is 
impossible to resolve and separate the effects of each of them, or gauge their relative 
strengths. Even more, since it is needed to use non-linear calculations, it is very difficult 
to trace the effect of co-adding several lines of sight in a single spectrum. Finally, it 
is mandatory to apply an aperture correction over the derived properties, based on assumed 
properties of galaxies \citep[e.g.][]{duarte17}. Adopting a different extinction law, 
a different SFR calibrator, a different selection for star-forming or non-starforming 
galaxies or different aperture corrections may lead to different results.

Finally, the derivation of the stellar masses are not free from biases and uncertainties. 
In the current analysis the stellar mass is derived spaxel by spaxel, spatially resolved, 
based on the stellar surface brightness in areas of $\sim$2kpc$^2$, following a similar 
procedure as the one we adopted for the analysis of the CALIFA and MaNGA datasets. However, 
SDSS based analyses adopt different approaches. In most cases the multi-band photometry 
is used, or single-band photometry with the mass-to-light derived using the single aperture 
spectra. This aperture mostly covers the center of the galaxies, and therefore, the 
mass-to-light ratio may be biased to the central values too. In general, the uncertainties 
in the derivation of the stellar masses 
\citep[$\sim$0.1 dex, e.g.][]{sanchez13,rosa14,Pipe3D_II} are of the order or even larger 
than the scatter around the mass-metallicity relation.

\subsection{Considered functional forms for the MZ-relation}

Throughout this article we have compared the results using two different functional forms 
to characterize the MZ distribution, a linear+exponential shape (MZR) already used in 
previous studies and a fourth order polynomial function (pMZR). We found clear differences 
between the two approaches. The first one produces large scatter (Tab. \ref{tab:val}), with 
a mild reduction of the scatter when introducing the generalized FMR (Tab. \ref{tab:FMR}), 
and a weak correlation of the residuals of the oxygen abundance and the sSFR, after removal 
of the dependence on the stellar mass (Tab. \ref{tab:salim2}). On the other hand, the 
second one produces lower scatter in all the explored relations, with a negligible reduction 
of the scatter when introducing the generalized FMR, and a very weak trend of the oxygen 
abundance residuals with the sSFR, that in any case does not improve the scatter. The 
differences between the two parametrizations of the mass metallicity relation highlight 
the importance of the adopted functional form in this kind of analysis. To our knowledge, 
this is the first analysis that performs such a comparison, showing that the strength of 
the possible secondary correlation with the SFR may depend on the adopted functional form. 
If the adopted functional form  does not describe well the observed MZ-distribution the 
residuals of that trend may still depend on the stellar mass, and therefore produce a 
weak secondary relation with the SFR. We do not know in detail the physical reason for 
the global MZ-relation, and as we will discuss later the reason for its bend and asymptotic 
value in the high mass range is still under debate. Therefore, it may still be the case 
that the imperfect description of the MZ-relation may contribute to the discrepant results.

\subsection{What do we understand by a secondary relation?}

A potential source of discrepancy in the results could be in the actual adopted analysis 
to explore the possible secondary relation with the SFR. For example, \citet{mann10} 
explored a modification of the stellar mass by a new parameter that depends on the SFR 
in the MZ-relation (the $\mu$ parameter), and reported a decrease of the global scatter 
around the best fit parametrization. Other studies explored linear correlations between 
the different involved parameters \citep{lara10a}. In some cases the existence of a 
secondary relation was suggested without analyzing the effects in the scatter after its 
introduction \citep[][]{2014ApJ...797..126S}, which is the main argument against the 
secondary relation by \citet{sanchez13}. Other studies are focused on the details of the 
systematics in the data themselves \citep[e.g.][]{telf16}. In the current study we tried 
to explore the secondary relation in the broadest possible way, adopting not only several 
different calibrators, but also different functional forms to describe the MZ-relation. 
Even more, we repeated the analysis by \citet{mann10}, \citet{lara10a}, \citet{sanchez13} 
and \citet{2014ApJ...797..126S}, and compared the results between the different methods. 
For none of the performed analyses we can report a significant secondary relation with 
the SFR that improves the scatter already found for at least one of the adopted functional 
forms of the original MZ-relation. This disagrees with the { primary} claim by 
\citet{mann10}, that the introduction of FMR reduces the global scatter of the relation 
significantly { (Sec. 4 of that articles)}. 

However, as indicated by \citet{sanchez17}, it is still possible that
the secondary relation does not produce a significant change in the
overall scatter, but improves the description of the distribution in
certain ranges, like the low-mass or low-SFR ranges {, or, as
  suggested by previous analysis, the main driver are extreme
  star-forming galaxies
  \citep[e.g.][]{mann10,2010ApJ...715L.128A,telf16}}. This
interpretation is supported by the apparent better description of the
data offered by the introduction of a generalized FMR, as discussed in
Sec.~\ref{sec:FMR}. Under this interpretation it is still possible to
reconcile the results. Nevertheless, in the current analysis, we do
not find any clear improvement in the description of the distributions
when introducing the SFR in the generalized pFMR functional form in
any of the explored ranges of stellar masses and SFRs. For the
generalized FMR we find just a mild improvement in the high
star-formation and low stellar mass ranges {, which agree qualitatively 
  with the claims by \citet{mann10}}. Therefore, although it is an
appealing scenario, we cannot { verify it with the current
  dataset.}

\subsection{Physical interpretation}

it is important to confirm, constrain or reject the presence or absence of a secondary 
relation with SFR, { in order to constrain the different proposed scenarios
outlined in the introduction} for the metal enrichment, recycling
and mixing in galaxies. {Those can be summarized in two:  (i) an evolution dominated by local
  processes, with little contribution from ouflows, and (ii) an
  evolution in which galactic-scale outflows are a fundamental ingredient in shaping the
  MZ-SFR relation. Based on our results we consider that the first
  interpretation is more likely. Under this assumption, the secondary
  relation is either dominated by effects in the central regions of
  galaxies and/or driven by some particular galaxies, and not in the
  general population of disk galaxies. \citet{mann10} already
  indicated that most of the reduction of the scatter is dominated by
  galaxies with high star formation rates. In the same line
  \citet{2010ApJ...715L.128A} and \citet{telf16} found that the secondary
  relation is mostly driven by galaxies well above the star-formation
  main sequence. These extremely star forming galaxies are rare in the
  local universe, and could be under-represented in the explored IFS surveys,
  and therefore, these surveys are less affected by the secondary relation with
  the SFR in its current form. { However, the fraction of extreme star-forming galaxies
in the SDSS subsample analyzed at least by \citet{mann10} is of the same order
as the one found in the current analyzed SAMI sample (Appendix \ref{sec:SDSS}), and the
SAMI sample covers a wider range of stellar masses (in particular at low M$_*$). 
Therefore, the presence of these particular
galaxies do not seem to be the driver for the differences in the results.}

Nevertheless, it could be possible to reconciliate the two
  views}, if we consider that outflows are concentrated
in the inner regions of galaxies that present strong central
star-formation densities \citep[e.g.][Lopez-Coba et
  al. submitted]{ho14,carlos16}. Under this interpretation the oxygen
abundance in the central regions may be more clearly affected,
although not in all the cases \citep[e.g.][]{jkbb15b}, while the
general distribution remains un-affected by strong outflows. In 
this scenario, maybe the secondary relation reported using data from
the SDSS survey is confined to the central regions, and it is not
visible when using the characteristic oxygen abundance of the
considered galaxies. { An alternative scenario is that shocks
induced by outflows could contaminate the line ratios in the central
regions, following the same argument as the one presented by \citet{davies16}. 
Under this circumstances, the derived oxygen abundances of extremely star 
forming galaxies may present an computational bias. We will explore that 
possibility using IFS data of better spatial resolution, like the one 
explored by \citet{laura18}, in future studies.}

\section{Conclusions}
\label{sec:con}

In summary, the proposed secondary dependence with the SFR of the
MZ-distribution is not confirmed by our analysis of the data provided
by the SAMI survey. { To gauge the presence of a secondary relation we used the 
  requirement that the introduction of that relation decreases the scatter in the
  distribution of data points around the relation}. The oxygen abundance is well described 
by a simple relation with the stellar mass, with a precision and accuracy
that depends mildly on the adopted functional form. The introduction
of the proposed secondary dependence with the SFR appears to represent
the data distribution in a better way only for some particular
functional forms and in no case we can report a general improvement of
the description of the data in terms of a significant decrease of the
scatter around the mean relation. { Only for extremely star-forming
  galaxies and/or a low masses there is a slight trend with the SFR,
  without a significant improvement of the overall scatter.} This
result agrees with previous ones based on the analysis of
integral-field spectroscopic surveys of galaxies. However, it
disagrees with different results based on single spectroscopic surveys
({ that we can reproduce using their data and our analysis}). This
disagreement could be the consequence of several observational
differences between the data, sample effects, procedures applied to
derive the involved parameters and even interpretation of the
results. Even more, there could be a scenario in which both results
agree, if the suggested secondary relation is limited to oxygen
abundances in the central regions.  We will explore that possibility
in future analysis.

\section*{Acknowledgements}

We would like to thank the referee for his/her comments that has improved the content of this manuscript.


We are grateful for the support of a CONACYT grant CB-285080 and funding from the PAPIIT-DGAPA-IA101217 (UNAM) project.

Parts of this research were conducted by the Australian Research Council Centre of Excellence for All Sky Astrophysics in 3 Dimensions (ASTRO 3D), through project number CE170100013.

The SAMI Galaxy Survey is based on observations made at the Anglo-Australian Telescope. The Sydney-AAO Multi-object Integral field spectrograph (SAMI) was developed jointly by the University of Sydney and the Australian Astronomical Observatory. The SAMI input catalogue is based on data taken from the Sloan Digital Sky Survey, the GAMA Survey and the VST ATLAS Survey. The SAMI Galaxy Survey is supported by the Australian Research Council Centre of Excellence for All Sky Astrophysics in 3 Dimensions (ASTRO 3D), through project number CE170100013, the Australian Research Council Centre of Excellence for All-sky Astrophysics (CAASTRO), through project number CE110001020, and other participating institutions. The SAMI Galaxy Survey website is http://sami-survey.org/.

This project makes use of the MaNGA-Pipe3D dataproducts. We thank the IA-UNAM MaNGA team for creating this catalogue, and the ConaCyt-180125 project for supporting them

Funding for the Sloan Digital Sky Survey IV has been provided by
the Alfred P. Sloan Foundation, the U.S. Department of Energy Office of
Science, and the Participating Institutions. SDSS-IV acknowledges
support and resources from the Center for High-Performance Computing at
the University of Utah. The SDSS web site is www.sdss.org.

SDSS-IV is managed by the Astrophysical Research Consortium for the 
Participating Institutions of the SDSS Collaboration including the 
Brazilian Participation Group, the Carnegie Institution for Science, 
Carnegie Mellon University, the Chilean Participation Group, the French Participation Group, Harvard-Smithsonian Center for Astrophysics, 
Instituto de Astrof\'isica de Canarias, The Johns Hopkins University, 
Kavli Institute for the Physics and Mathematics of the Universe (IPMU) / 
University of Tokyo, Lawrence Berkeley National Laboratory, 
Leibniz Institut f\"ur Astrophysik Potsdam (AIP),  
Max-Planck-Institut f\"ur Astronomie (MPIA Heidelberg), 
Max-Planck-Institut f\"ur Astrophysik (MPA Garching), 
Max-Planck-Institut f\"ur Extraterrestrische Physik (MPE), 
National Astronomical Observatories of China, New Mexico State University, 
New York University, University of Notre Dame, 
Observat\'ario Nacional / MCTI, The Ohio State University, 
Pennsylvania State University, Shanghai Astronomical Observatory, 
United Kingdom Participation Group,
Universidad Nacional Aut\'onoma de M\'exico, University of Arizona, 
University of Colorado Boulder, University of Oxford, University of Portsmouth, 
University of Utah, University of Virginia, University of Washington, University of Wisconsin, 
Vanderbilt University, and Yale University.

This study uses data provided by the Calar Alto Legacy Integral Field Area (CALIFA) survey (http://califa.caha.es/).

Based on observations collected at the Centro Astron\'omico Hispano Alem\'an (CAHA) at Calar Alto, operated jointly by the Max-Planck-Institut f\"ur Astronomie and the Instituto de Astrof\'isica de Andaluc\'ia (CSIC).

\appendix

\section{Photometric and spectroscopic masses}
\label{sec:Mass}

The current study of the MZ-relation uses the stellar masses derived by the spectroscopic 
analysis performed by {\sc Pipe3D} on the datacubes provided by the SAMI galacy survey. 
However, due to the limited FoV of the SAMI IFU fiber-bundle, these stellar masses may 
be affected by aperture effects that could alter our results. The SAMI galaxy survey 
provides stellar masses for all the observed galaxies \citep{bryant2015} derived from 
a multi-band spectral energy distribution fitting of optical photometric data extracted 
from the GAMA survey, following the procedure described in \citet{taylor2011}. Those 
stellar masses are not affected by aperture effects. However, being based on photometric 
data they are affected by other biases. In particular, the estimation of the dust 
extinction is less accurate than the one provided by full spectroscopic fitting (due 
to the limited number of photometric bands used). Those stellar masses are derived 
adopting a different IMF \citep{chabrier2003} than the one adopted here 
\citep{Salpeter:1955p3438}, and a different dust attenuation law \citep{calz01}. 
Therefore,  a simple one-to-one agreement between both quantities is not expected. 

\begin{figure}
\includegraphics[width=1.0\linewidth,clip,trim=120 30 130 50]{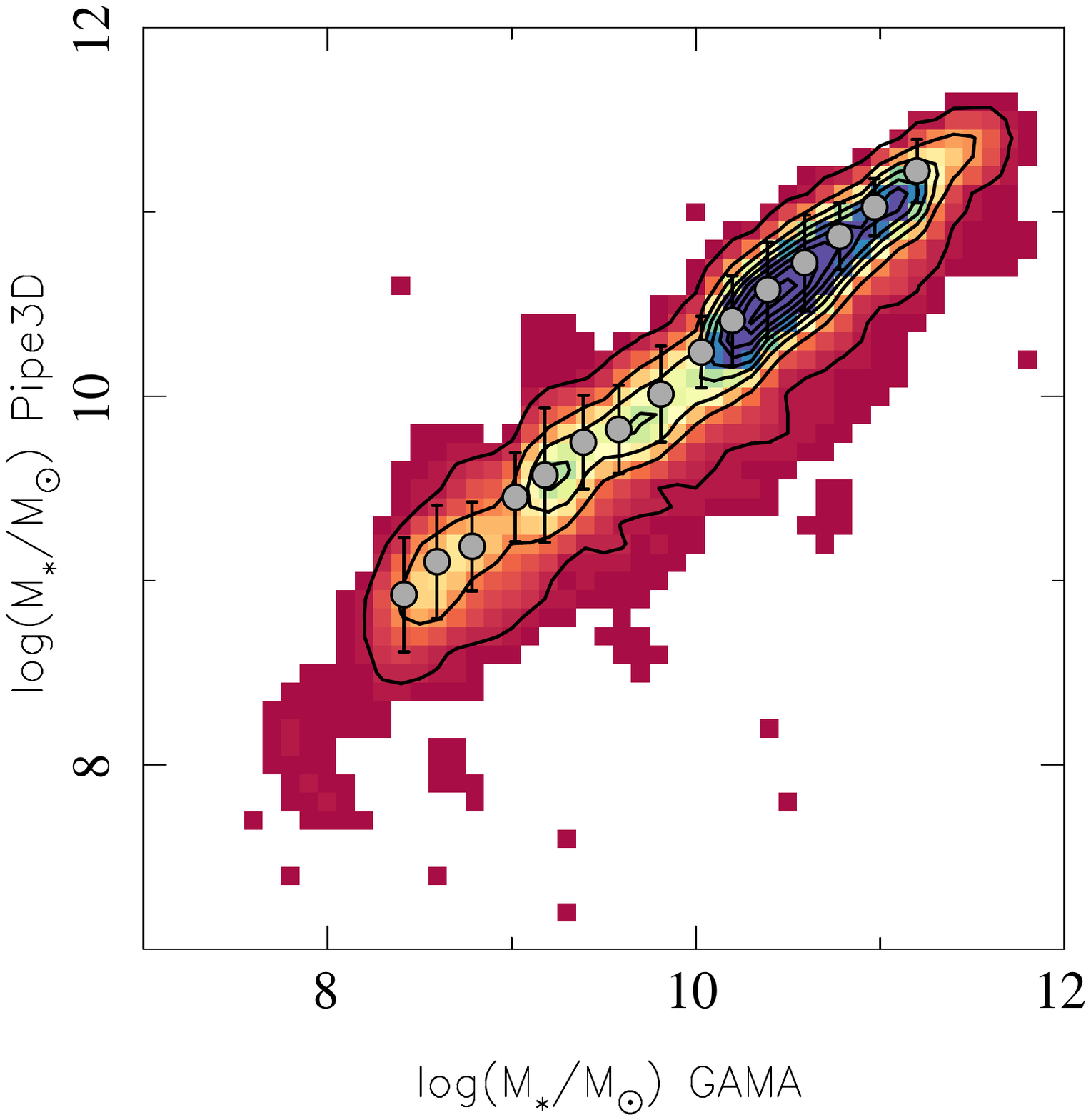}
\caption{Comparison between the stellar masses derived using the photometric information provided by the GAMA survey and those derived by {\sc Pipe3D} extracted from the SAMI datacubes.}
 \label{fig:Mass}
\end{figure}

Figure \ref{fig:Mass} shows the comparison between both stellar masses for the 2263 
galaxies for which both quantities are available. There is a clear relation between 
both quantities, near to a one-to-one relation. On average the stellar masses derived 
by {\sc Pipe3D} are 0.10$\pm$0.23 dex larger than the ones estimated based on the 
photometric data. This offset is a consequence of the combined effect of the different 
adopted IMFs, which will make the {\sc Pipe3D} stellar masses larger by 0.21 dex 
(on average), and the aperture effect, that would make the photometric stellar masses 
slightly larger, compensating partially the IMF difference. Even more, the correlation 
between the two stellar masses deviates from a one-to-one relation, having a slope of 
$\sim$0.82 (when considered, the standard deviation of the difference between the two 
stellar masses decreases to $\sim$0.18 dex). On average, the stellar masses agree 
better for the high stellar mass range (M$_*>$10$^{10}$M$_\odot$), with an offset of 
0.05$\pm$0.15 dex, than for the low stellar mass range (M$_*<$10$^{10}$M$_\odot$), 
with an offset of 0.26$\pm$0.30 dex. Since aperture effects may be similar at any 
stellar mass, due to the sample selection (that does not involve any aperture matching 
dependent on the mass), these differences should be attributed to the way the mass-to-light 
ratio is derived and how the dust attenuation is treated. In both cases, low mass galaxies, 
with a large diversity of stellar populations \citep[e.g.][]{ibarra16} and more relative 
dust content \citep[e.g.][]{sanchez18}, are more likely to be affected by differences 
in both estimates.

Due to these differences we repeated the complete analysis shown in this article 
using the stellar masses derived by the SAMI galaxy survey. This limits the sample to 
993 galaxies, since the filler targets (see Sec. \ref{sample}) do not present stellar 
masses in the SAMI catalogs. Apart from the number of galaxies, we found no significant 
differences in the results when using the photometric based stellar masses. We do not 
reproduce all the plots and tables here for the sake of brevity

\section{Comparison between Pipe3D and LZIFU}
\label{sec:LZIFU}

\begin{figure*}
 \minipage{0.995\textwidth}
\includegraphics[width=0.47\linewidth,clip,trim=10 30 20 30]{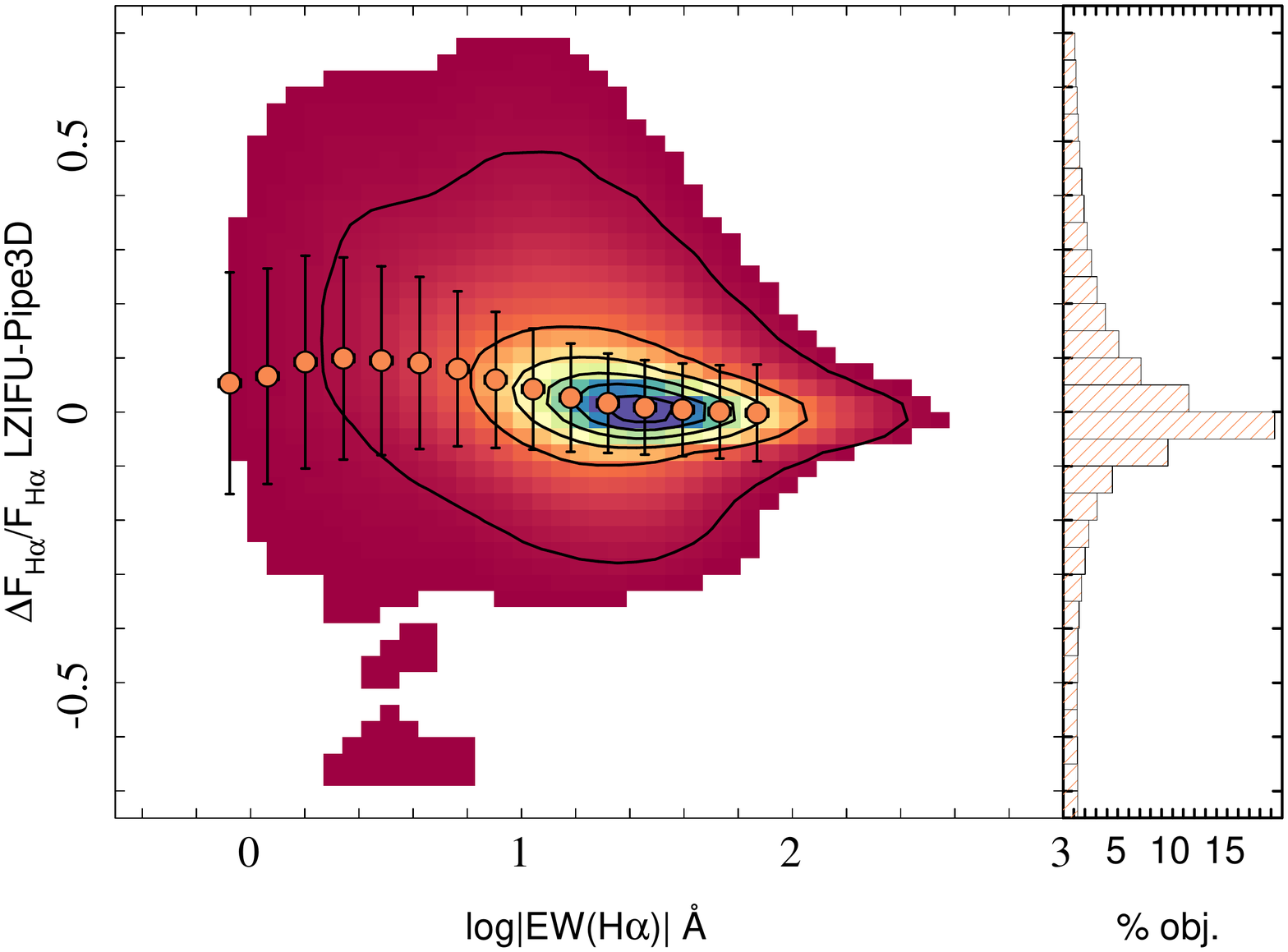}\includegraphics[width=0.47\linewidth,clip,trim=10 30 20 30]{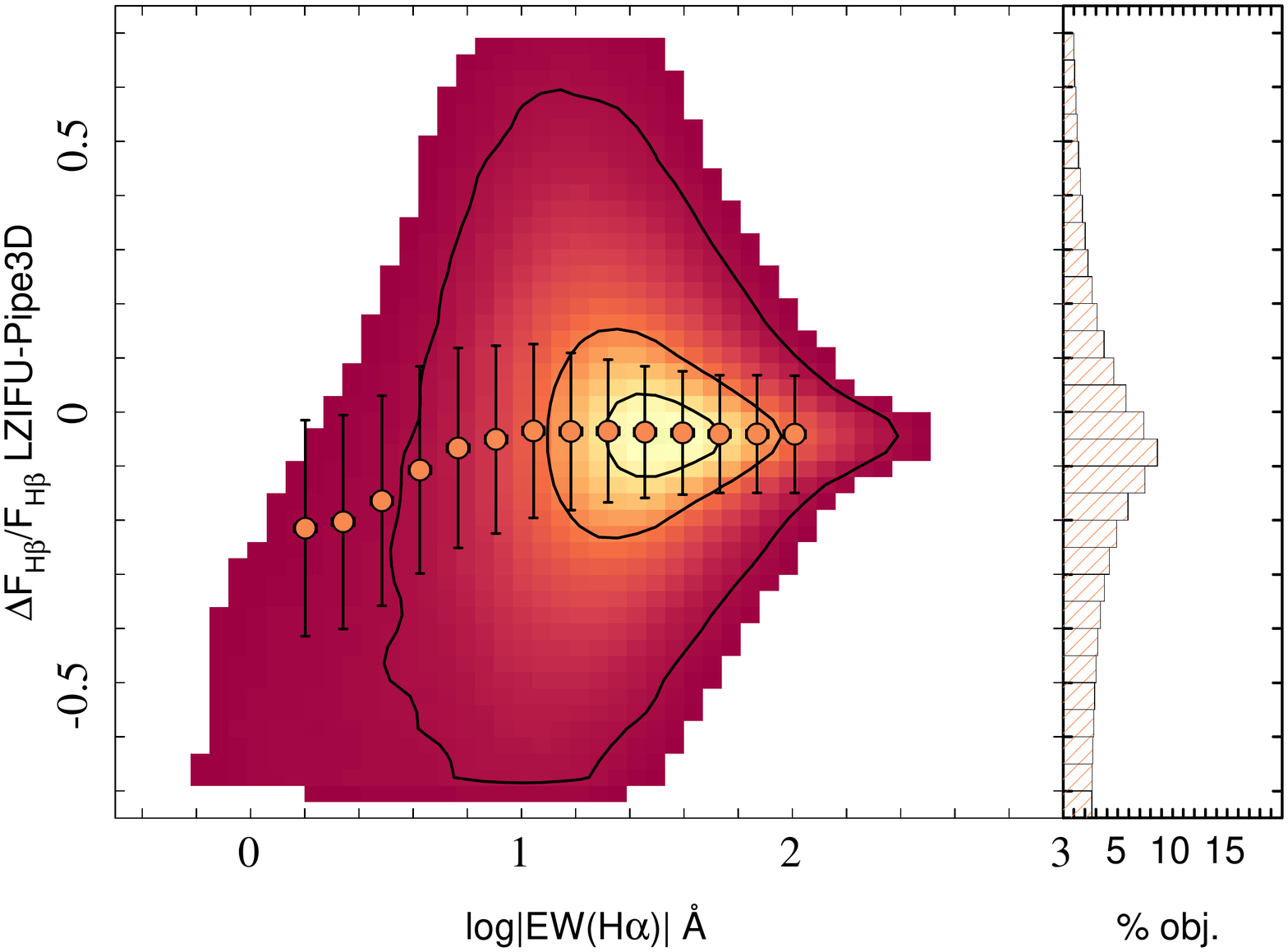}
\includegraphics[width=0.47\linewidth,clip,trim=10 30 20 30]{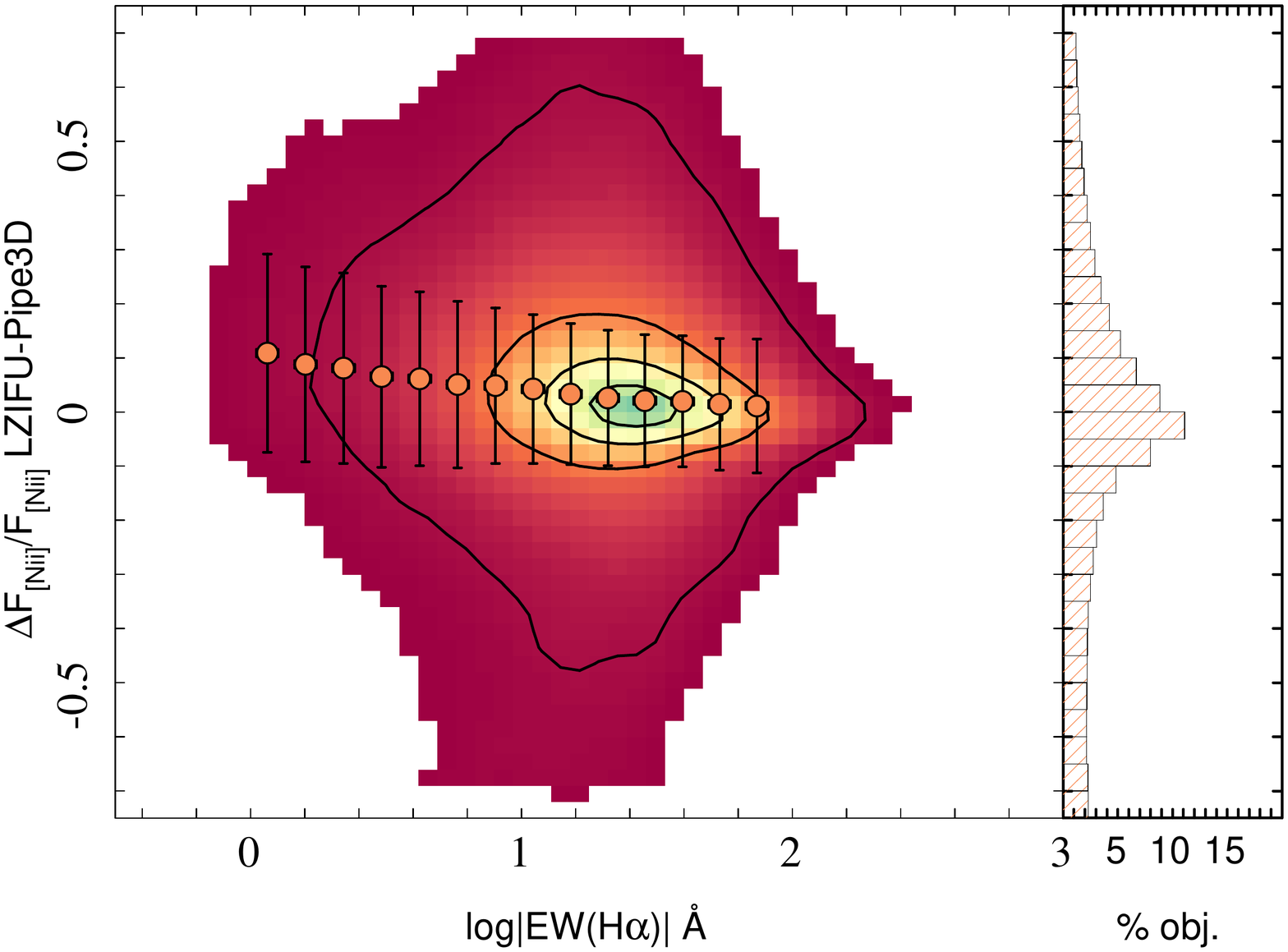}\includegraphics[width=0.47\linewidth,clip,trim=10 30 20 30]{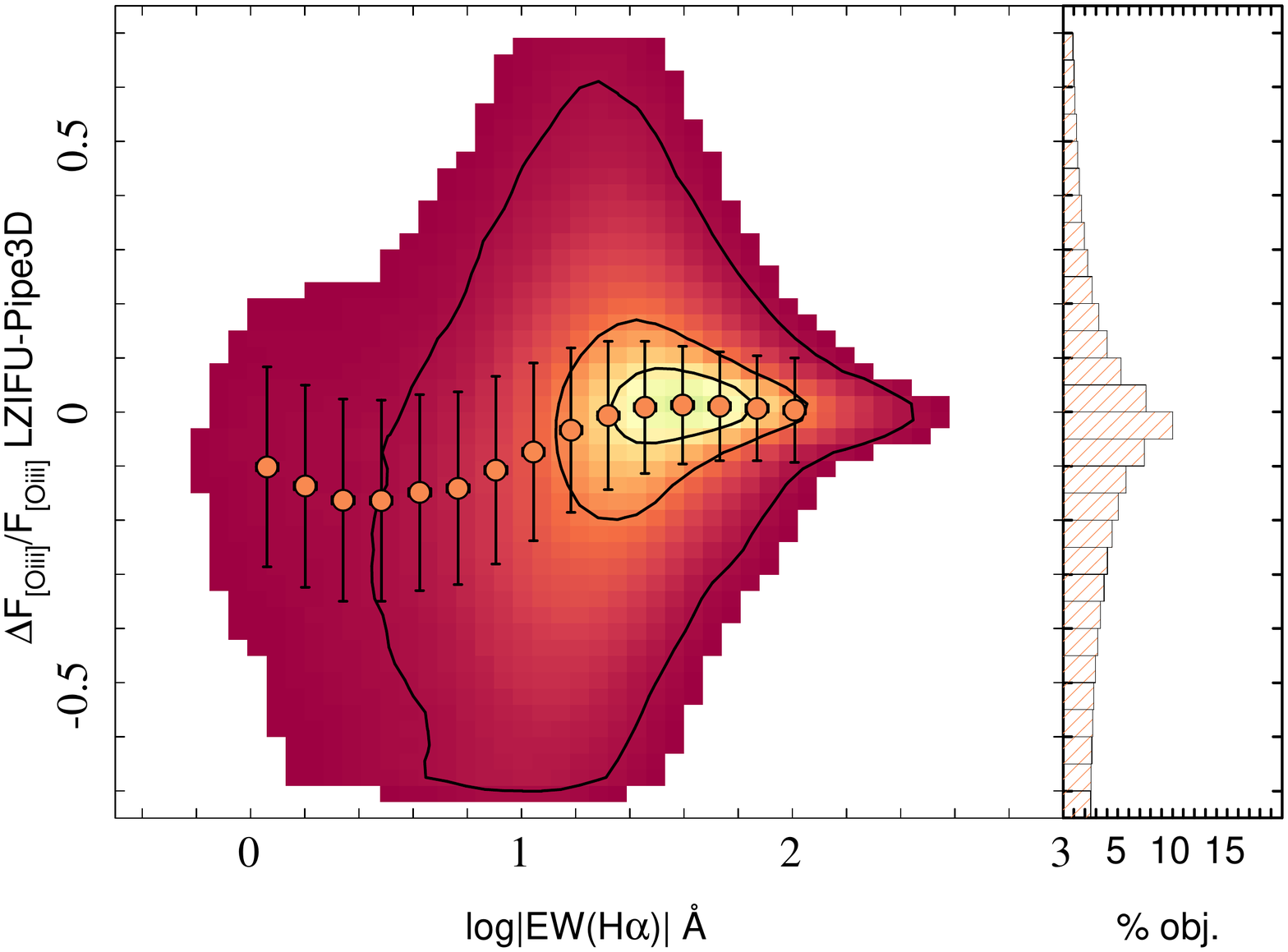}
\includegraphics[width=0.47\linewidth,clip,trim=10 30 20 30]{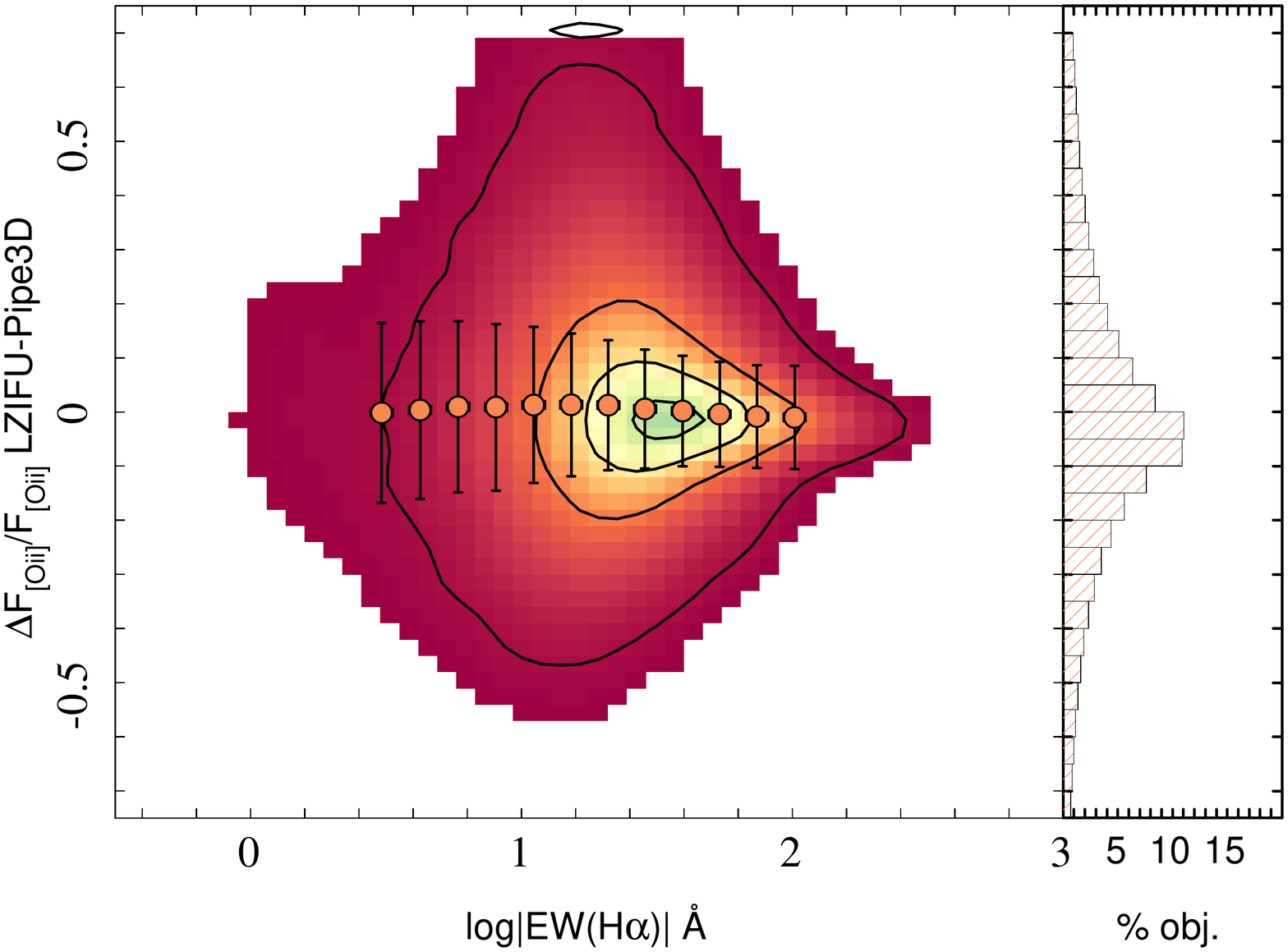}\includegraphics[width=0.47\linewidth,clip,trim=10 30 20 30]{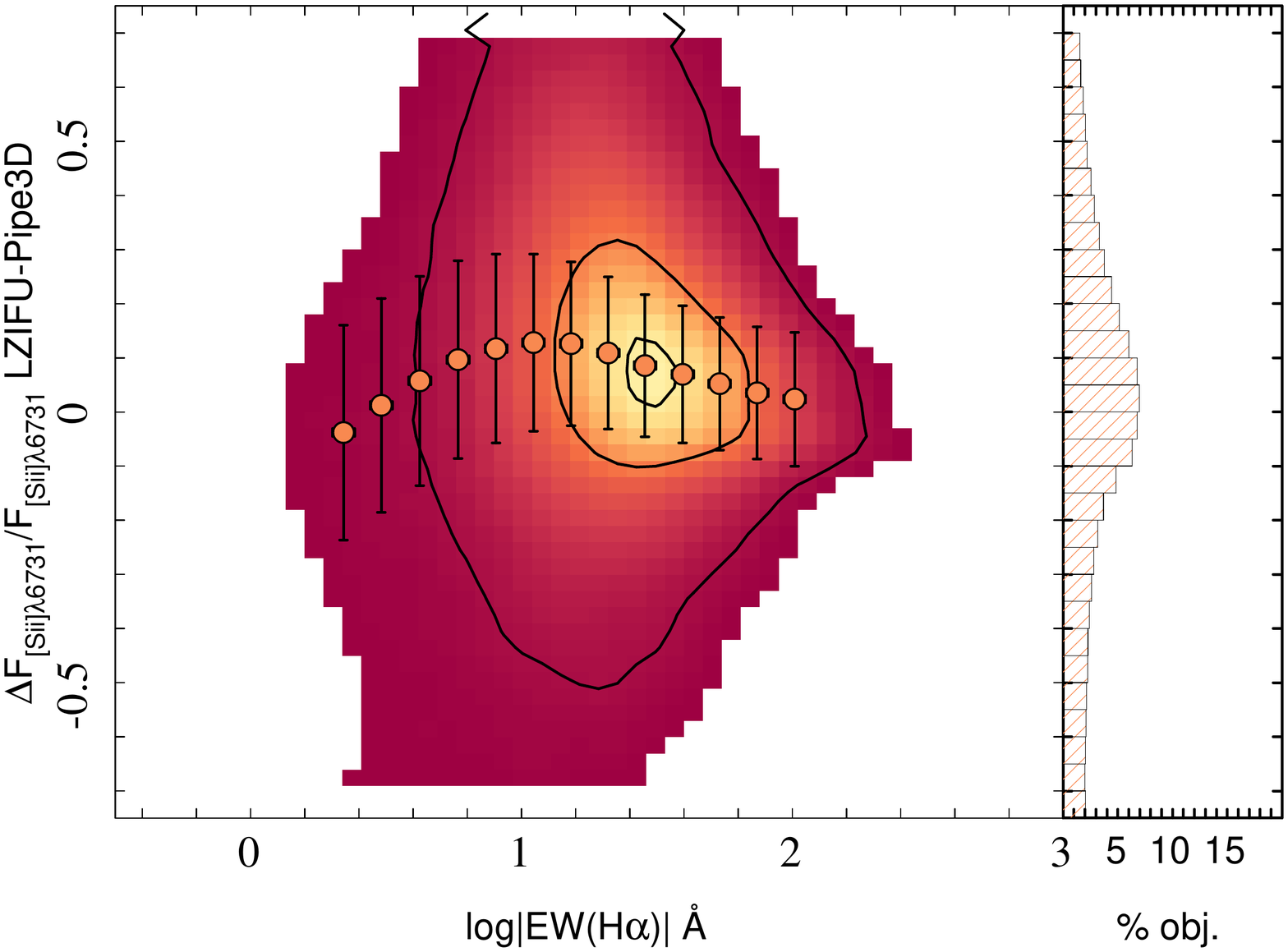}
\endminipage
\caption{Relative difference between the flux intensities derived for the strongest emission lines in the considered wavelength range as measured by LZIFU with respect to {\sc Pipe3D} over the EW(H$\alpha$) for more than 500,000 individual spectra/spaxels analyzed here. Colour maps show the density of points, represented by the black-solid contours (with each contour encircling 95\%, 65\%, 35\% and 20\% of the points). The orange solid circles indicate the average value of the relative difference in bins of 0.15 dex in EW(H$\alpha$), with the error bars indicating the standard deviation with respect to this mean value. For each emission line a histogram  is included showing the distribution of relative differences. From top-left to bottom-right the relative differences are shown for  H$\alpha$, H$\beta$, [\ion{N}{ii}]$\lambda$6583, [\ion{O}{iii}]$\lambda$5007, [\ion{O}{ii}]$\lambda$3737 and [\ion{S}{ii}]$\lambda$6731, as labeled in each panel. } 
 \label{fig:LZIFU}
\end{figure*}

In this article we have used the data-products delivered by {\sc Pipe3D} for the 
current dataset provided by the SAMI survey, a pipeline broadly tested for different
IFU datasets, including the ones provided by the CALIFA and MaNGA surveys 
\citep[e.g.][]{Pipe3D_II,sanchez17,sanchez18} and more complex datasets, like the ones 
provided by MUSE \citep[e.g][]{laura18}. However, the SAMI collaboration has developed 
its own pipeline, LZIFU \citep[][]{ho2016b}, focused on the analysis
of multiple components of the emission lines, taking the advantage of the super-high 
resolution of the SAMI red-arm datasets (although it has been applied to other datasets). 
The approach of LZIFU is slightly different from the one adopted for {\sc Pipe3D}. 
First, the blue and red arms of the SAMI data are combined in order to produce a single 
spectrum. Then, emission lines are masked, and the stellar continuum is fitted using 
{\sc pPXF} \citep{cappellari2017}. As we showed in \citet{Pipe3D_II} the kinematic 
analysis provided by {\sc pPXF} is one of the best, however, the stellar population 
analysis suffers from both the regularization and the multi-polynomial fitting included 
in the analysis. The approach of using {\sc pPXF} to derive only the stellar kinematics 
was adopted by other previous studies \citep[e.g.][]{patri14a,patri14b}. For this analysis 
LZIFU uses the MILES \citep[][]{miles} stellar templates library, with a set of ages 
and metallicities, comprising a total of 75 individual SSPs. 

After subtraction of the underlying stellar population, without a {\it dezonification} 
\citep[the procedure to create a spaxel-wise stellar spectrum model from the spatially 
binned analysis, e.g.][]{cid-fernandes13,Pipe3D_II}, the emission lines are fitted 
using a single or multi-gaussian modeling. The main differences with respect to 
{\sc Pipe3D} is therefore: (1) the adopted fitting tool to analyze the stellar 
population; (2) the adopted SSP-library; (3) the lack of dezonification; (4) the 
lack of an iterative process of fitting the stellar population, unmasked, once the 
model of the emission lines is subtracted; and (5) the addition of multi-Gaussian modeling, 
when required. On top of that, the currently available LZIFU used the SAMI datacubes 
uncorrected for galactic extinction. Therefore, emission line fluxes should be corrected 
for that effect. For all these reasons we have included in here a comparison between 
the emission line fluxes derived using LZIFU and those derived using {\sc Pipe3D}, 
to show the differences.

Figure \ref{fig:LZIFU} shows the comparison between the flux intensities spaxel-by-spaxel 
derived using LZIFU and Pipe3D for the sub-set of emission lines used in the current study, 
for the entire analyzed dataset (i.e., $\sim$2,000 galaxies), showing the relative difference 
in the flux intensities compared to the equivalent width of H$\alpha$, used as a proxy of 
the contrast of the emission lines over the continuum. On average there is a good agreement 
between both measurements, with the standard deviation of the difference ranging between 
25\%\ and 31\%\ for H$\alpha$ and [\ion{S}{ii}]$\lambda$6731, respectively, for all 
the considered values. These differences decrease with the EW(H$\alpha$), being of the 
order of 16\% for most of the considered emission lines for EW(H$\alpha$)$>$30\AA. 
For the stronger emission lines, in particular for H$\alpha$, [\ion{N}{ii}] and 
[\ion{O}{ii}] we found no systematic offset between both analyses, with offsets below 1\%. 
The strongest systematic difference is found for H$\beta$, an emission line strongly 
affected by the subtraction of the underlying stellar population. For this emission line 
the offset is of the order of 7\%, with {\sc Pipe3D} deriving larger values than 
{\sc LZIFU}. A visual inspection of Fig. \ref{fig:LZIFU} clearly shows that the main 
difference is in the regime of the low EWs, i.e, for the retired regions of galaxies, 
that in any case are not considered in the current study.

\begin{figure}
\includegraphics[width=1.0\linewidth,clip,trim=10 30 20 30]{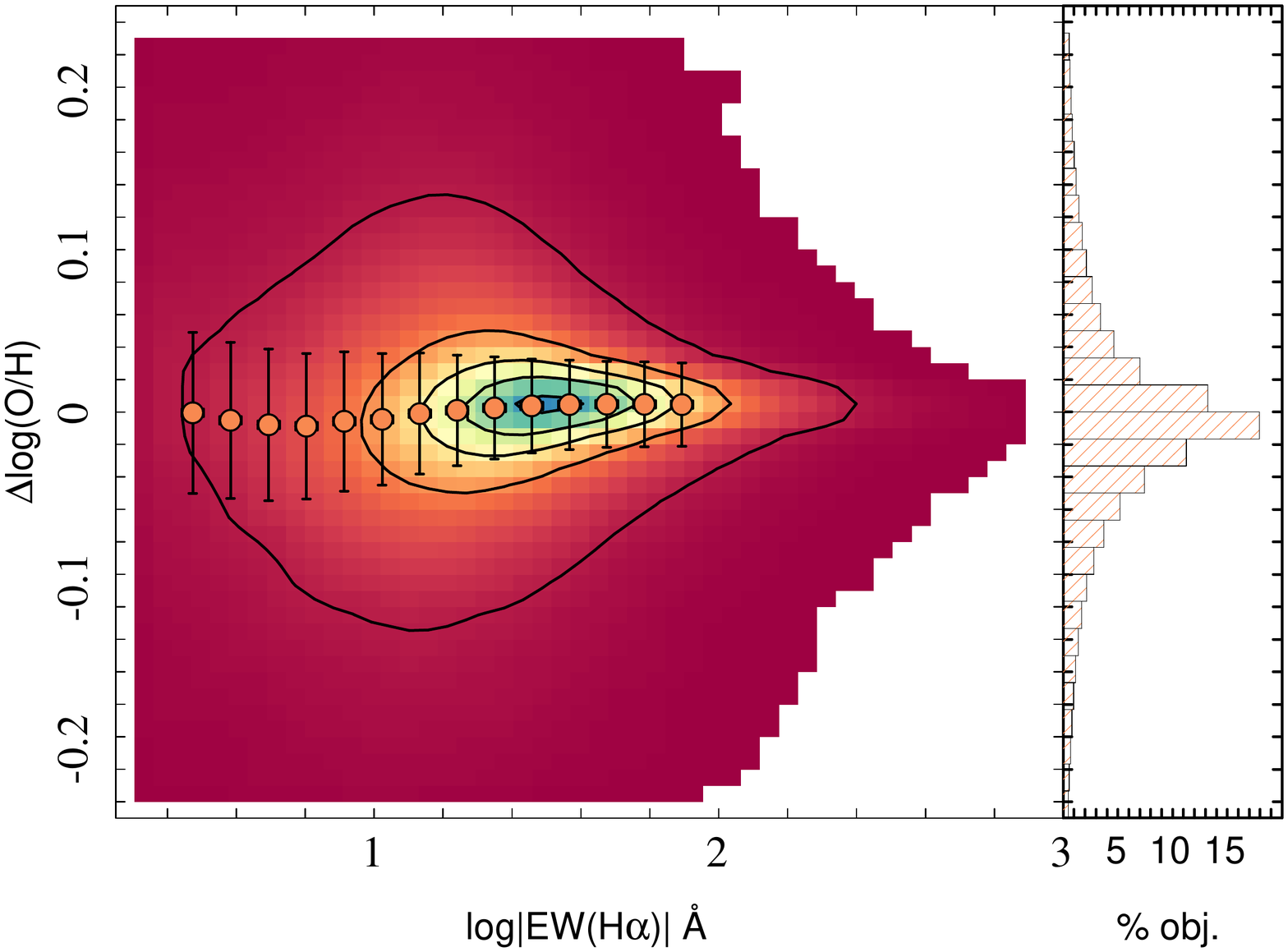}
\caption{Relative difference between the oxygen abundance derived using the O3N2-M13 calibrator based on the emission line intensities measured by LZIFU with respect to those measured by {\sc Pipe3D}, plotted over the EW(H$\alpha$) for the more than 500,000 individual spectra/spaxels analyzed here. Colour map shows the density of points, represented by the black-solid contours (with each contour encircling a 95\%, 65\%, 35\% and 20\% of the points). The orange solid circles indicate the average value of the relative difference in bins of 0.15 dex in EW(H$\alpha$), with the error bars indicating the standard deviation with respect to this mean value. }
 \label{fig:LZIFU_OH}
\end{figure}

The main goal of the current comparison is to determine how the results are affected by 
the use of the dataproducts provided by {\sc Pipe3D} instead of those provided by {\sc LZIFU}. 
Of the different involved parameters in the current analysis the more sensitive one is the 
oxygen abundance. A detailed comparison would require that we repeat the full analysis using 
the new dataset, which it is clearly beyond the scope of the current analysis. However, 
it is still possible to compare the oxygen abundance for the different involved calibrators. 
Figure \ref{fig:LZIFU_OH} illustrates this analysis, showing the difference between the 
oxygen abundances derived using the emission line intensities estimated using {\sc LZIFU} 
and those ones derived using the values provided by {\sc Pipe3D}, for the O3N2-M13 calibrator, 
as a function of the EW(H$\alpha$), for those spaxels with EW(H$\alpha$)$>$3\AA\ 
($\sim$200,000 points). We do not include the remaining calibrators since in all the cases 
we found similar results. Despite the differences reported in the individual emission lines, 
we found a good agreement in the derived oxygen abundances, with an average systematic 
offset of $\sim$0.01 dex, and a scatter of $\sim$0.07 dex. These offsets decrease to 
$\sim$0.005 dex, with a dispersion of $\sim$0.05 dex for EW(H$\alpha$)$>$10\AA, i.e., 
the stronger star-forming regions. These differences are only slightly larger than the 
nominal errors of the considered oxygen abundances in this article ($\sim$0.03 dex), 
and clearly lower than the systematic error of the calibrators. As indicated before we 
found similar results for the remaining calibrators. In summary, we do not consider that 
our results are significantly affected by the use of either the {\sc LZIFU} or the 
{\sc Pipe3D} dataproducts.

\section{The MZ Relation from different IFS galaxy surveys}
\label{sec:comp_IFS}

\begin{figure}
\includegraphics[width=1.0\linewidth,clip,trim=10 10 10 5]{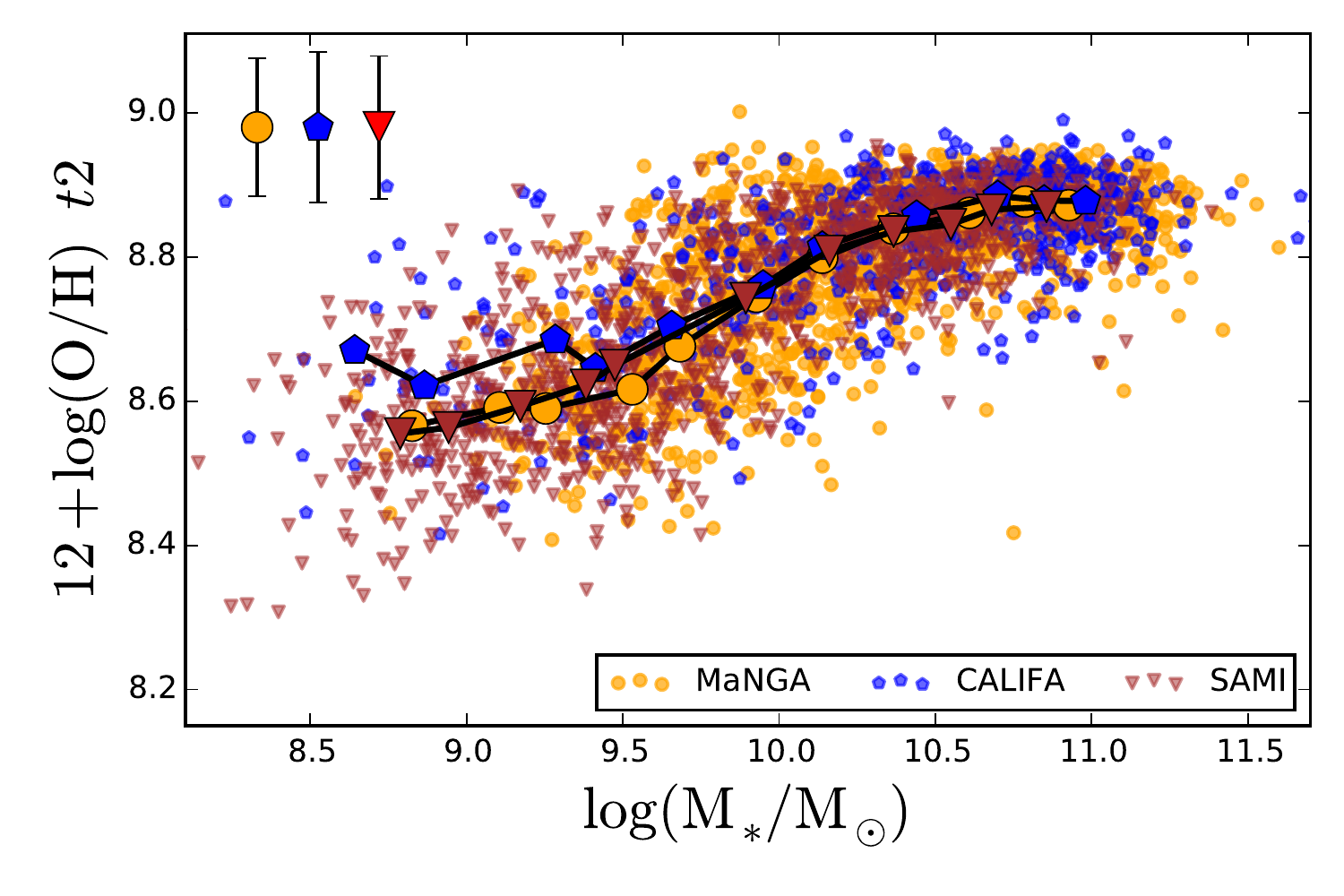}
\caption{Mass-metallicity relation derived using the {\it t2} oxygen abundance calibrator for the 1044 galaxies extracted from the SAMI survey analyzed in this article (red triangles), together with the same distribution for the 1729 galaxies analyzed by \citet{jkbb18}, extracted from the MaNGA survey (orange circles), and the 612 galaxies analyzed by \citet{sanchez18}, extracted from the CALIFA survey (blue pentagons). Line-connected symbols represent median values at a given mass bin for each different dataset, as indicated in the inset, with their average standard deviations represented with an errorbar.}
 \label{fig:comp_IFS}
\end{figure}

We claimed in Sec. \ref{sec:MZ} and subsequent ones that the MZ relations
found using the SAMI galaxy survey dataset are similar to that found by previous 
explorations using other IFS surveys, in particular CALIFA \citep{sanchez17} and 
MaNGA \citep{bb17}. It is beyond the goal of the current study to make a detailed 
comparison between the different results reported using those surveys. However, to 
illustrate how the different datasets compare to each other we show in Figure 
\ref{fig:comp_IFS} the MZ-distribution for the {\it t2} calibrator for the three 
quoted datasets. The values for the CALIFA galaxies were extracted from the published 
table by \citet{sanchez17}, which values were derived using the {\sc Pipe3D} pipeline. 
On the other hand, the MaNGA data were extracted from the publicly available 
{\sc Pipe3D} Value Added 
Catalog\footnote{\url{https://www.sdss.org/dr14/manga/manga-data/manga-pipe3d-value-added-catalog/}}, 
distributed as part of the SDSS-IV DR14 \citep{abol18}. THose are the same date used by 
\citet{bb17}. As already noticed by \citet{sanchez17} the range of stellar masses covered 
by the three surveys is very similar, despite the fact that the redshift range is 
considerably more narrow for the CALIFA survey. On the other hand, this survey is 
incomplete below M$_*<$10$^{9.5}$M$_\odot$ \citep{walcher14,DR3}, a regime much better 
sampled by SAMI than by any of the other three surveys. Despite these differences, 
both the distribution of individual points and the mean values in the difference stellar 
mass bins are remarkably similar, in particular in the mass ranges where the three 
surveys are complete. Therefore, despite the clear differences in the selection of 
targets, redshift ranges, coverage of the galaxies and spatial resolutions, when 
analyzed using the same pipeline they produce very similar results.

Similar distributions are found for the remaining calibrators included in the current 
article. We do not include a plot for each of them for the sake of brevity.

{ 

\section{Comparison with the FMR derived using the SDSS}
\label{sec:SDSS}

\begin{figure*}
\includegraphics[width=0.5\linewidth,clip,trim=0 0 10 5]{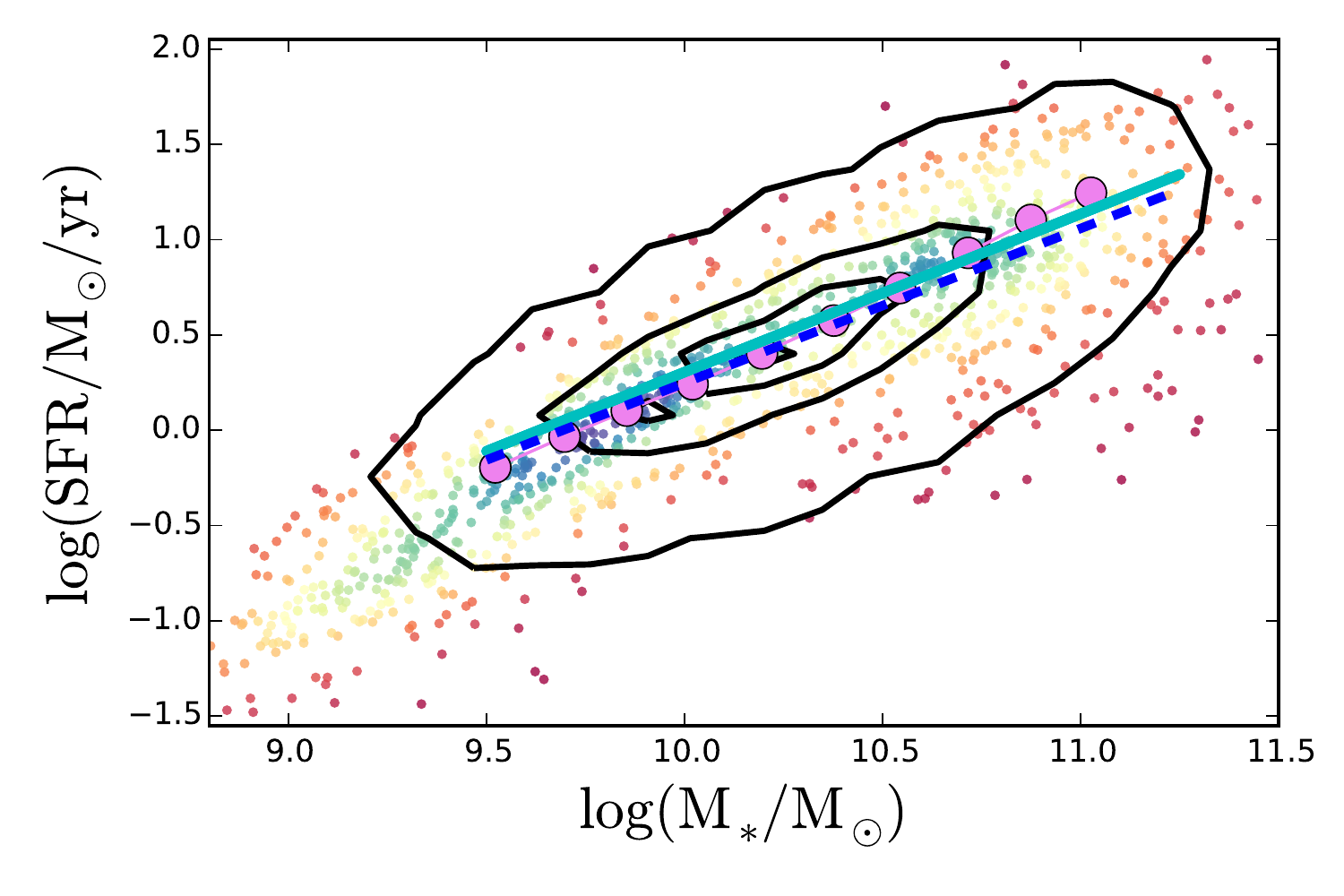}\includegraphics[width=0.5\linewidth,clip,trim=0 0 10 5]{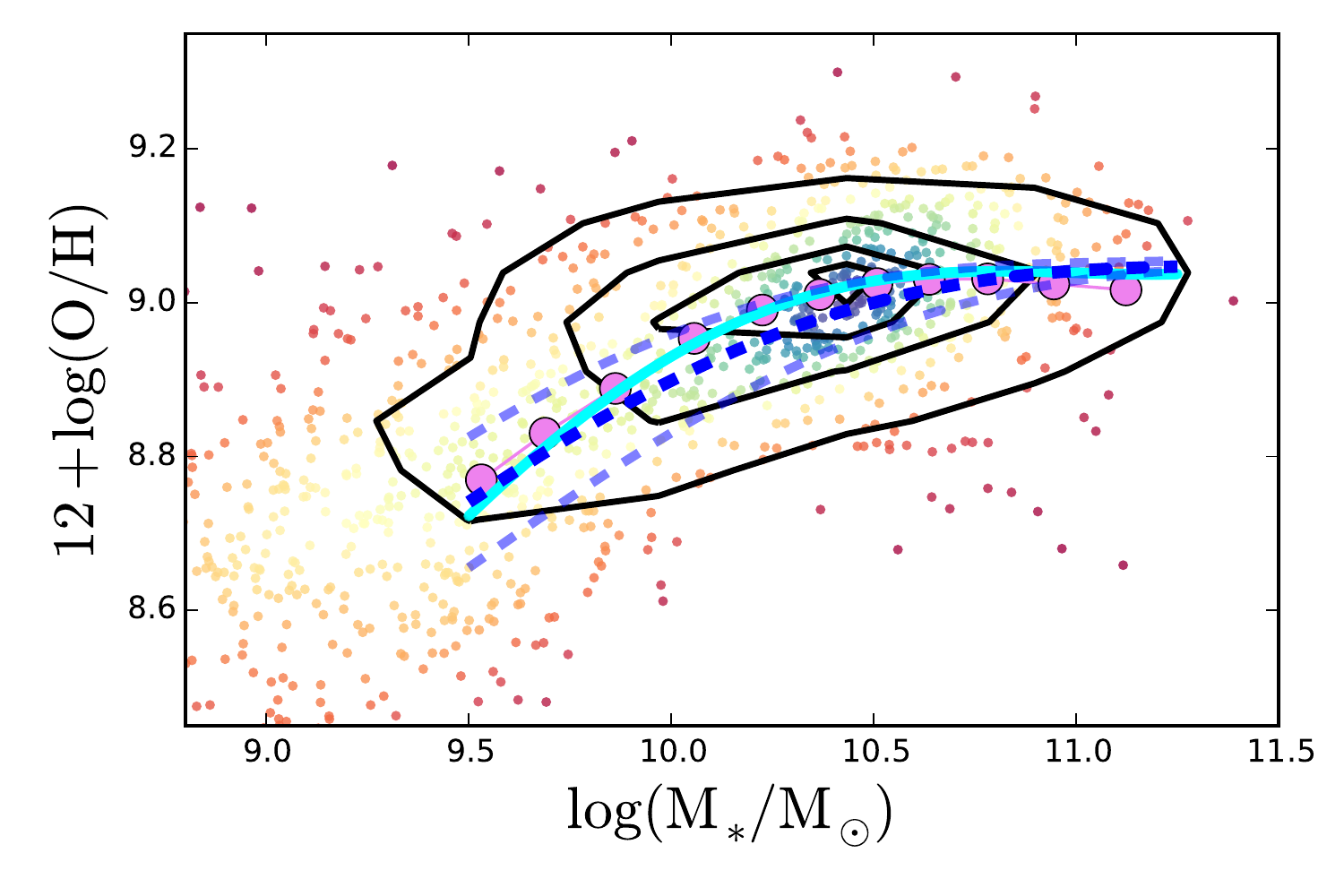}
\caption{ {\it left-panel:} Density contourplot of the distribution along the SFR-M$_*$ diagram of the SDSS galaxies selected following the criteria described in \citet{mann10} (black contours), with each contour encircling 95\%, 60\%, 30\% and 10\% of the points respectively. The SFR and M$_*$ were both extracted from the MPIA/JHU catalog. The violet solid-circles represent the median value of the SFR in bins of M$_*$ of 0.2 dex, ranging between 10$^{9.3}$ and 10$^{11.3}$ solar masses. The cyan solid line corresponds to the best fit linear relation between the two quantities, while the dashed-blue line corresponds to the adopted SFMS along this article \citep[extracted from][]{mariana16}. {\it right-panel:} Density contour plot of the distribution along the M$_*$-O/H diagram for the same galaxies shown in the left panel, with the same level for the contours. The oxygen abundance was derived using the prescriptions described in \citet{mann10}, adopting the emission line intensities extracted from the MPIA/JHU catalog. The violet solid-circles represent the median value of the oxygen abundances in the same bins described for the left panel. The opaque blue-dashed line corresponds to the expected location of the FMR parametrized as described in \citet{mann10} (Eq. 4 of that article) for galaxies following exactly the SFMS, with the two transparent blue-dashed lines indicating the location for galaxies at $\pm$2$\sigma$ the SFMS. The cyan-solid line corresponds to the MZR estimated by \citet{mann10} (Eq. 1 of that article). For comparison purposes, we include the distribution of values for the galaxies observed by the SAMI survey with oxygen abundance derived using the M08 calibrator as solid-circles, whith colour representing the density of points. The stellar masses and SFRs were shifted considering the differences in the IMF and the evolution along the SFR-M$_*$ diagram due to the difference in redshift, following the results by \citet{sanchez18b}.}
 \label{fig:MZ_M10}
\end{figure*}

We compared the distribution of residuals of the oxygen abundances, after 
subtraction of the best fit MZR and pMZR, derived using the adopted calibrators 
for the sample of galaxies studied in this article with that expected adopting 
the FMR as described by \citet{mann10} in Sec.~\ref{sec:residuals}. We did that 
by describing that location in the diagram of residuals as the distribution 
obtained by subtracting the FMR using the adopted parametrization derived by 
those authors, from the distribution obtained adopting the pure MZR 
(Fig. \ref{fig:dMZ} and \ref{fig:dMZ_M}. To justify that this approximation is 
indeed valid, we explore the same distributions for the SDSS galaxies using 
the same selection criteria and calibrators presented in \citet{mann10}.

We selected more than 100,000 galaxies from the SDSS DR7
\citet{abazajian09}, using the values included in the MPA/JHU catalog,
described in \citet{kauffmann03} and \citet{brinchmann04} following
the same criteria listed in \citet{mann10} (Sec. 2.1). Those include
basically (i) galaxies within a redshift range between 0.07 and 0.3;
(ii) with clear detections of the strong emission lines within the
considered wavelength range; (iii) with a limit in the dust
attenuation; (iv) with line ratios compatible with being ionized by
star-formation, within the limitations of single aperture results; and
(iv) with a compatible measurement of the oxygen abundance using two
different calibrators, the one based on the R23 line ratio proposed by
\citet{maio08} (known as M08 in this article), and the one based on
the N2 line ratio proposed by \citet{nagao06}. In summary, we mimic as
much as possible the selection of galaxies and oxygen abundance
derivation described by \citet{mann10}.

Fig. \ref{fig:MZ_M10} shows the distribution of SDSS selected galaxies 
following the criteria outlined before in the SFR-M$_*$ and M$_*$-O/H
diagrams.  In general, the distributions are similar to the ones
described by \citet{mann10} (e.g., their Fig. 1). In addition, we show
the distributions of values extracted from our analysis of the SAMI
galaxies, including only star-forming galaxies (according to their
sSFR limit adopted in Eq. \ref{eq:sSFR_cut}). We shifted
the stellar masses derived for the SAMI survey to consider the
differences in IMF and redshift, following the results by
\citet{sanchez18b}, considering that \citet{mann10} adopted a Chabrier
IMF \citep{chabrier2003}, instead of the Salpeter one adopted in the
current article, and that the average redshift of the SDSS subsample
selected by those authors correponds to the double of that of the SAMI
galaxy survey. For the oxygen abundance, we adopted the values derived
using the M08 calibrator, which are the most similar ones to those derived
by these authors, as indicated before. Once considered those
corrections, both samples present similar distributions in the
considered diagrams. As expected, the SAMI galaxy survey
expands over a wider range of stellar masses, in particular below
10$^{9.5}$, and therefore it samples a regime of lower star-formation
rates and lower oxygen abundances. However, for the range in which they
overlap they present a very similar distribution in both diagrams.

\begin{figure*}
\includegraphics[width=0.5\linewidth,clip,trim=0 0 10 5]{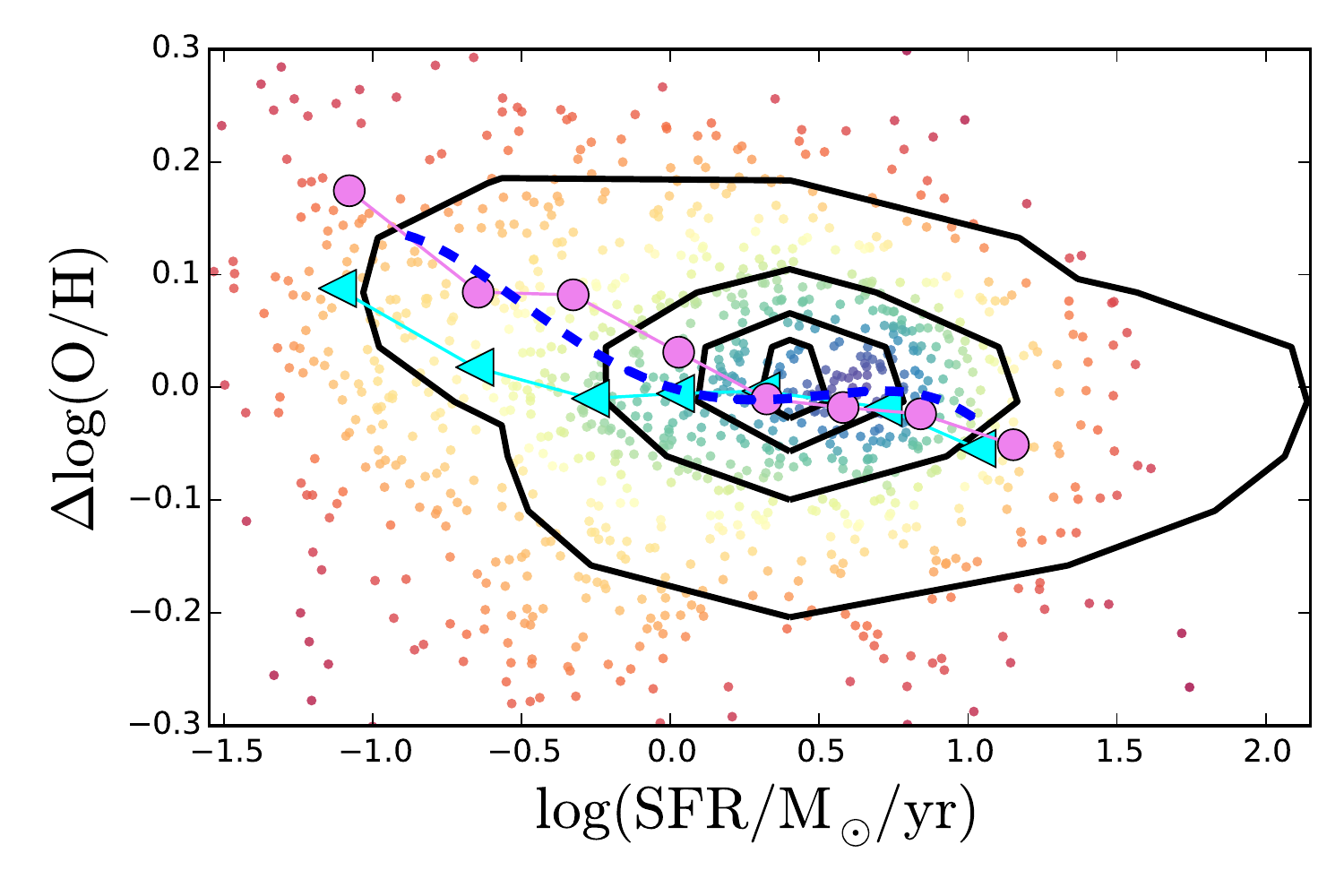}\includegraphics[width=0.5\linewidth,clip,trim=0 0 10 5]{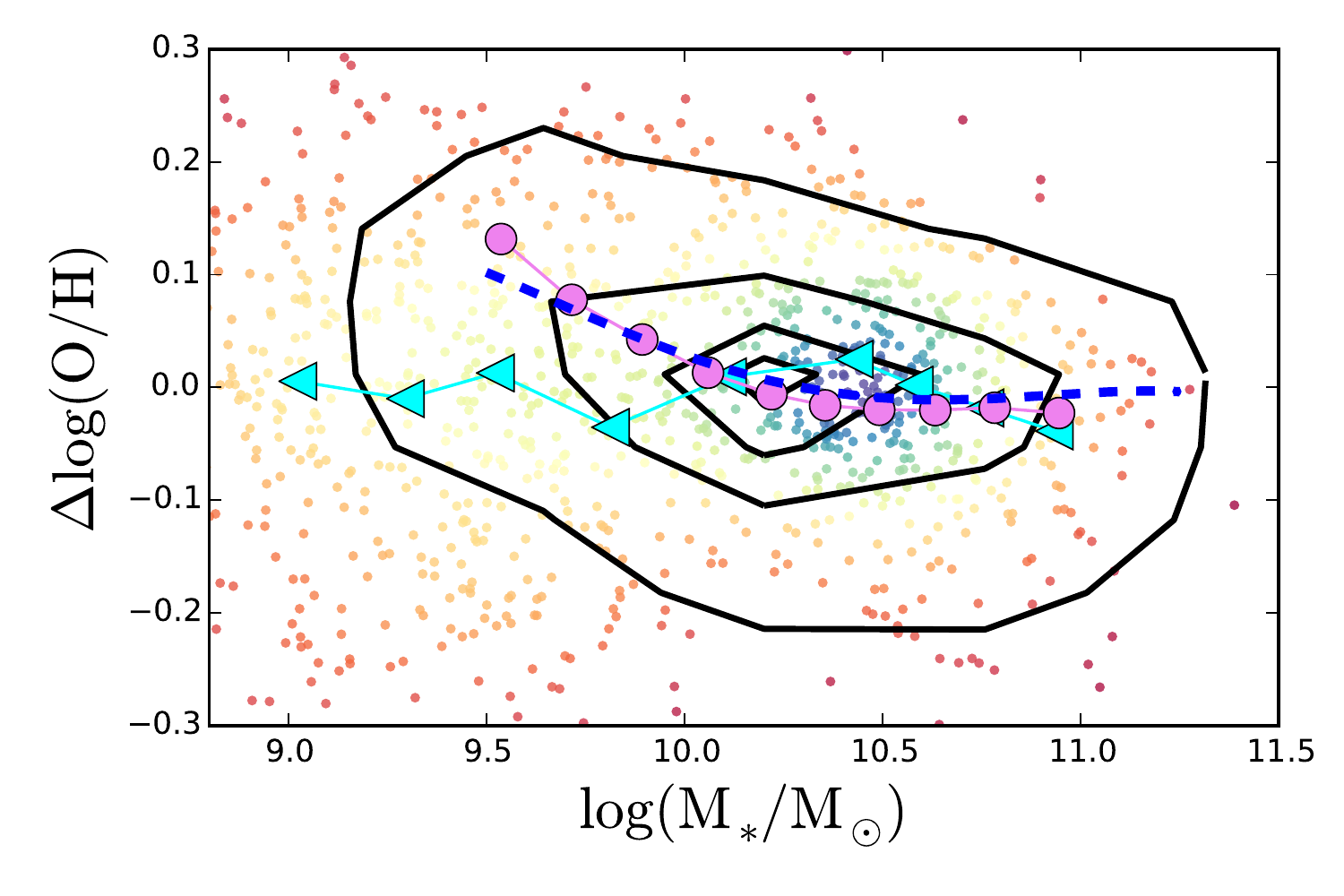}
\caption{ Density contourplot of the distribution residuals of the oxygen abundance, once subtracted the MZR (solid line in Fig. \ref{fig:MZ_M10}, right panel), for the SDSS galaxies selected the criteria described in \citet{mann10} (black contours), as a function of the SFR (left panel) and the stellar mass (right panel). The violet-solid circles corresponds to the median values of the oxygen abundance residuals for the SDSS galaxies in bins of 0.5 dex and 0.2 dex in SFR and stellar masses, respectively. The blue-dashed line corresponds to the expected residuals once subtracted the FMR relation described by \citet{mann10} (dashed-blue line in Fig. \ref{fig:MZ_M10} to the MZR relation described by those authors, for galaxies following exactly the SFMS. It is clearly shown that the blue-dashed line describe very well the average distribution of residuals, represented by the violet-solid line in both panels. For comparison purposes it was included the distribution of values extracted for the galaxies corresponding to the SAMI survey with oxygen abundance measured using the M08 calibrator as solid-circles, with colours indicating the density of points. The blue-solid triangles correponds to the median values of the residuals of the oxygen abundance for these SAMI galaxies (in the case of the stellar masses, the bins have been increased to 0.3 dex, due to the lower number of objects). }
 \label{fig:DMZ_M10}
\end{figure*}

Besides that, we have derived the SFMS (Fig. \ref{fig:MZ_M10}, left
panel, cyan-solid line), and we compare it with the parametrization
adopted in this article (Fig. \ref{fig:MZ_M10}, left panel,
dashed-blue line), finding a very good agreement. Therefore, our
parametrization of the SFMS is fully compatible with the sample
analyzed by \citet{mann10}. In both cases, the dispersion along the
SFMS for star-forming galaxies is of the order of $\sim$0.27 dex, with
a very similar fraction of galaxies with SFR larger than 2$\sigma$, 
that range between $\sim$1.2\% (for the SDSS
subsample) and $\sim$2.5\% (for the SAMI galaxies). 

In the same line, we show that the adopted functional form for the MZR
and FMR, as described by \citet{mann10}, describe well the
distribution of both SDSS and SAMI galaxies within the M$_*$-O/H
diagram. Therefore, they can be used to trace the location of residuals
of the oxygen abundance, after removal of the MZR, the parameter $\Delta log(O/H)$
shown in Fig. \ref{fig:dMZ} and \ref{fig:dMZ_M}. To demonstrate this 
more clearly we show in Fig. \ref{fig:DMZ_M10} the distribution of that
parameter for the considered subsample of SDSS galaxies as a function
of the SFRs and the stellar masses, as a contourplot, together with the same distribution for
the SAMI galaxies adopting the M08 calibrator. In addition we show the
expected location of the residuals characterized by the subtraction of the FMR,
as proposed by \citet{mann10}, to the same functional form assuming no dependence 
on stellar mass, as described in Sec. \ref{sec:residuals}. It is evident
from the figures that the proposed characterization of the effect of the FMR 
in the distrubution of residuals does actually trace the residuals of the SDSS
dataset. Confirming indeed the results by \citet{mann10} for that subsample
of galaxies, when adopting the same selection criteria and calibrator. However,
when comparing with the same distribution for the analyzed SAMI sample, the
dependence with the SFR is much weaker than the one proposed by \citet{mann10}. 
This is particularly evident in the right-panel of Fig. \ref{fig:DMZ_M10}.
Only in the left-panel of Fig. \ref{fig:DMZ_M10} we find a weak trend, as
already discussed in Sec. \ref{sec:residuals}.}

\section{Dataset}
\label{sec:dataset}

The stellar masses, star-formation rates and oxygen abundances for a
sub-set of the 1044 galaxies extracted from the SAMI galaxy survey
analyzed in the present article are listed in Table \ref{tab:dataset}. The
derivation of the different parameters is described in 
Sec. \ref{sec:ana}. For each parameter we include its nominal error,
without taking into account the systematic errors of each calibrator,
although they have been considered in the analysis. We include in the
table the eleven oxygen abundances derived for each galaxy as
described in Sec. \ref{sec:ana}, for those galaxies for which it was
possible to derive this parameter at the effective radius
(characteristic oxygen abundance). For those galaxies for which it was
not possible to derive one of them we marked the value as a {\sc
  nan}. In the case of the oxygen abundances those errors correspond
to those derived by the linear regression over the radial
distributions. Therefore, they are much smaller than the ones usually 
derived from studies based on single aperture spectroscopic data. The
full table for the 1044 galaxies is available in electronic format
at the following HTTP address:
\url{http://132.248.1.15:9154/SAMI/DR_v10/tables/published_table.SAMI.fits}.

\include{published_table_short}




\bibliographystyle{mnras}
\bibliography{main,CALIFAI,sami_dr2}

\end{document}